\documentclass[pra,aps,epsf,epsfig,preprint,nofootinbib,showkeys]{revtex4-2}
\usepackage{mathrsfs}
\usepackage{dcolumn}
\usepackage{bm}
\usepackage{hyperref}
\usepackage{epsfig,graphicx,color,appendix}
\usepackage{multirow}
\usepackage{tabularx}
\usepackage{capt-of}
\usepackage{subfigure}
\usepackage[countmax]{subfloat} 
\usepackage{threeparttable}
\usepackage{amssymb}
\usepackage{array}
\usepackage{amsmath}
\usepackage{booktabs}
\usepackage{longtable}
\usepackage{mathrsfs} 
\usepackage{placeins}

\begin{document}

\title{Spectroscopy and Annihilation Decay Widths of Charmonium and Bottomonium Excitations}
	
\author{Christas \surname{Mony A.}}
\email{christasmony@gmail.com}
\author{Rohit Dhir}
\email[Corresponding author : ]{dhir.rohit@gmail.com}
\affiliation{Department of Physics and Nanotechnology, SRM Institute of Science and Technology, Kattankulathur 603203, India.}
\date{\today}

\begin{abstract}
We study the charmonium ($c\bar{c}$) and bottomonium ($b\bar{b}$) spectra, using Cornell potential with additional constant potential term in a non-relativistic framework, with spin-dependent corrections corresponding to the spin-spin, spin-orbit, and tensor interactions added perturbatively. We predict the masses of low-lying and excited states of charmonium and bottomonium up to $n=6$ and $L=2$. We analyze the radial and orbital Regge trajectories of both the systems and investigate their departure from linearity. Further, we estimate the wave function at the origin and, consequently, predict the decay widths of charmonium and bottomonium states annihilating to leptons and photons. Also, we investigate the effect of scale dependence of the wave function at the origin and the strong coupling constant on the predictions of annihilation decay widths. We compare our predictions of heavy quarkonium spectra and decay widths with experimental results and other theoretical models.
\end{abstract}

\maketitle

\newpage

\section{Introduction}
Heavy quarkonium, composed of a heavy quark and a heavy antiquark, provides an exemplary system to study the interplay of the physics in the perturbative and non-perturbative regimes of quantum chromodynamics (QCD). The analysis of mass spectroscopy and the properties of heavy quarkonia provide a robust test for the proposed theoretical models in these regimes. The precise spectroscopic measurements of the heavy quarkonium states, especially for the low-lying charmonium ($c\bar{c}$) and bottomonium ($b\bar{b}$), have been listed in the particle data group (PDG) \cite{ParticleDataGroup:2022pth}. In addition, rigorous experimental efforts of CLEO-c, BESIII, the B-factories, ATLAS, CMS, and LHCb are ongoing to accumulate data on the decays of heavy quarkonia. In the last decade, multiple observations of the heavy quarkonium spectra, particularly higher excitations and exotics, have provided more challenging problems for physicists \cite{Brambilla:2019esw, Brambilla:2010cs}. Recently, in $2019$, the charmonium resonance $\psi_{3}(3842)$ was observed by LHCb \cite{LHCb:2019lnr} and has properties consistent with the expectations of $\psi_{3}(1^3D_3)$. In $2021$, LHCb observed $X(4630)$ and $\chi_{c1}(4685)$ \cite{LHCb:2021uow}. These states exhibit properties different from conventional charmonia and are currently viewed as exotic (XYZ) states \cite{ParticleDataGroup:2022pth}. Among bottomonium, Belle discovered the $\Upsilon(10753)$ state in $2019$ \cite{Belle:2019cbt}, which is expected to be either a candidate for $\Upsilon(3D)$ state or an exotic structure \cite{ParticleDataGroup:2022pth}. Although a handful of unexpected states (including exotics) have been discovered over the years, many of the excited charmonium and bottomonium states are yet to be observed. Additionally, the decays of charmonium excitations, such as $h_c$, $\chi_{cJ}$ and $\eta_c(2S)$, are currently being studied \cite{BESIII:2021ktv, BESIII:2022olx, BESIII:2022hcv}; further, plans are underway to experimentally study the heavy quarkonium spectra with higher precision \cite{Chapon:2020heu}. It may be noted that these experimental studies provide crucial inputs to formulate theoretical frameworks and calibrate multiple parameters to bridge the gap between theory and experiment.

The production mechanism of heavy quarkonium involves both short and long distance interactions corresponding to perturbative and non-perturbative effects, respectively. However, non-perturbative dynamics in the low energy region are difficult to estimate. The precise experimental observations expose the complexity inherent to charmonium and bottomonium systems, renewing interest in the theoretical studies of the same. Coincidentally, the heavier mass of charm and bottom quarks, along with their smaller velocities inside the bound system, encourages the study of charmonium and bottomonium systems in a non-relativistic framework. The non-relativistic potential model formulations have so far been successful in describing heavy quarkonium spectroscopy, where the potential usually incorporates a Coulomb term at short distances and a linear confining term at large distances. In fact, the Coulomb term also employs the strong coupling constant ($\alpha_s$), which crucially contributes to the mass spectra of heavy quarkonium at leading order. Further, spin-dependent potentials at order $1/m_Q^2$ (where $m_Q$ is the mass of the heavy quark) corresponding to spin-spin, spin-orbit, and tensor interactions are perturbatively included, which in turn leads to the spin-dependent multiplet splittings of the complete mass spectra. Although the mass spectra of charmonium and bottomonium are well explained numerically, the distinction between the perturbative and non-perturbative regimes, due to variation of $\alpha_s$, is barely understood. It is important to note that the heavy quarkonia are well explained by one gluon exchange via the short distance part of the potential, and the non-perturbative contribution is expected to be minimal for low-lying states. Nonetheless, the asymptotic behavior of $\alpha_s$ pose serious challenges. Note that with increasing value of $\alpha_s$ asymptotic freedom diminishes; therefore, at lower energies, non-perturbative corrections become important. In addition, non-perturbative effects also influence the quarkonium wave function, which is nonzero at long distance; however, these effects are non-trivial to calculate. Therefore, understanding of scale dependence of both $\alpha_s$ and wave function of quarkonium is necessary for precise predictions of the mass spectra and properties. Furthermore, the bound system is solved using the Coulomb plus linear confining potential to obtain the wave function, which depends upon potential model parameters including the scale dependent strong coupling. Thus, it is intriguing to explore and better understand the impact of both the strong coupling constant and the wave function estimates on charmonium and bottomonium systems. 

The main focus of the present work is to investigate the scale dependent effects on the decay widths of heavy quarkonium, which is propagated through the wave function at the origin and the strong coupling constant. Beyond the calculation of the mass spectra of charmonium and bottomonium, the strong coupling constant and the estimates of the square of the wave function at the origin ($|\Psi (0)|^2$) also play a decisive role in their annihilation decays. $|\Psi (0)|^2$ being the probability density is necessary to understand the annihilation, and $\alpha_s$ governs the QCD corrections to the decays. Interestingly, the choice of scale for different QCD processes may seem reasonable for one type of process and may be questioned for the other; for example, the mass of the bound system and annihilation decays of quarkonia have different scale dependence. Therefore, we should expect $\alpha_s$ and $|\Psi (0)|^2$ to exhibit different behavior in the decay processes, which is obvious from the discrepancies between the predicted decay widths and experimental observations. 

In this work, we study the mass spectra of charmonium and bottomonium by calculating their excitations up to $D$-wave states and investigate the scale dependence in their leptonic and electromagnetic annihilation decays. We first obtain the spectroscopy of charmonium and bottomonium by adding an extra constant potential term to the Cornell potential in a non-relativistic framework. We perturbatively include the spin-spin, spin-orbit, and tensor interactions to obtain the non-degenerate masses for the heavy quarkonium states. We optimize the numerical values of the input parameters to obtain suitable mass predictions throughout the entire spectrum. We thoroughly assessed the incorporation of the constant term in the potential to achieve better optimization of spectra for both low-lying and excited states, simultaneously. In addition, we use the predicted mass states to analyze the Regge trajectories of both charmonium and bottomonium systems in the radial and orbital planes. Consequently, we look for the departure from the linearity shown by the $c\bar{c}$- and $b\bar{b}$-systems through the obtained Regge trajectories. We also calculate the numerical estimates of the radial wave function of the charmonium and bottomonium states to evaluate the square of the wave function at the origin, and consequently, predict the leptonic and diphoton decay widths of heavy quarkonia for both at lowest order and with QCD corrections. Aforementioned, we explore the possible scale dependence of $\alpha_s$ and $|\Psi (0)|^2$ on these decay processes. We study the effects of scale dependence on the leptonic and diphoton decay widths of the corresponding charmonium and bottomonium states. Subsequently, we provide estimates of the scale dependent factor to improve the agreement between the theoretical decay width predictions and the experimental results. Throughout our analyzes, we compare our predictions with those of other theoretical works.

The paper is arranged as follows. In Sec.~\ref{sec:2_methodology}, we provide the methodology for the calculation of the mass spectra, Regge trajectories, wave function at the origin, annihilation decay widths, and scale dependent factor. In Sec.~\ref{sec:3_ND}, we discuss our numerical results. We summarize and present our conclusions in the last section.
\section{Methodology}
\label{sec:2_methodology}
\subsection{Potential}
In the present work, we study the $c\bar{c}$ and $b\bar{b}$ bound states, collectively known as heavy quarkonium, in the non-relativistic potential model approach. The heavy quarkonium state commonly exhibits an interquark potential that simulates Coulombic characteristics over short distances and linear behavior over longer distances in order to account for the confinement of the quarks. We use the most generic form of the potential referred to as the Cornell potential given as \cite{Eichten:1974af, Eichten:1979ms}
\begin{equation}
\label{eqn:1_cornel_pot}
    V(r) = -\frac{4\alpha_s}{3r} + \sigma r + V_c ;
\end{equation}
where $\alpha_s$ (strong coupling constant) and $\sigma$ (string tension) are phenomenological parameters. The constant $V_c$ is a dimensional quantity corresponding to the normalization of the low-lying as well as excited energy levels of the $Q\overline{Q}$ system. The potential model approach is established on the phenomenological depiction of constituent quark interaction, which is non-relativistic in nature. This representation works well even for light quarks; however, it can only be justified from fundamental principles when dealing with heavy quarks. The rationale behind adopting a non-relativistic approach is that one can consider interactions among constituents as resulting from potential when their typical energy is significantly smaller than their relative momentum. Therefore, their respective masses should be larger as compared to $\Lambda_{QCD}$ (the QCD confinement scale). Since $m_Q$ supersedes $\Lambda_{QCD}$, $m_Qv$, and $m_Q v^2$ scales, the potentials can be expressed through an expansion corresponding to $1/m_Q$~\cite{Brambilla:2019esw}.

We proceed to calculate the energy eigenvalues of the bound system by solving the Schrodinger wave equation, which is given in its most general form as $[T + V]\Psi = E\Psi$, where $T$ is the kinetic energy of the constituents and using the potential $V$ as given in Eq.~\eqref{eqn:1_cornel_pot}. Naively, the total wave function $\Psi(\Vec{r})$ is written as radial and angular components, $\Psi(\Vec{r}) = R_{nl}(r) Y_{lm}(\theta,\phi)$, where $R_{nl}(r)$ is the radial wave function, $Y_{lm}(\theta,\phi)$ denotes the spherical harmonics, and the symbols $n$ and $l$ have their usual meaning. It may be noted that $R_{nl}(r)$ is usually represented in terms of $u_{nl}(r)$, the reduced radial wave function, with $R_{nl}(r) = u_{nl}(r)/r$ and is normalized as $\int_0^\infty |R_{nl}(r)|^2 r^2 dr = \int_0^\infty |u_{nl}(r)|^2 dr = 1$. These substitutions lead to the Schrodinger radial wave equation for a bound system given by
\begin{equation}
\label{eqn:2_radial_SE}
    u''_{nl}(r) + 2 \mu \bigg[ E_{nl} - V(r) - \frac{l(l+1)}{2\mu r^2} \bigg] u_{nl}(r) = 0.
\end{equation}
$\mu$ is the reduced mass of the quarks and $E_{nl}$ is the binding energy of the bound state. We solve Eq.~\eqref{eqn:2_radial_SE} numerically using the Runge-Kutta method. Thereby, we obtain the binding energy $E_{nl}$, from which the degenerate mass of the bound state can be obtained as $M = m_Q + m_{\bar{Q}} + E_{nl}$, where $m_Q(m_{\bar{Q}})$ is the mass of the heavy quark (antiquark).

The potential given in Eq.~\eqref{eqn:1_cornel_pot} does not include the spin of the heavy quarkonium state. Consequently, the masses obtained from Eq.~\eqref{eqn:1_cornel_pot} for the bound states are degenerate. Therefore, it is mandatory to include spin dependence in order to break the degeneracy seen in masses. As prior mentioned, the heavy mass $m_Q$ is much larger than characteristic scales and, thus, allows $1/m_Q$ expansion of the potential. However, in non-relativistic framework, $ 1/m_Q$ components of the spin-dependent potential are suppressed due to the heavy quark spin symmetry \cite{Brambilla:2019esw}. Nevertheless, spin-dependent corrections of the order $1/m_Q^2$ namely the spin-spin, spin-orbit, and tensor interactions can be added as perturbations to $E_{nl}$ \cite{DeRujula:1975qlm, Godfrey:1985xj, Kwong:1987mj, QuarkoniumWorkingGroup:2004kpm}.

The spin-spin interaction contributes to hyperfine splitting and is given by
\begin{equation}
    V_{SS} = \frac{32\pi\alpha_s}{9m_Q^2}\textbf{S}_1.\textbf{S}_2 \delta(r) \label{eqn:3_spin-spin},
\end{equation}
with the analytical expression for the spin-spin operator as,
\begin{equation}
    \langle \textbf{S}_1.\textbf{S}_2 \rangle = \frac{1}{2} s(s+1) - \frac{3}{4}.
\end{equation}
We choose the Dirac delta function, $\delta(r)$, in $V_{SS}$ term to be given as the smeared Gaussian 
\begin{equation}
\label{eqn:5_smearfn}
    \delta(r) = \Big(\frac{\rho}{\sqrt{\pi}}\Big)^{3} e^{-\rho^2r^2},
\end{equation}
where $\rho$ (the smearing constant) is a phenomenological parameter that governs the shape of the overlap and Eq.~\eqref{eqn:5_smearfn} incorporates the relativistic correction of $\mathcal{O}(v^2/c^2)$ as stated in Ref.~\cite{Barnes:2005pb}.

In addition to spin-spin interaction, there are spin-orbit and tensor interactions that contribute only to the orbitally excited states with $L > 0$. The spin-orbit term is given as
\begin{equation}
    V_{LS} = \frac{2\alpha_s}{m_Q^2r^2}\textbf{L.S} - \frac{\sigma}{2m_Q^2r}\textbf{L.S} \label{eqn:6_spin-orbit},
\end{equation}
where the spin-orbit operator is analytically stated as
\begin{equation}
    \langle \textbf{L.S} \rangle = \frac{1}{2} \Big[j(j+1) - l(l+1) - s(s+1)\Big].
\end{equation}
The tensor term is 
\begin{equation}
    V_{T} = \frac{\alpha_s}{3m_Q^2r^3}S_{12}, \label{eqn:8_tensor}
\end{equation}
where the tensor operator of the $Q\overline{Q}$ bound state $S_{12}$ is defined as 
\begin{equation}
    S_{12} = 2(\textbf{S}_1 \cdot \hat{r}) (\textbf{S}_2 \cdot \hat{r}) - (\textbf{S}_1 \cdot \textbf{S}_2)
\end{equation}
and contributes only for $L>0$. This is evident from its analytical expression,
\begin{equation}
    \langle S_{12} \rangle = \frac{4}{(2l + 3)(2l - 1)} \Bigg[ s(s+1)l(l+1) - \frac{3}{2}\langle\textbf{L.S}\rangle - 3(\langle\textbf{L.S}\rangle)^2 \Bigg].
\end{equation}
In the above expressions, the symbols $l, s$, and $j$ are the quantum numbers of $Q\overline{Q}$ state. Eqs.~\eqref{eqn:3_spin-spin}, \eqref{eqn:6_spin-orbit}, and \eqref{eqn:8_tensor} are treated as perturbations and solved to break the degeneracy, resulting in the mass spectra of heavy quarkonium. After obtaining the masses we proceed to analyze the Regge trajectories of the charmonium and bottomonium states.

Regge theory \cite{Regge:1959mz, Regge:1960zc} is a widely used approach to study the hadron spectra. Here, the analytical properties of the scattering amplitude are studied by considering unphysical complex angular momentum $l$. The singularities of the amplitude in the complex $l$ plane are termed Regge poles, which correspond to bound states or resonances for physical angular momenta. These Regge poles generally arise in the $t$ channel; hence, the Mandelstam variable $t$ is used. Interestingly, the Regge trajectory $\alpha(t)$ is expected to be linear in $t$: $\alpha(t) = \alpha(0) + t \alpha'$. Thus, in a Regge trajectory corresponding to physical particles and resonances, the square of mass of a hadron can be linearly related to its angular momentum. It is also worth noting that, the slope of the Regge trajectory is inversely proportional to the string tension, which is related to the linear part of the potential. The fact that mesons and baryons lie on linear trajectories on the radial and orbital planes, with the hadrons differing only in their masses and radial (or orbital) excitation occupying the same trajectory, makes Regge trajectories well suited for hadron classification. In general, the  linear Regge trajectory relating the mass of a hadron ($M$) with its orbital angular momentum ($l$) and radial quantum number ($n_r$) can be expressed as
\begin{equation}\label{eqn:11_regge_linear}
    M^2 = \alpha_{lin} l + \beta_{lin} n_r + C_{lin},
\end{equation}
where $\alpha_{lin}$ and $\beta_{lin}$ are the Regge slopes and $C_{lin}$ is a constant. 

However, it has been suggested by various authors \cite{Martin:1986rtr, Lucha:1991vn, Chen:2018hnx} that the hadron Regge trajectories may not always follow a linear pattern. For example, in Refs.~\cite{Martin:1986rtr, Lucha:1991vn}, the authors note that the use of a linear potential in non-relativistic treatments does not necessarily lead to a linear Regge trajectory. Similarly, in the relativistic formulation of Ref.~\cite{Chen:2018hnx}, Regge trajectories are obtained in the parameterized form
\begin{equation} \label{eqn:12_regge_chennonlinear}
    M^2 = \beta {(c_l l + \pi n_r + c_0)}^{2/3} + c_1 ,
\end{equation}
where $\beta$ and $c_l$ are universal parameters,  $c_0$ and $c_1$ vary with different trajectories. We employ the above equation to study the non-linear behavior of Regge trajectories along with Eq.~\eqref{eqn:11_regge_linear} for the linear case. To fit the various parameters in Eqs.~\eqref{eqn:11_regge_linear} and \eqref{eqn:12_regge_chennonlinear} we utilize the iminuit Python package \cite{iminuit} based on the MINUIT algorithm \cite{James:1975dr}. Subsequently, we analyze the linear and non-linear Regge trajectories acquired in both the radial and orbital planes, which are discussed in Sec.~\ref{subsubsec:A3_ND_Regge}.

\subsection{Square of Radial Wave Function at the Origin}

The square of radial wave function at the origin $|R_{nl}^{(l)}(0)|^2$ is a crucial quantity for studying many of the physical properties of charmonium and bottomonium. The square of $Q\overline{Q}$ wave function at zero separation is related to the point where the particles must coincide if they are to annihilate. Thus, this quantity is present as a factor in all processes. Say, for the vector meson $Q\overline{Q}$ state to annihilate into lepton pairs via a photon, the decay width will be directly proportional to $|R_{nS}(0)|^2$, a probability density with dimensions of (length)${}^{-3}$ or (mass)${}^{3}$. Thus, $|R_{nl}^{(l)}(0)|^2$ is given in units of GeV${}^{3+2l}$ \cite{Close:1979bt, LeYaouanc:1988fx}. However, $|R_{nl}^{(l)}(0)|^2$ being a dimensional quantity is difficult to estimate numerically. The dimensional dependence of the square of wave function at the origin makes it sensitive to the choice of parameters of the potential and the quark masses \footnote{Note that for a potential $V(r) \sim r^N$ the $|R_{nS}(0)|^2$ scales as $m_Q^{3/2+N}$ \cite{Close:1979bt, LeYaouanc:1988fx}.}. Therefore, a change in the values of the quark masses (corresponding to their flavors) or the potential will lead to different estimates of the wave function at the origin. These complexities and the lack of consensus on the estimates of the square of radial wave function at the origin makes it intriguing and worthwhile to investigate.

In this work, the radial wave function $R_{nl}(r)$ is estimated along with the energy eigenvalue $E_{nl}$, when solving the radial Schrodinger equation. Using which, its value at the origin 
\begin{equation}
\label{eqn:13_radial_wfo}
    R_{nl}^{(l)}(0) = \frac{d^lR_{nl}(r)}{dr^l}\bigg|_{r=0},
\end{equation}
can be determined for $l = 0$ and $l \ne 0$ cases. To estimate $|R_{nS}(0)|^2$ for the $S$-wave case ($l = 0$), we extrapolate the numerical values of the radial wave function to a region near $r = 0$ and then square it. Alternatively, the same can be calculated from the expression~\cite{Lucha:1991vn},
\begin{equation}
\label{eqn:14_wfo_swave}
    |R_{nS}(0)|^2 = m_Q\langle V'(r) \rangle,
\end{equation}
where $m_Q$ is the mass of the heavy quark and $\langle V'(r) \rangle$ is the expectation value of the derivative of the potential. Both procedures to obtain $|R_{nS}(0)|^2$ are reliable and are employed together as a check. The total wave function is related to the radial wave function at origin, for $l = 0$ states, as $ \Psi(0) = \frac{R_{nS}(0)}{\sqrt{4\pi}}$.

For $l \ne 0$, the first non-vanishing derivative of the normalized eigenfunction is directly extrapolated to a region near $r = 0$. In the case of $P$-wave with $l = 1$, the first derivative of the radial wave function is extrapolated near $r = 0$ and then squared to obtain $|R'_{nP}(0)|^2$. Similarly, for the $D$-wave with $l = 2$, the second derivative of the radial wave function is extrapolated and subsequently squared to obtain $|R''_{nD}(0)|^2$.

In addition to this, a challenge to the clear estimation of the wave function at the origin is its close relationship with the strong coupling constant ($\alpha_s$). Notably, these quantities appear together even in the determination of bound state masses through the correction terms. Among them, the spin-spin interaction term stands to be more dominant and is the only term that contributes to the $S$-wave states. Furthermore, the mass difference between the vector meson state ($M_V$) and the pseudoscalar meson state ($M_P$) is given by
\begin{equation}
    M_V - M_P = \frac{32\pi}{9m_Q^2}\alpha_s|\Psi(0)|^2. \label{eqn:15_masssplitting_asnwfo}
\end{equation}
Thus, from the above expression, it is seen that both quantities in question have to be varied simultaneously or individually (with the other fixed) in order to obtain the necessary mass splitting. Further, we know that $\alpha_s$ affects the wave function due to its occurrence in the potential. Therefore, $\alpha_s$ is fixed and the wave function at origin is estimated such that both the mass of respective states and their corresponding splittings are reliable. Additionally, since $\alpha_s$ is also scale dependent, it contributes to the scale dependence in wave function estimates. We provide our estimates of $|R_{nl}^{(l)}(0)|^2$ in units of GeV${}^{3+2l}$ and discuss its behavior in Sec.~\ref{subsec:B_ND_wfo}.
\subsection{Annihilation Decay Widths}
The annihilation decay widths of quarkonia can be computed using the estimates of $|R_{nl}^{(l)}(0)|^2$ in GeV${}^{3+2l}$, as stated earlier. In this work, we specifically examine the annihilation of charmonium and bottomonium into leptons and photons.

\subsubsection{Leptonic Decay Widths}
The quarkonium vector states annihilate into lepton pairs through a single virtual photon. The expressions for the leptonic decay width, along with the first-order QCD corrections, of $S$- and $D$-wave quarkonia are given as \cite{VanRoyen:1967nq, Kwong:1987ak, Bradley:1980eh, Segovia:2016xqb}

\begin{equation}
\label{eqn:16_leptonic_S}
    \Gamma({n}^3S_1 \rightarrow e^+e^-) = \frac{16\pi e^2_Q \alpha^2}{M_{nS}^2}|\Psi(0)|^2 \big( 1 - \frac{16\alpha_{sc}}{3\pi} \big),
\end{equation}

\begin{equation}
\label{eqn:17_leptonic_D}
    \Gamma({n}^3D_1 \rightarrow e^+e^-) = \frac{25 e^2_Q \alpha^2}{2 m_Q^4 M_{nD}^2}|R''_{nD}(0)|^2 \big( 1 - \frac{16\alpha_{sc}}{3\pi} \big),
\end{equation}
where $e_Q$ ($m_Q$) is the charge (mass) of the quark, $\alpha$ and $M_{nL}$ are the fine structure constant and the mass of the decaying state, respectively. $\alpha_{sc}$ is essentially the strong coupling constant. The decay widths for lowest order (without QCD corrections) can be obtained by setting $\alpha_{sc} = 0$ in the above expressions. As mentioned before, the scale dependence of the strong coupling constant can be crucial for the evaluation of masses and decay widths. Therefore, $\alpha_{sc}$ should be considered at the mass scale of the charm and bottom quarks for the annihilation decays of their respective states. However, it is difficult to estimate the strong coupling constant at these scales. Nevertheless, we observe a range of numerical values for the strong coupling constant corresponding to charm and bottom quark mass scale in the literature \cite{Kwong:1987ak, McNeile:2010ji, Bazavov:2014soa, Chakraborty:2014aca, Maezawa:2016vgv, Petreczky:2019ozv, Komijani:2020kst}. Based on these available estimates from lattice QCD (LQCD), we fix $\alpha_{sc} = 0.3$ for charmonium and $\alpha_{sc} = 0.2$ for bottomonium at their respective mass scales. Utilizing these values of $\alpha_{sc}$ in Eqs.~\eqref{eqn:16_leptonic_S} and \eqref{eqn:17_leptonic_D}, we obtain the decay widths of charmonium and bottomonium states inclusive of the QCD corrections\footnote{For a better comparison of the scale dependence of $\alpha_s$, one can also extract it from the available experimental information. We found that the ratio of the experimentally determined decay widths of $\Gamma({1}^3S_1 \rightarrow ggg)$ and $\Gamma({1}^3S_1 \rightarrow e^+e^-)$, favors smaller value of $\alpha_s$ than employed in the case of masses in various models.}.

We would like to point out that the product of $\alpha_{sc}$ and $|\Psi(0)|^2$, which appears in the decay width expressions, is usually treated as constant. Nevertheless, we treat them independently. We are aware that $|\Psi(0)|^2$ itself has a scale dependent contribution corresponding to that of the $\alpha_{s}$ used in the potential. However, it could be seen as a correction to the constant product of $\alpha_{sc}$ and $|\Psi(0)|^2$, which is required for understanding the experimental results. Further, a second correction can come from the experimental results of the decay widths to determine $|\Psi(0)|^2$ while treating $\alpha_{sc}$ constant at a particular scale. We used the experimentally available value of $\Gamma({1}^3S_1 \rightarrow e^+e^-)$ to estimate the $|\Psi_{expt}(0)|^2$ for $1^3S_1$ state of charmonium and bottomonium. A ratio of the experimentally obtained $|\Psi_{expt}(0)|^2$ to the theoretical $|\Psi_{theo}(0)|^2$ would then allow us to get the correction factor denoted by $f$, \textit{i.e.},
\begin{equation}
\label{eqn:18_f_flav_fac}
    f = \frac{\Gamma_{expt}({1}^3S_1 \rightarrow e^+e^-)}{\Gamma_{theo}({1}^3S_1 \rightarrow e^+e^-)} = \frac{|\Psi_{expt}(0)|^2}{|\Psi_{theo}(0)|^2}.
\end{equation}
This factor could also be scale dependent, corresponding to the scale at which $|\Psi(0)|^2$ is determined. We expect that the scale dependent correction factor $f$ can improve our predictions for the decay widths in the corresponding charmonium and bottomonium systems. We have discussed these corrections in detail in Sec.~\ref{subsec:C_ND_Anni}.

\subsubsection{Diphoton Decay Widths}

The diphoton decay width formulae with first-order QCD corrections for pseudoscalar (${}^1S_0$), scalar (${}^3P_0$), and tensor (${}^3P_2$) quarkonia states are expressed as \cite{Kwong:1987ak}

\begin{equation}
\label{eqn:19_diphoton_S}
    \Gamma({n}^1S_0 \rightarrow \gamma\gamma) = \frac{12\pi e_Q^4 \alpha^2} {m_Q^2}|\Psi(0)|^2 \big( 1 - \frac{3.4\alpha_{sc}}{\pi} \big),
\end{equation}
\begin{equation}
\label{eqn:20_diphoton_3P0}
    \Gamma({n}^3P_0 \rightarrow \gamma \gamma) = \frac{27 e_Q^4 \alpha^2} {m_Q^4}|R'_{nP}(0)|^2 \big( 1 + \frac{0.2\alpha_{sc}}{\pi} \big),
\end{equation} 
\begin{equation}
\label{eqn:21_diphoton_3P2}
    \Gamma({n}^3P_2 \rightarrow \gamma \gamma) = \frac{36 e_Q^4 \alpha^2} {5m_Q^4}|R'_{nP}(0)|^2 \big( 1 - \frac{16\alpha_{sc}}{3\pi} \big),
\end{equation} 
where the quantities have their usual meaning. Similar to the leptonic decay of $S$-wave states, we have studied the corrections pertaining to the strong coupling constant and the square of the wave function in these decays in the next section.

\section{Numerical Results and Discussion}
\label{sec:3_ND}

Employing the methodology outlined in the previous section, we computed the masses and square of radial wave function at the origin for the ground and excited states of charmonium and bottomonium. First, we solved the Schrodinger radial equation with the Cornell potential for $nL$ states of charmonium and bottomonium up to $n = 6$ and $L = 2$. Thereby, we obtained the energy eigenvalues to predict the degenerate masses of the states, and subsequently, obtained the radial wave function. Further, we estimate the square of radial wave function at the origin. As mentioned before, the Cornell potential alone is insufficient to explain the hyperfine splitting between the respective $S$-, $P$-, and $D$-wave states; thus, we added the spin-dependent correction terms perturbatively. The inclusion of spin-dependent correction terms break the degeneracy in the masses of the ground and excited heavy quarkonium states. Furthermore, we studied the radial and orbital Regge trajectories for the ground and excited states of charmonium and bottomonium. Later, we focused on the predictions of the leptonic and diphoton annihilation decay widths of these states, where we included scale dependent effects through the strong coupling constant and the square of the wave function at the origin.

In this work, we focused on predictions of the masses of the charmonium and bottomonium excited states and reliable estimation of the wave function at the origin to predict their annihilation decay widths. As stated before, the strong coupling constant plays a crucial role in predicting the masses of states and their hyperfine splittings. Along with $\alpha_s$, the values of the other parameters $\sigma$, $\rho$, and $V_c$ are optimized in order to improve our numerical predictions. The string tension $\sigma$ constrains the mass of the bound state, which is more prominently seen in the higher excited states. The parameter $\rho$ in the spin-spin interaction term is adjusted to govern the hyperfine splitting between the states. $V_c$ is used to normalize the energy levels corresponding to low-lying and excited states, simultaneously, which compensates for the scale dependence of various terms in the potential. The masses of the quarks are initialized at half of the ground state mass. The finely optimized values for the parameters in both flavor systems are listed in Table~\ref{tab:1_parameters}. It may be noted that these parameters are also used in multiple other approaches available in the literature. We notice that the parameter values are observed to span a range, \textit{e.g.}, $\alpha_s$ ranges from $0.33 - 0.55$ for charmonium \cite{Radford:2007vd, Barnes:2005pb, Lakhina:2006vg} and $0.23 - 0.37$ for bottomonium \cite{Soni:2017wvy, Li:2009nr}. In addition, for most of the cases, the value of $\alpha_s$ is either computed based on a renormalization scale or is fixed based on the available experimental masses. The string tension ($\sigma$) falls in the range $0.14 - 0.25$ GeV${}^2$ \cite{Barnes:2005pb, Soni:2017wvy}. However, the input masses for charm and bottom quarks follow a stricter range $1.3 - 1.6$ GeV \cite{Soni:2017wvy, Barnes:2005pb} and $4.4 - 4.9$ GeV \cite{Li:2009nr, Kher:2022gbz}, respectively. Note that our optimized inputs for the parameters lie comfortably within the observed ranges, with a few exceptions. Furthermore, we found that the values of our input parameters ($m_Q$, $\alpha_s$ and $\sigma$) are similar to some of the other models, which ensures the reliability of our approach. We follow $n^{2S+1}L_{J}$ notation to represent the states, where $n$ is the principal quantum number and is related to the radial quantum number $n_r$ as $n = n_r + 1$ \cite{Griffiths:2008zz}.

\subsection{Mass Spectra}
\label{subsec:A_ND_Mass_spectra}
In this subsection, we tabulate our predictions for the mass spectra of $c\bar{c}$- and $b\bar{b}$-systems and compare our results with the available experimental and theoretical works. We first enlist our predictions and result analysis for charmonium followed by bottomonium.

\subsubsection{Charmonium Spectroscopy}
Our predictions for the mass spectra of charmonium are given in column $2$ of Table~\ref{tab:2_c_mass_long}. We list the results from experiment and other theoretical works in columns $3-12$ of Table~\ref{tab:2_c_mass_long}. Mostly, we compare our results with experiment \cite{ParticleDataGroup:2022pth}, however, in the absence of experimental results, we compare with a range of theoretical works, namely, the Godfrey-Isgur relativized potential model (GI) \cite{Barnes:2005pb}, semi-relativistic approach (SR) \cite{Radford:2007vd}, various non-relativistic potential models that employ the Cornell potential (NR) \cite{Barnes:2005pb, Deng:2016stx, Soni:2017wvy}, screening effect (SP) \cite{Li:2009zu, Deng:2016stx}, $\mathcal{O}(\frac{1}{m})$ corrections to the Cornell potential (NRC) \cite{Kher:2018wtv}, and LQCD \cite{Kalinowski:2015bwa}. We observe the following.

\begin{enumerate}
    \item The calculated masses of $S$-wave charmonium states are found to be in excellent agreement with the experimental masses. Note that the experimental averages for the masses of low-lying $S$-wave states are known very precisely. However, the $\psi(4040)$ state is not assigned as $\psi(3^3S_1)$, for which we observe a deviation of $34$ MeV compared with the theoretically assigned $\psi(3^3S_1)$ state. It may also be noted that we have not used experimental masses as inputs for our predictions.

    \item We observe the hyperfine splitting between pseudoscalar and vector states as $M_{J / \psi(1S)} - M_{\eta_c(1S)} = 112.94$ MeV, which is in very good agreement with the experimental value ($113.0 \pm 0.4$) MeV \cite{ParticleDataGroup:2022pth}. Likewise, we found the mass difference between the ground and first excited vector states of charmonium ($M_{\psi(2S)} - M_{J / \psi(1S)}$) as $580.21$ MeV, which is also in good agreement with the experimental expectations ($589.188 \pm0.028$) MeV~\cite{ParticleDataGroup:2022pth}.

    \item Furthermore, our predictions for the masses of $P$-wave charmonia are also in very good agreement with the available experimental results, where we observe a variation as small as $10$ MeV. We calculated the masses for states up to $\chi_{c2}(6^3P_2)$ as shown in Table~\ref{tab:2_c_mass_long}. Apart from our predictions, only NR \cite{Soni:2017wvy} predicted masses of the excited states beyond $\chi_{c2}(3^3P_2)$. Their results are in general larger than ours for excitations higher than $\chi_{c2}(2^3P_2)$.

    \item Similarly, our predictions for $D$-wave states of charmonium are in good agreement with the available experimental masses within $30$ MeV. It maybe noted that we predicted the masses for states up to $n = 6$, until the bottom meson mass threshold, consistently for all $S$-, $P$-, and $D$-waves, as shown in Table~\ref{tab:2_c_mass_long}. Here again, except NR \cite{Soni:2017wvy}, all other theoretical works went only up to $\psi_{3}(3^3D_3)$.
\end{enumerate}

We would like to remark that our predictions for the masses agree well with the experimental results, and in addition, our results lie well within the predictions of the other theoretical approaches. Aforementioned, all the theoretical works, including ours, follow a very close approach in optimization of the said parameters. It may be noted that $\alpha_s$ and $\sigma$, along with the quark masses, are crucial for making reliable mass predictions for the states. We noticed that apart from $\alpha_s$, the existing models comply with a very small range of $\sigma$ and quark masses, and their numerical values are limited to certain range, as can be observed from the literature. In addition, further optimization of the masses of states is provided through the values of the parameters $\rho$ and $V_c$. Due to the nature of the contact term (Eq.~\eqref{eqn:5_smearfn}) in the spin-spin interaction, plausible spin-spin splitting is observed for the case of $S$-wave states, whereas $P$- and $D$-wave states are minimally affected by a few MeV \cite{Kwong:1987mj}. The smearing constant ($\rho$) further refines the spin singlet-triplet splitting. On the other hand, $V_c$ normalizes the ground and excited energy eigenvalues for the entire spectrum to match well with the experimental values. As mentioned earlier, the restricted numerical values of $\alpha_s$, $\sigma$ and the quark masses do not leave much scope for their optimization with respect to experimental masses. However, $\rho$ and $V_c$ provide a reasonable optimization corresponding to both ground and excited states in order to obtain reliable predictions for the mass spectra, which agree well with the experimental results.

Furthermore, while comparing our results with those of other theoretical models, we found that our predictions are in good agreement with their results. As noted earlier, we have listed our mass predictions for $nL$ states up to $n = 6$ for $S$-, $P$-, and $D$-wave states; on the other hand, the SP \cite{Li:2009zu}, NR \cite{Soni:2017wvy}, and NRC \cite{Kher:2018wtv} approaches extend their predictions to $n=6$ for $S$-wave states alone. A comparison with these models reveals that our predictions for radial excitations beyond $\psi{(3 {}^3S_1)}$ are particularly larger than those of SP \cite{Li:2009zu} and NRC \cite{Kher:2018wtv}, while they are smaller than those of NR \cite{Soni:2017wvy}. A similar pattern can also be observed in $P$- and $D$-wave predictions. Further, our predictions agree well with the available NR \cite{Barnes:2005pb, Deng:2016stx} results  for both ground and excited states. This is due to the nearly similar values of most of the parameters in all three models for the case of charmonium. Furthermore, our results show a variation of up to $\sim 30$ MeV in the absence of $V_c$ with respect to the current values. In general, our predictions are in good agreement with the relativistic model (GI) \cite{Barnes:2005pb}; however, they are roughly smaller by $\sim 50$ MeV in the case of $S$- and $D$-wave states. In the case of $P$-wave states, a maximum difference of $\sim 100$ MeV is noted between the models. Comparison of various theoretical models, relativistic and non-relativistic, with experiment reveals that both types of models agree well with experiment and that the relativistic effects appear to be smaller in magnitude. Further, our results are comparable with the screened potential models (SP) \cite{Deng:2016stx,Li:2009zu} for $n \le 2$ states; however, we notice that their results are generally smaller than most of the theoretical works for higher states. Finally, our predictions are mostly comparable to the SR \cite{Radford:2007vd} approach with a maximum variation of $\sim 33$ MeV. For comparison, we have also listed the results of LQCD \cite{Kalinowski:2015bwa} and found that our predictions for $P$-wave states are in good agreement with their results.

\subsubsection{Bottomonium Spectroscopy}
The numerical predictions for the mass spectra of bottomonium from our work are given in column $2$ of Table~\ref{tab:3_b_mass_long}. We compare our results with available experimental masses (PDG) \cite{ParticleDataGroup:2022pth} and, in their absence, with other theoretical works, namely, GI \cite{Godfrey:2015dia}, modified Godfrey-Isgur model (MGI) which takes account of color screening effects \cite{Wang:2018rjg} and non-relativistic approaches such as constituent quark model (CQM) \cite{Segovia:2016xqb}, SP \cite{Deng:2016ktl, Li:2009nr}, NR \cite{Soni:2017wvy}, instanton induced potential with confinement term (IMWC) \cite{Pandya:2021vby} and NRC \cite{Kher:2022gbz}, as shown in columns $3-11$ of Table~\ref{tab:3_b_mass_long}. We list our observations below.

\begin{enumerate}
    \item We calculated the $S$-, $P$-, and $D$-wave $b\bar{b}$ states till $n = 6$. Similar to charmonium, our predicted masses for $S$-wave bottomonium states are in excellent agreement with the available experimental masses for the lowest lying states ($n=1$). For $n=2$ states, we notice a deviation of $\sim 15$ MeV from the experimental results. It is interesting to note that all the theoretical models show a similar deviation in either of the low-lying $1S$ and $2S$ states, except for the MGI \cite{Wang:2018rjg} model. Our predictions for the remaining $S$-wave excited states are also in very good agreement with experiment, except for $\Upsilon(4^{3}S_{1})$ and $\Upsilon(6^{3}S_{1})$. The comparison of experimental results with those of other theoretical models demonstrates similar behavior for these states, although there are deviations in some cases.

    \item Notably, our predictions for the hyperfine mass splittings, \textit{i.e.}, $M_{\Upsilon (1S)} - M_{\eta_b(1S)} = 61.16$ MeV and $M_{\Upsilon(2S)} - M_{ \eta_{b}(2S)} = 26.12$ MeV, match exceptionally well with the experimental averages ($62.3 \pm3.2$) MeV and ($24 \pm4$) MeV \cite{ParticleDataGroup:2022pth}, respectively. Furthermore, we obtain the mass difference between the second and first radially excited vector states ${M}_{\Upsilon(3S)} - {M}_{\Upsilon(2S)}$ as $344.13$ MeV, which agrees well with the experimental observation ($331.50 \pm0.13$) MeV \cite{ParticleDataGroup:2022pth}.

    \item In the case of $P$-wave bottomonium states, our results are in very good agreement with the experimental observations for states up to $n=2$, \textit{i.e.}, within $10$ MeV. Our predictions for $\chi_{b1}(3^3P_1)$ and $\chi_{b2}(3^3P_2)$ deviate from the experiment by a maximum of only $30$ MeV. We predicted the masses of $P$-wave states up to the $\chi_{b2}(6^3P_2)$. In addition to our approach, MGI \cite{Wang:2018rjg} has also made predictions up to the same state. However, their fitting results in numerical predictions smaller than our results for the higher excited states.

    \item Experimentally, there exists only one observation for $D$-wave, $\Upsilon_2(1^3D_2)$ state, having mass ($10163.7 \pm 1.4$) MeV, which is larger than our prediction of $10141.35$ MeV roughly by $22$ MeV. The IMWC \cite{Pandya:2021vby} approach also gives predictions for higher $D$-wave excitations, and our results match well with theirs until $n = 3$. However, beyond that, our predictions are generally larger than those of the other models.
\end{enumerate}

In general, our predictions are in good agreement with the experimental values, with a few exceptions. As discussed in the case of charmonium, the numerical values of the input parameters for bottomonium in our model are much different from those of other theoretical models. On comparison with other theoretical models, we observe that all models are considerably different in their predictions, where agreement can be seen only in the low-lying states. Predictions of all the models are generally larger than the experimental expectations, and disagreement between the models increases with increasing mass of the excited states.

These substantial differences in the predictions of the theoretical models can be attributed to the addition of various terms in the potential or entirely different potentials. However, in our case we see that the Cornell potential (along with the correction terms) is sufficient to compute the mass of bound states with a higher degree of accuracy. In order to exploit the predictability of our approach, we do not rely on the experimental numbers as inputs. In addition, our focus is to analyze the wave function at the origin and its role in heavy quarkonium decays. Nevertheless, since our calculated masses of both charmonium and bottomonium match well with the available experimental masses, it validates that the employed methodology and our predictions are reliable. Further, we use our calculated masses to obtain the charmonium and bottomonium Regge trajectories. After which, we extend our work to estimate the wave function at the origin.

\subsubsection{Regge Trajectories}
\label{subsubsec:A3_ND_Regge}

\begin{figure}%
     \centering
     	\subfigure[$\eta_c (n^1S_0)$ states.]{\includegraphics[width=.472\textwidth]{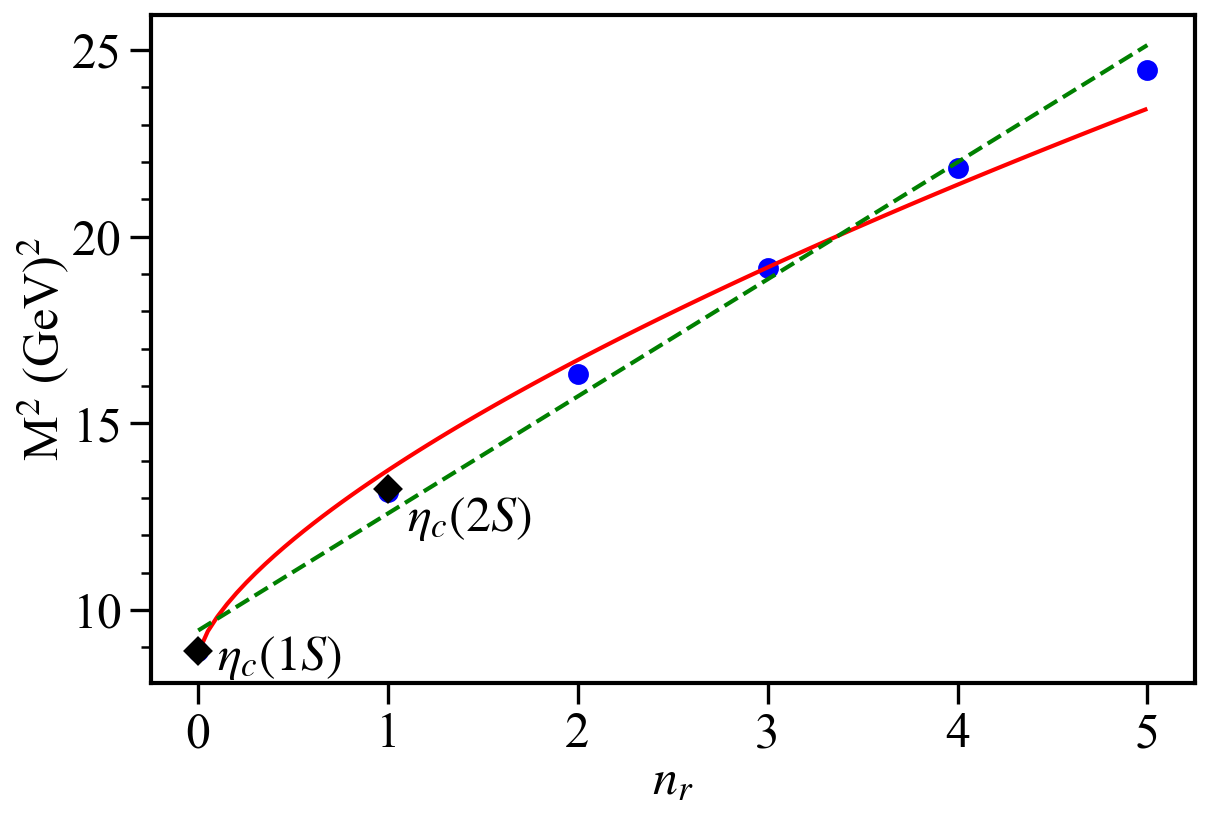}\label{fig:Charm_nMsq_S_etac}}
     \qquad 
     \subfigure [$\psi(n^3S_1)$ states.]{\includegraphics[width=.472\textwidth]{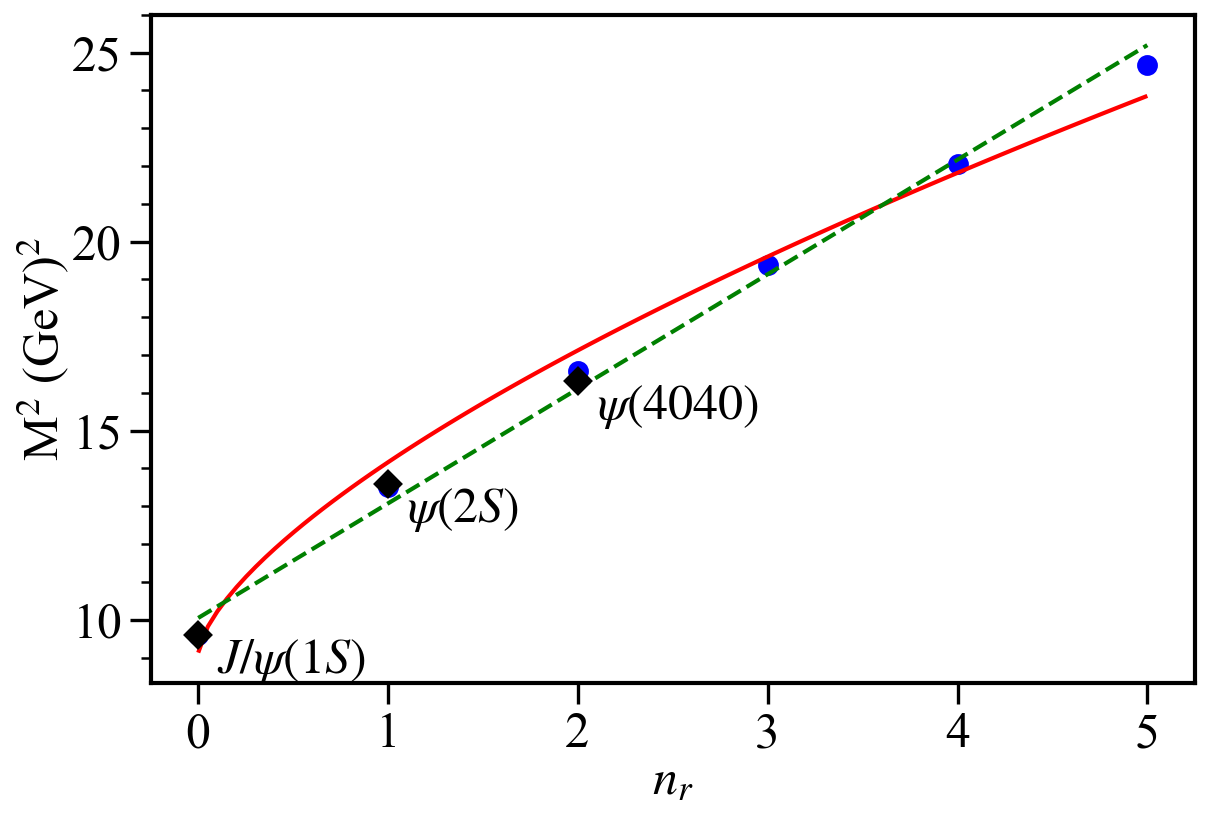}\label{fig:Charm_nMsq_S_psi}}
         
        \caption{Radial Regge trajectories of $S$-wave charmonium. The designated experimental states are given by solid diamonds and our calculated masses are given by solid dots. The dashed line and the solid curve represent the linear and non-linear fit, respectively. The same legend is followed throughout.}
        \label{fig:1_Charm_radial_regge_S}
\end{figure}

\begin{figure}
     \centering
     \subfigure[$h_c(n^1P_1)$ states.]{\includegraphics[width=.472\textwidth]{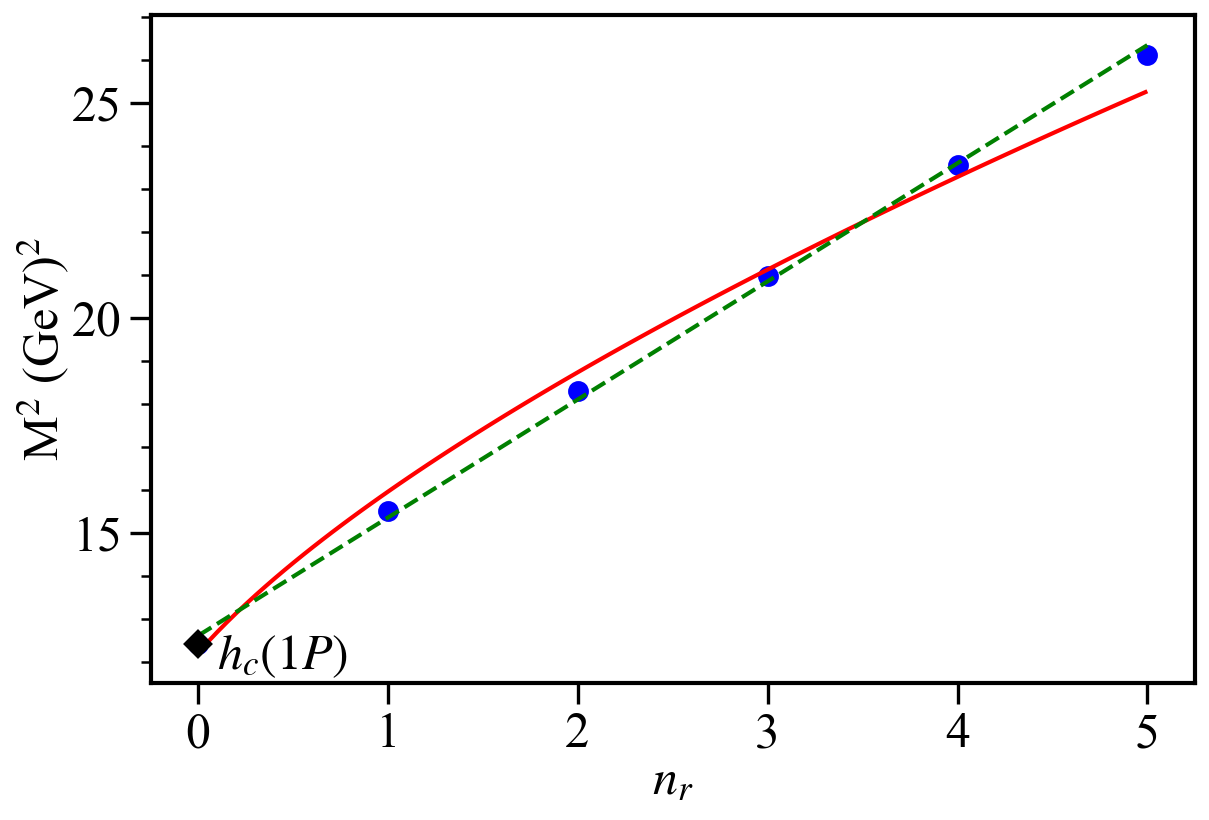}\label{fig:Charm_nMsq_P_hc}}
     \qquad
     \subfigure[$\chi_{c0}(n^3P_0)$ states.]{\includegraphics[width=.472\textwidth]{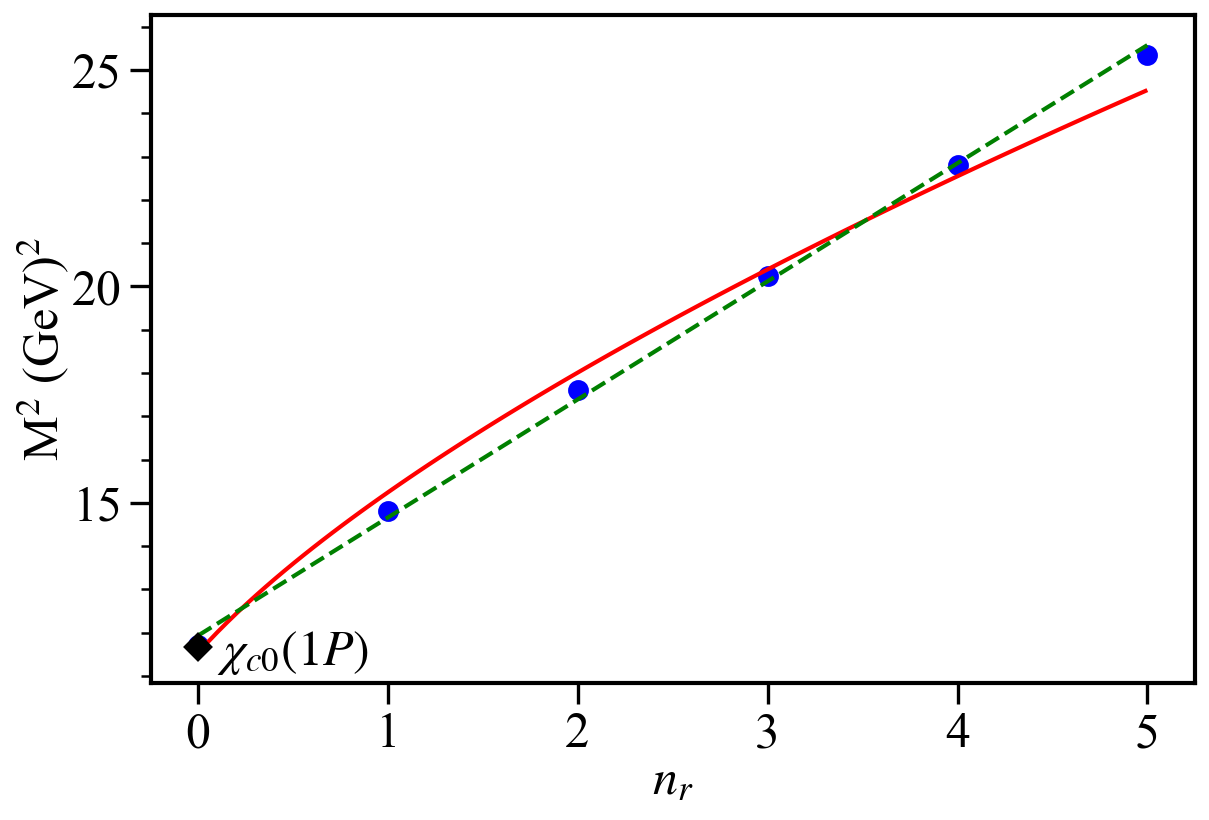}\label{fig:Charm_nMsq_P_Xc0}}

     \subfigure[$\chi_{c1}(n^3P_1)$ states.]{\includegraphics[width=.472\textwidth]{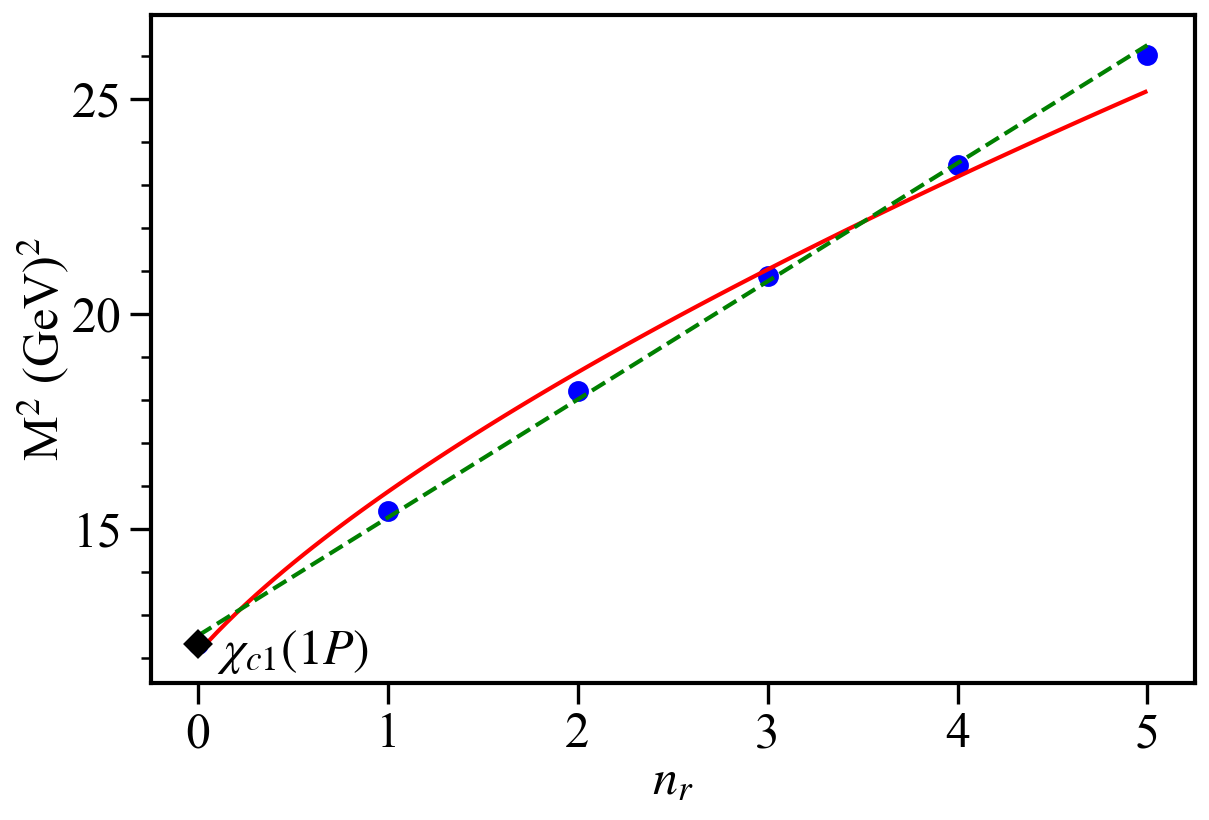}\label{fig:Charm_nMsq_P_Xc1}}
     \qquad
     \subfigure[$\chi_{c2}(n^3P_2)$ states.]{\includegraphics[width=.472\textwidth]{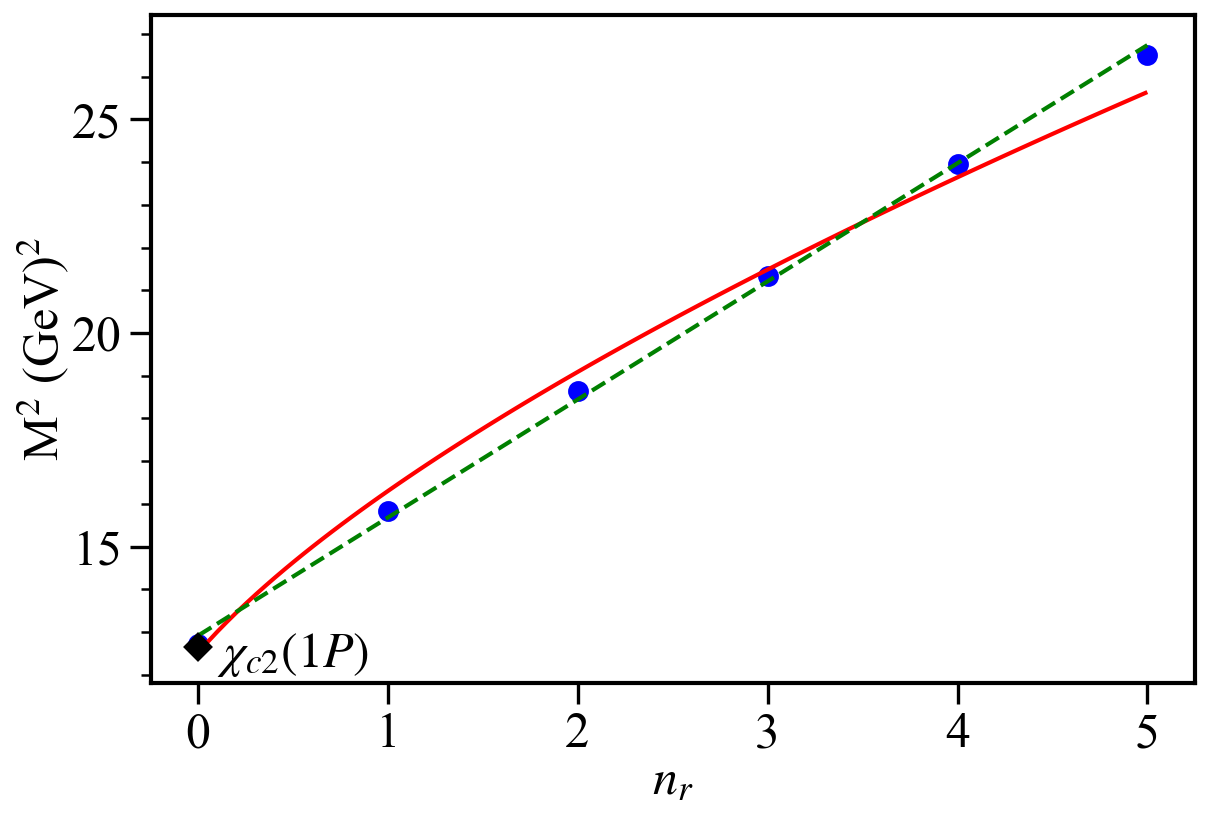}\label{fig:Charm_nMsq_P_Xc2}}
        \caption{Radial Regge trajectories of $P$-wave charmonium. Legend same as in Fig.~\ref{fig:1_Charm_radial_regge_S}.}
        \label{fig:2_Charm_radial_regge_P}
\end{figure}

\begin{figure}
     \centering
     \subfigure[$\eta_{c2}(n^1D_2)$ states.]{\includegraphics[width=.472\textwidth]{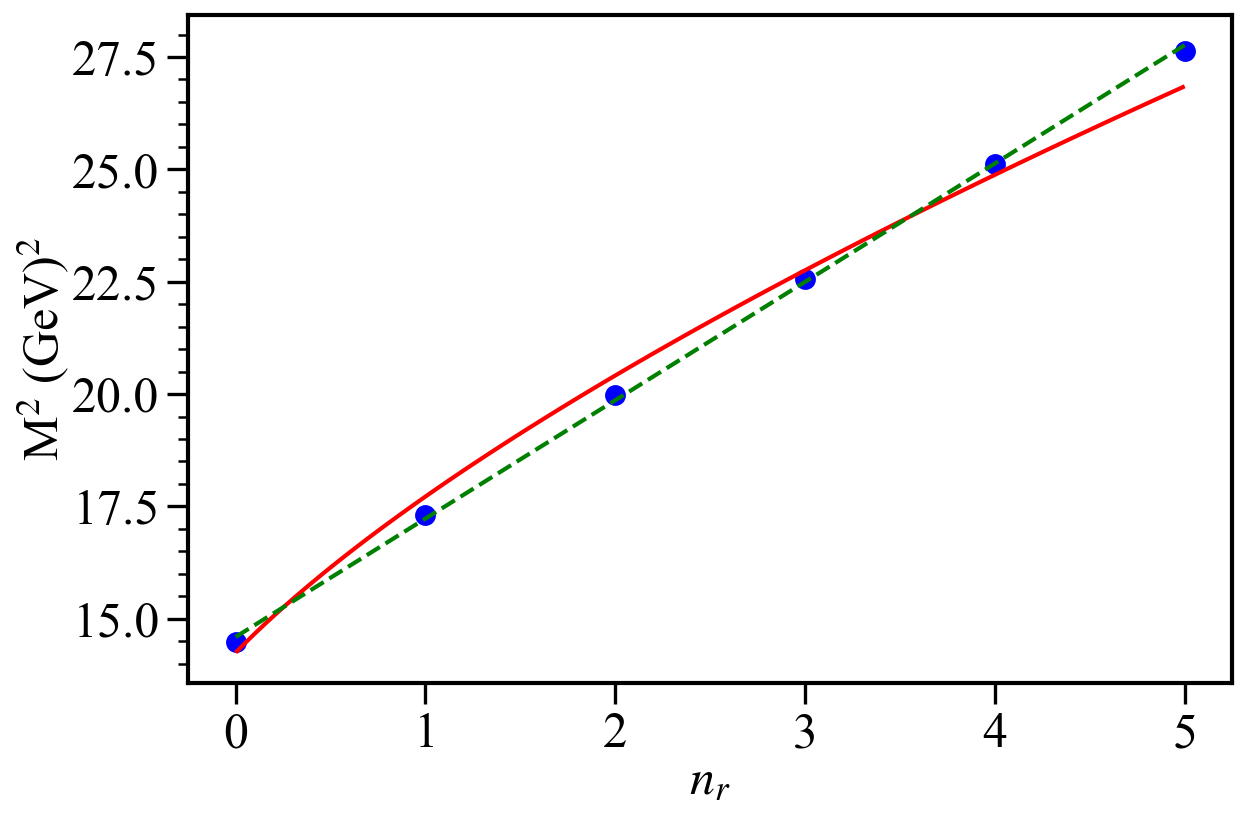}\label{fig:Charm_nMsq_D_etac2}}
     \qquad
     \subfigure[$\psi_{1}(n^3D_1)$ states.]{\includegraphics[width=.472\textwidth]{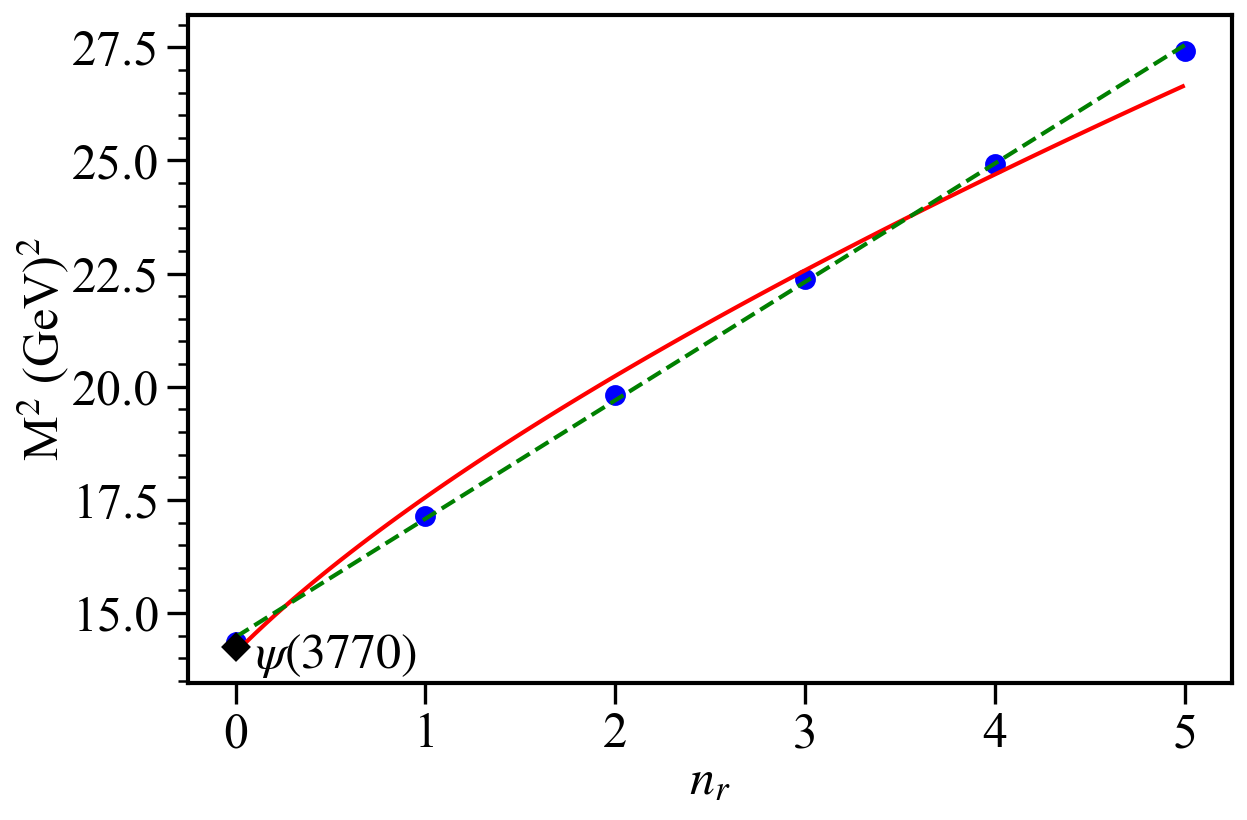}\label{fig:Charm_nMsq_D_psi1}}
     \subfigure[$\psi_{2}(n^3D_2)$ states.]{\includegraphics[width=.472\textwidth]{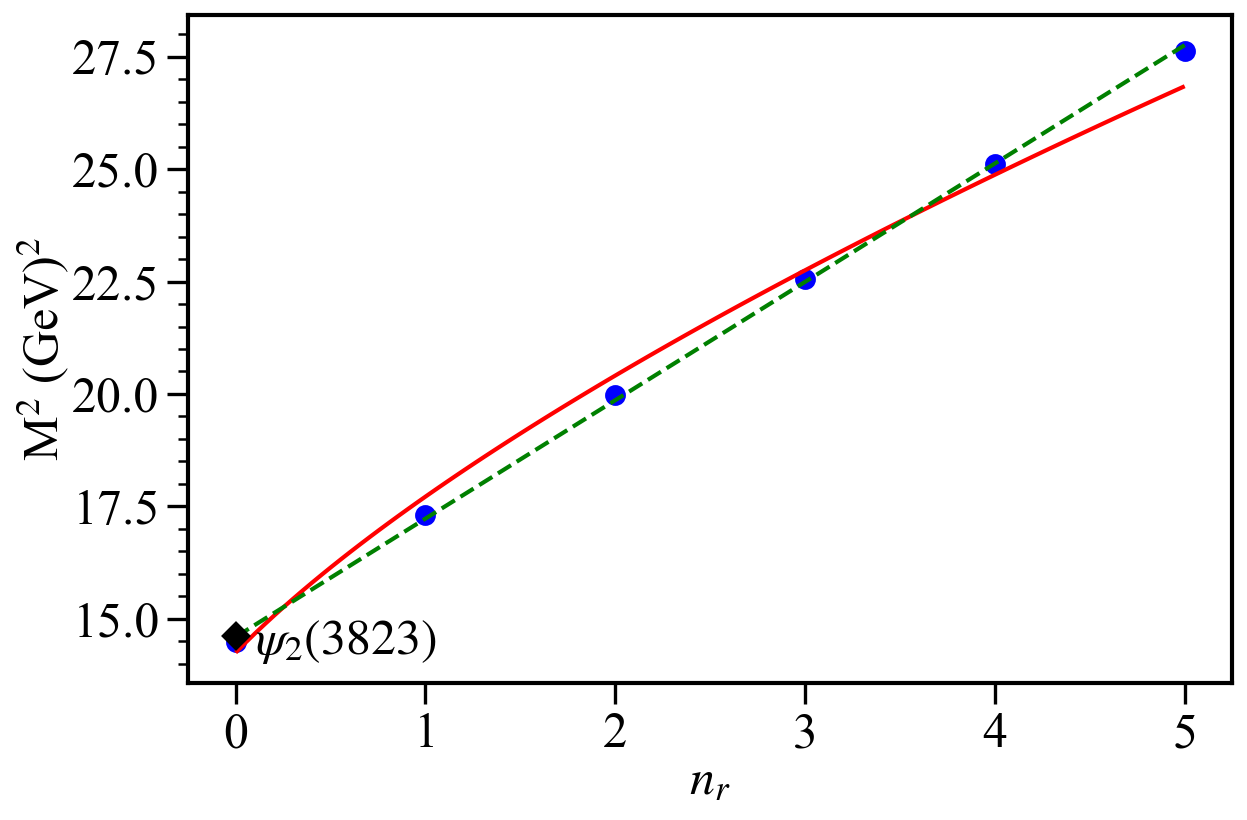}\label{fig:Charm_nMsq_D_psi2}}
     \qquad
     \subfigure[$\psi_{3}(n^3D_3)$ states.]{\includegraphics[width=.472\textwidth]{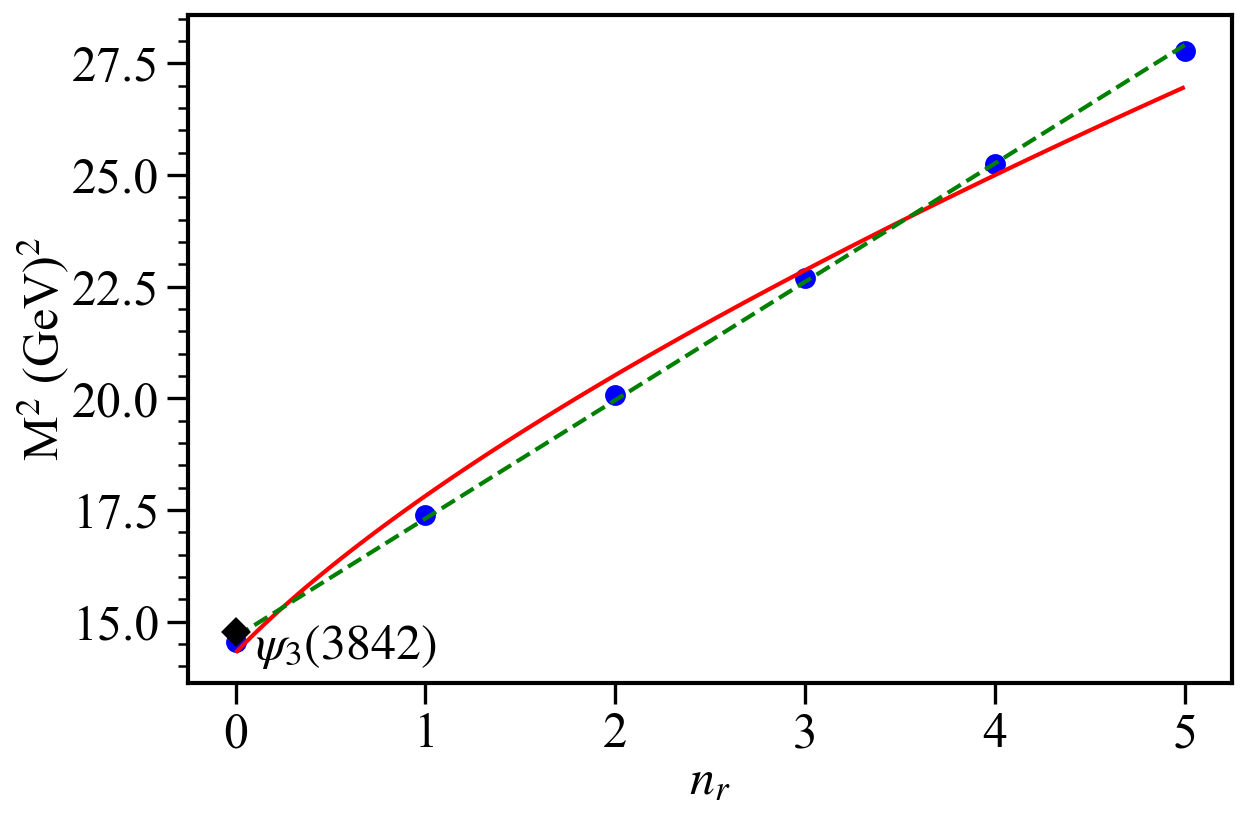}\label{fig:Charm_nMsq_D_psi3}}
        \caption{Radial Regge trajectories of $D$-wave charmonium. Legend same as in Fig.~\ref{fig:1_Charm_radial_regge_S}.}
        \label{fig:3_Charm_radial_regge_D}
\end{figure}

\begin{figure}
     \centering
     \subfigure[States with $J^P = 0^-, 1^+$, and $2^-$.]{\includegraphics[width=.472\textwidth]{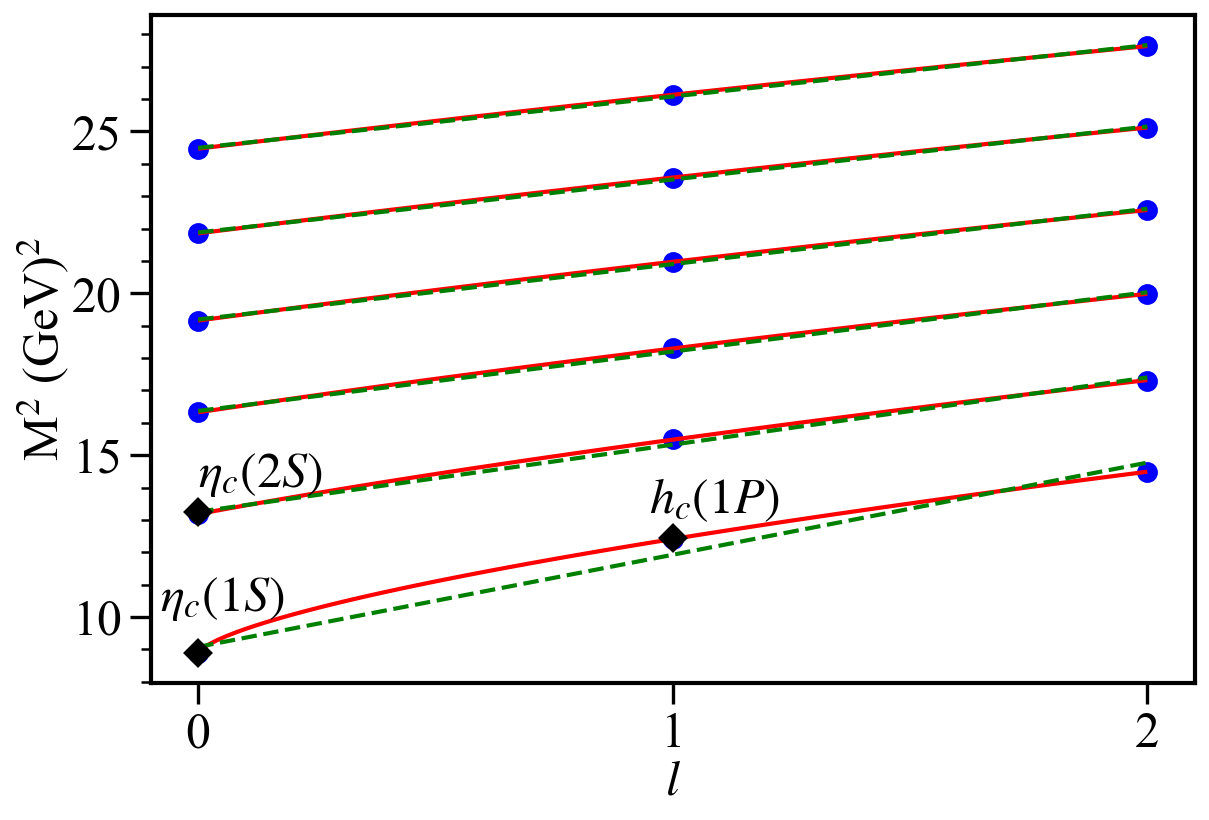}\label{fig:Charm_lMsq_singlets}}
     \qquad
     \subfigure[States with $J^P = 1^-, 2^+$, and $3^-$.]{\includegraphics[width=.472\textwidth]{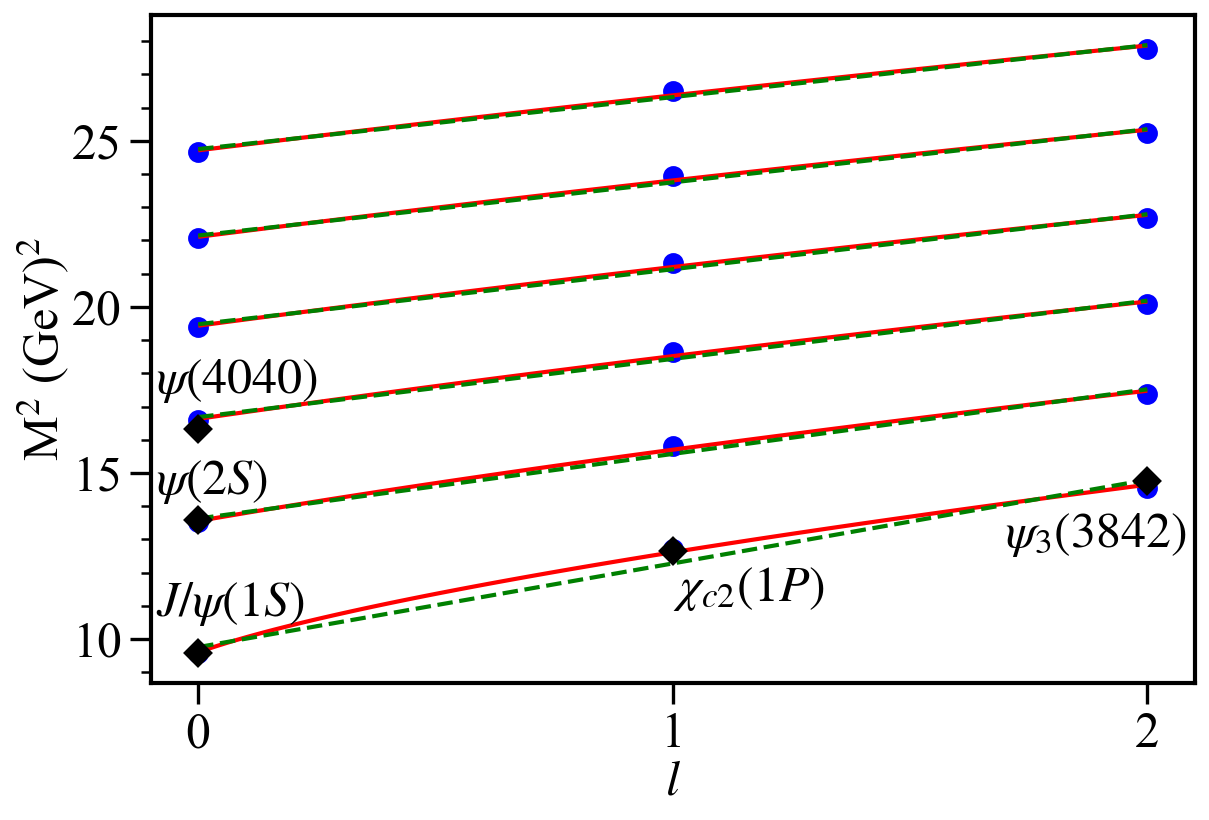}\label{fig:Charm_lMsq_triplets}}
        \caption{Parent and daughter orbital Regge trajectories of charmonium. Legend same as in Fig.~\ref{fig:1_Charm_radial_regge_S}.}
        \label{fig:4_Charm_orbital_regge}
\end{figure}

\begin{figure}
     \centering
     \subfigure[$\eta_b(n^1S_0)$ states.]{\includegraphics[width=.472\textwidth]{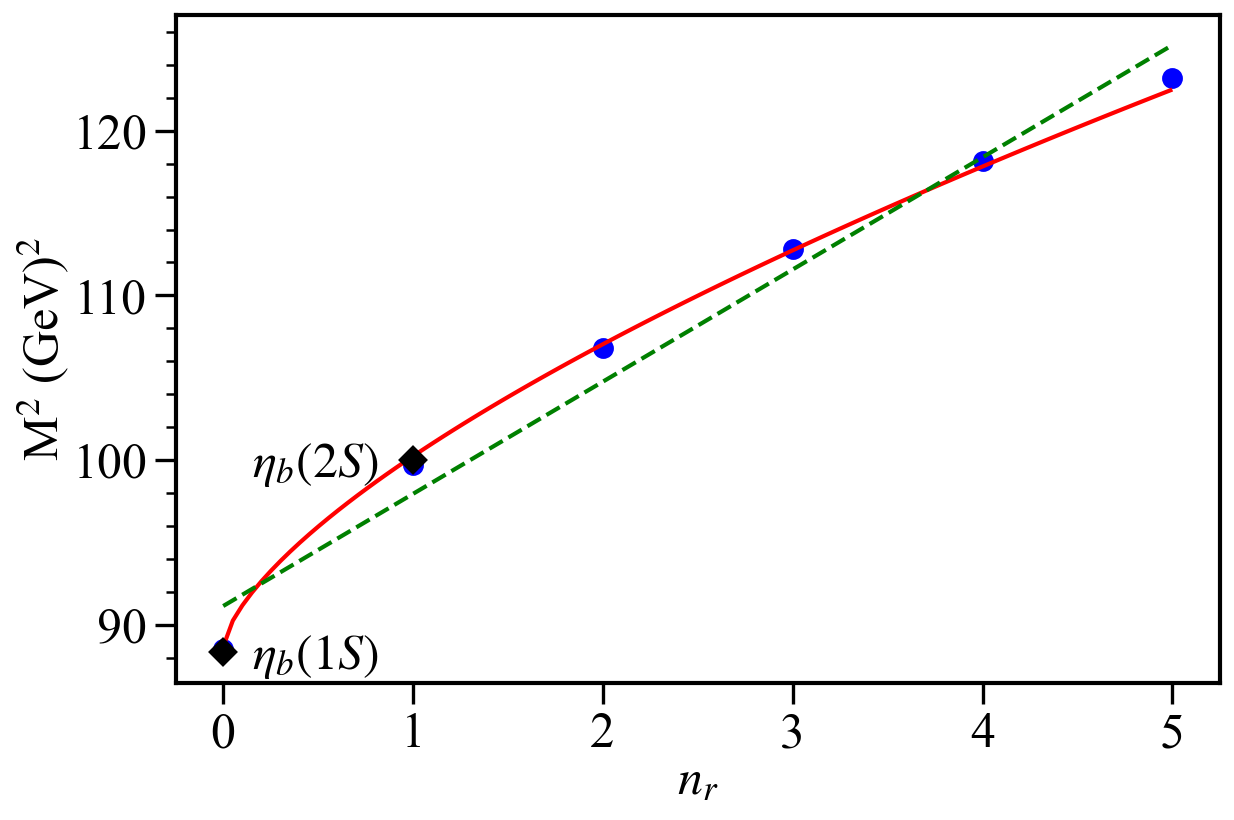}\label{fig:Bottom_nMsq_S_etab}}
     \qquad
     \subfigure[$\Upsilon(n^3S_1)$ states.]{\includegraphics[width=.472\textwidth]{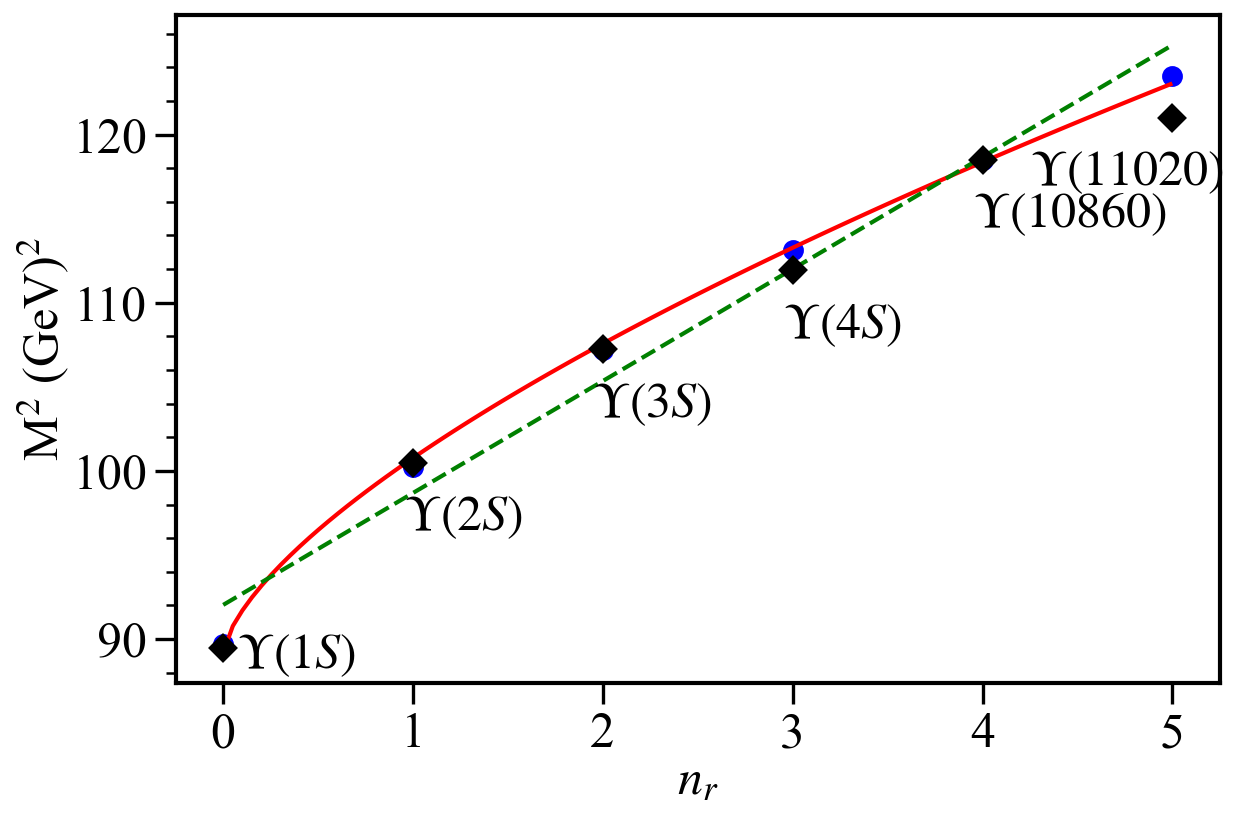}\label{fig:Bottom_nMsq_S_upsilon}}
        \caption{Radial Regge trajectories of $S$-wave bottomonium. Legend same as in Fig.~\ref{fig:1_Charm_radial_regge_S}.}
        \label{fig:5_Bottom_radial_regge_S}
\end{figure}

\begin{figure}
     \centering
     \subfigure[$h_b(n^1P_1)$ states.]{\includegraphics[width=.472\textwidth]{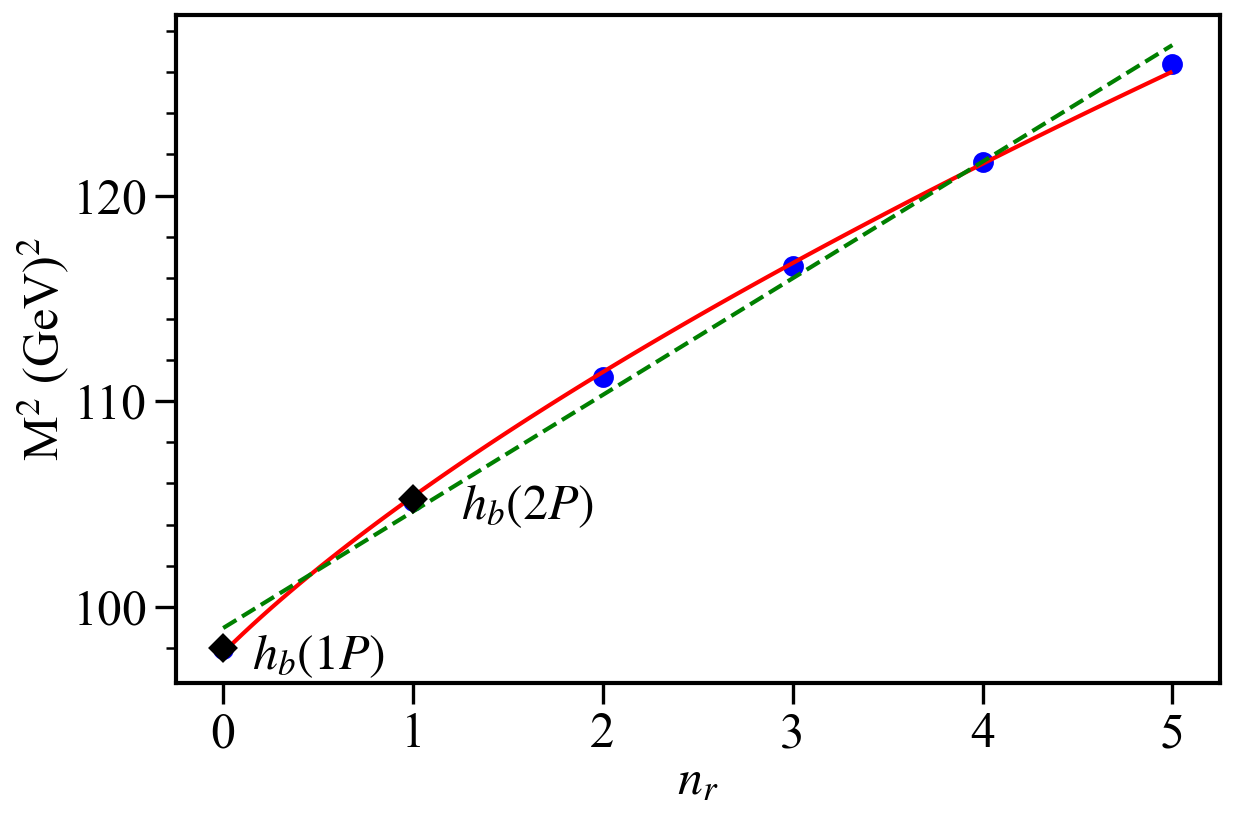}\label{fig:Bottom_nMsq_P_hb}}
     \qquad
     \subfigure[$\chi_{b0}(n^3P_0)$ states.]{\includegraphics[width=.472\textwidth]{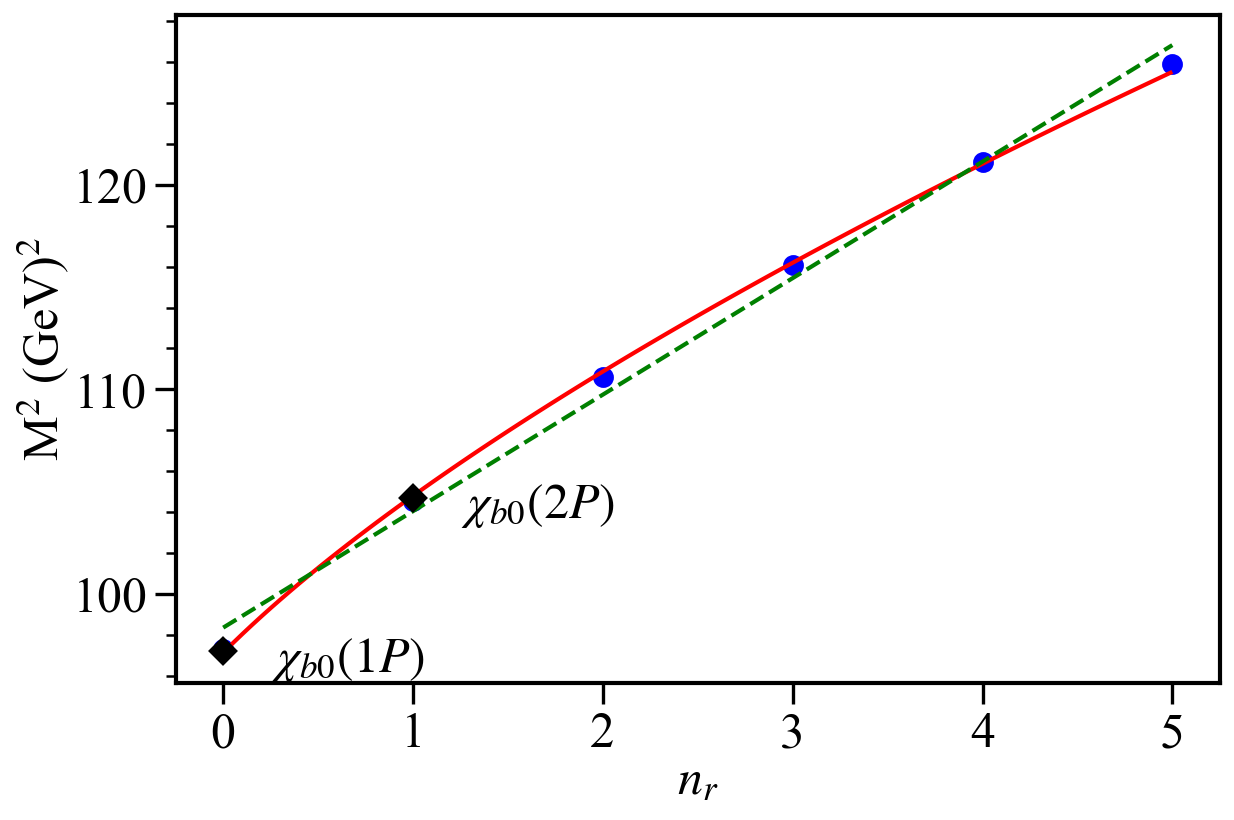}\label{fig:Bottom_nMsq_P_Xb0}}
     \subfigure[$\chi_{b1}(n^3P_1)$ states.]{\includegraphics[width=.472\textwidth]{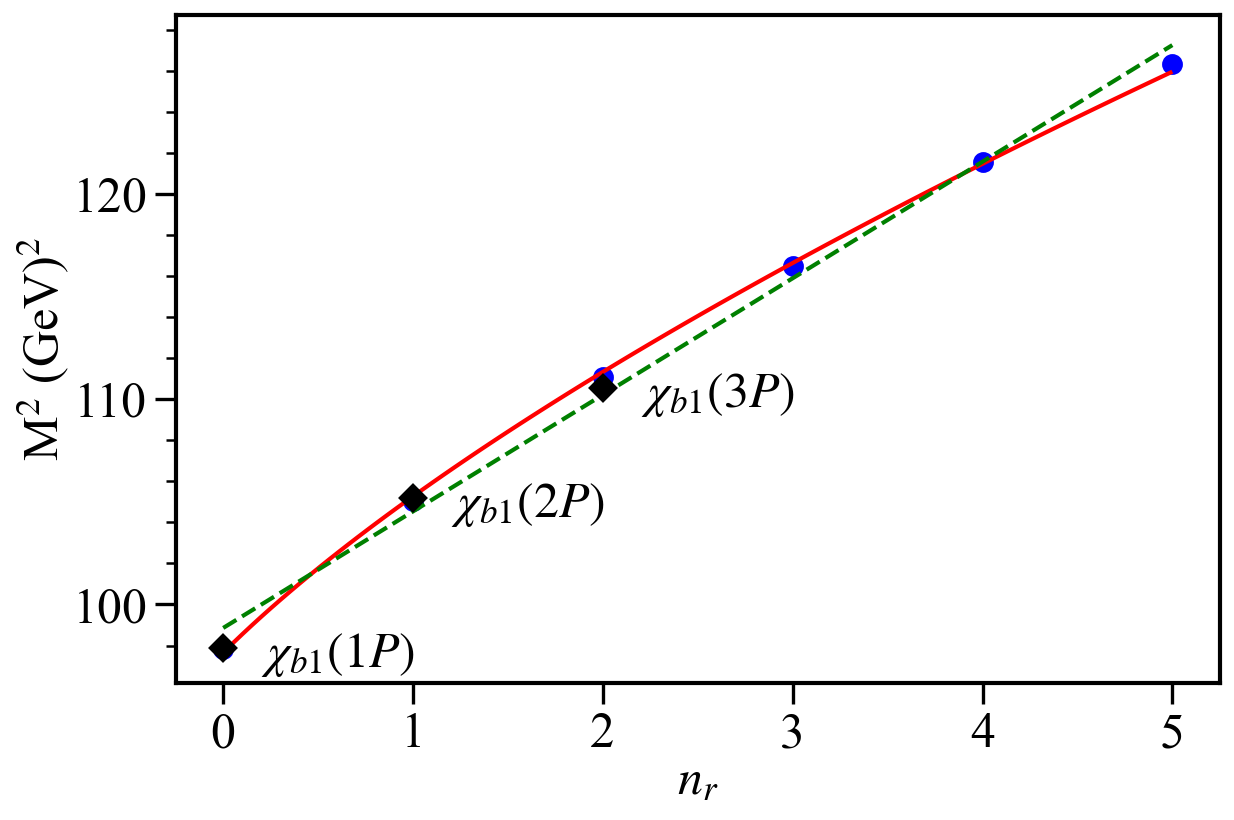}\label{fig:Bottom_nMsq_P_Xb1}}
     \qquad
     \subfigure[$\chi_{b2}(n^3P_2)$ states.]{\includegraphics[width=.472\textwidth]{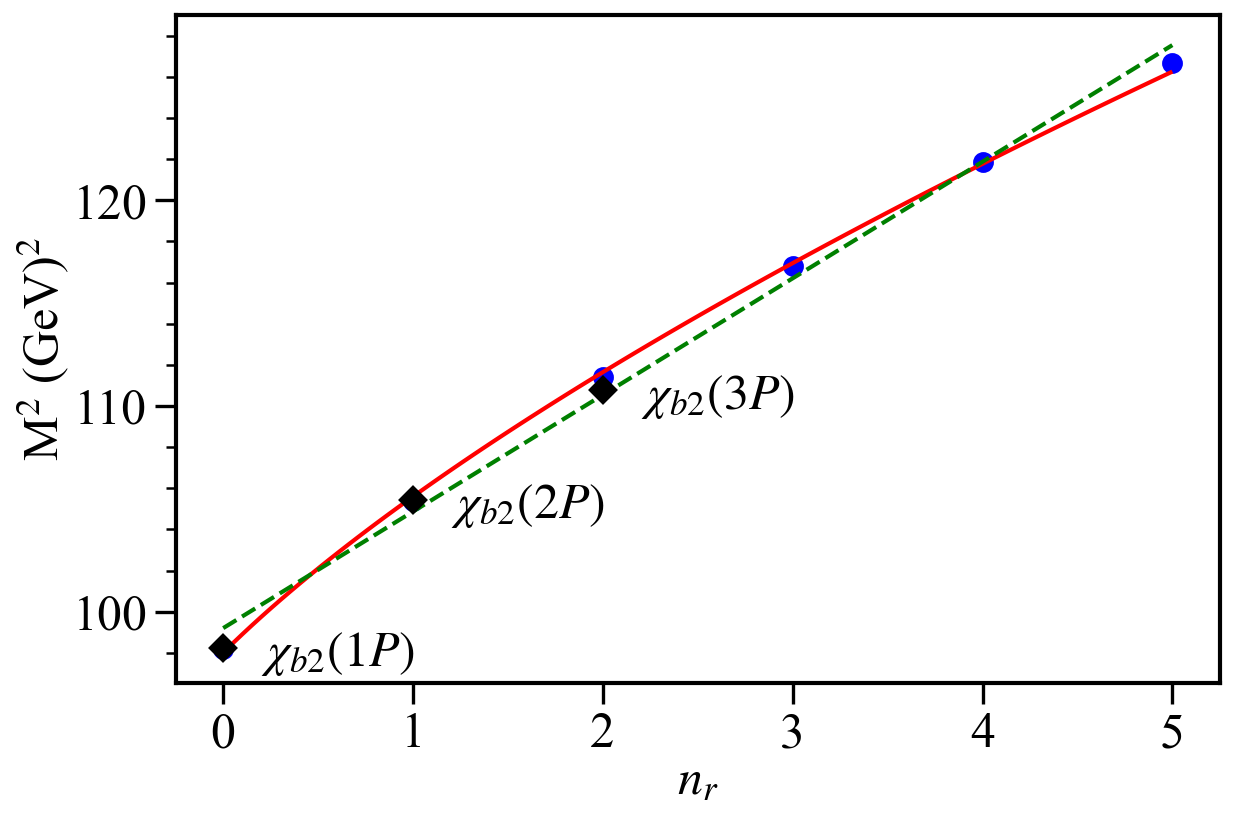}\label{fig:Bottom_nMsq_P_Xb2}}
        \caption{Radial Regge trajectories of $P$-wave bottomonium. Legend same as in Fig.~\ref{fig:1_Charm_radial_regge_S}.}
        \label{fig:6_Bottom_radial_regge_P}
\end{figure}

\begin{figure}
     \centering
     \subfigure[$\eta_{b2}(n^1D_2)$ states.]{\includegraphics[width=.472\textwidth]{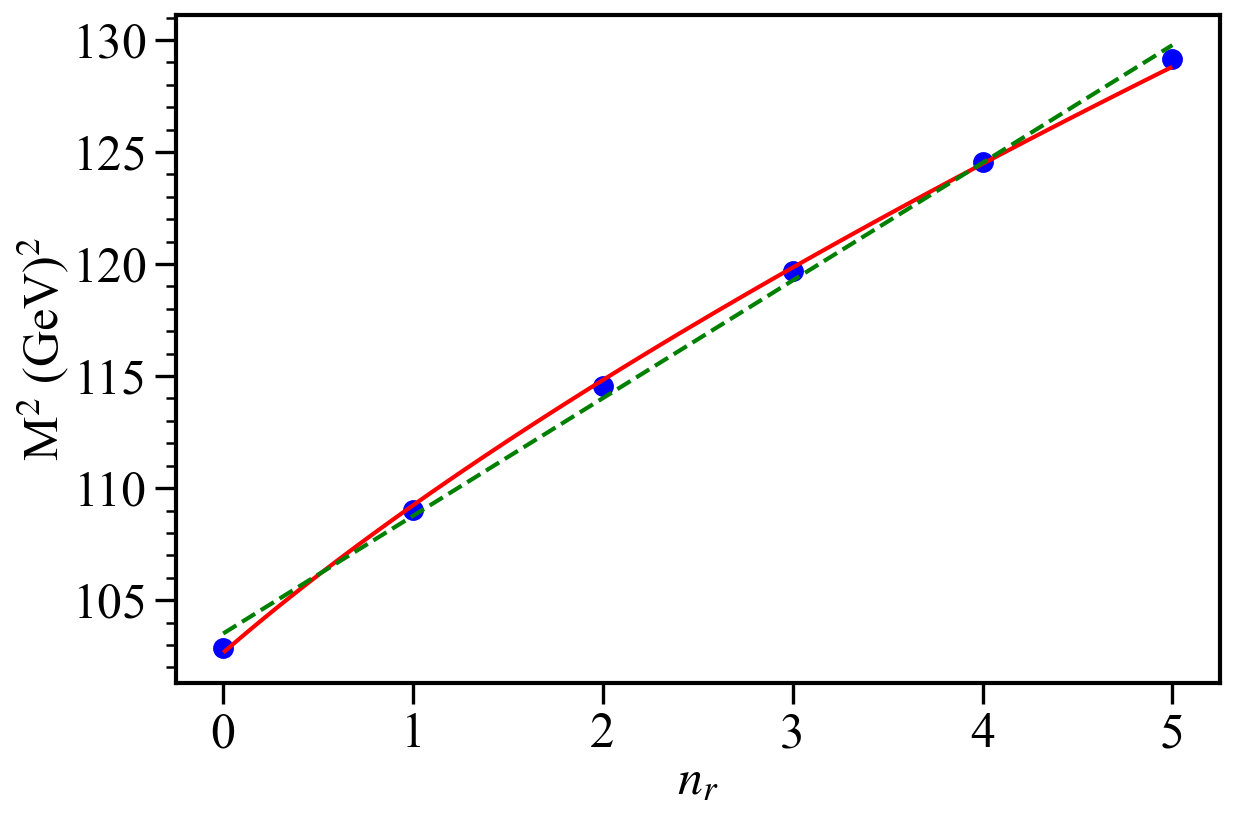}\label{fig:Bottom_nMsq_D_etab2}}
     \qquad
     \subfigure[$\Upsilon_{1}(n^3D_1)$ states.]{\includegraphics[width=.472\textwidth]{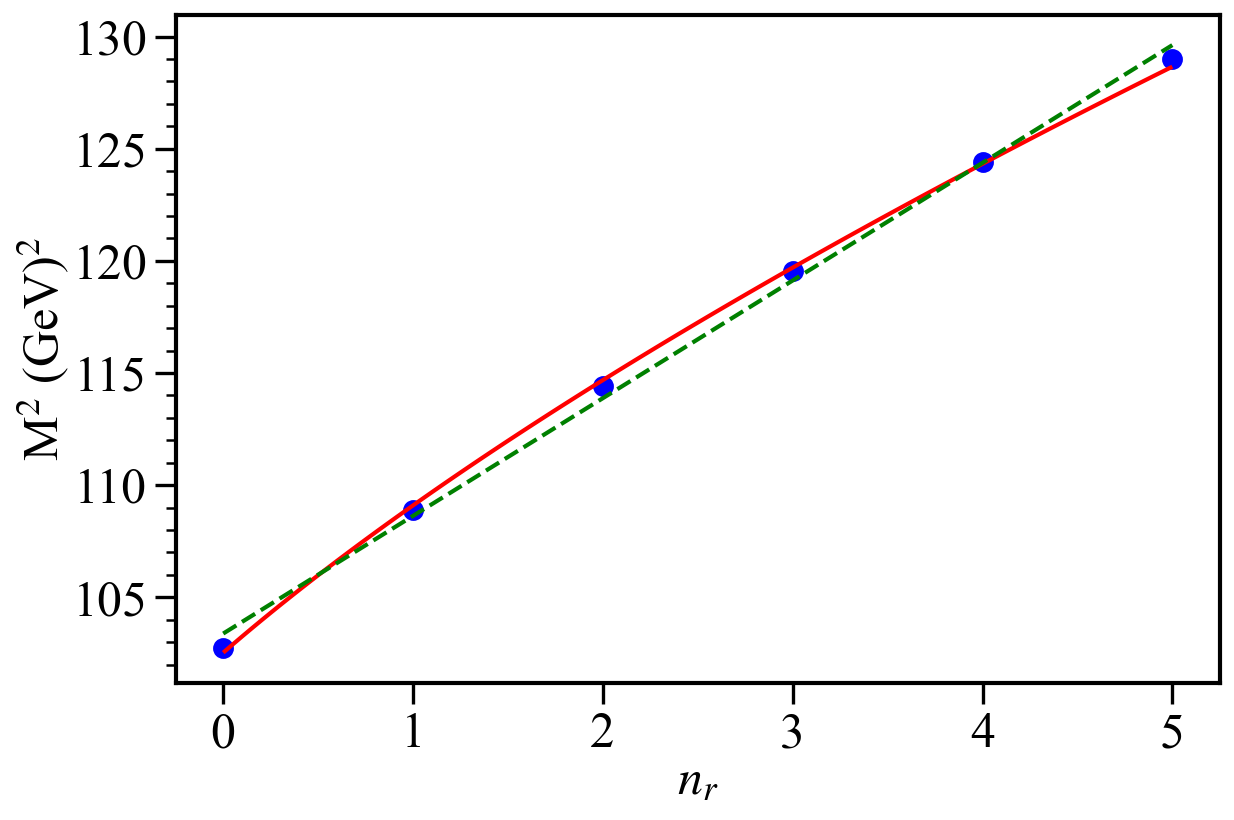}\label{fig:Bottom_nMsq_D_upsilon1}}
     \subfigure[$\Upsilon_{2}(n^3D_2)$ states.]{\includegraphics[width=.472\textwidth]{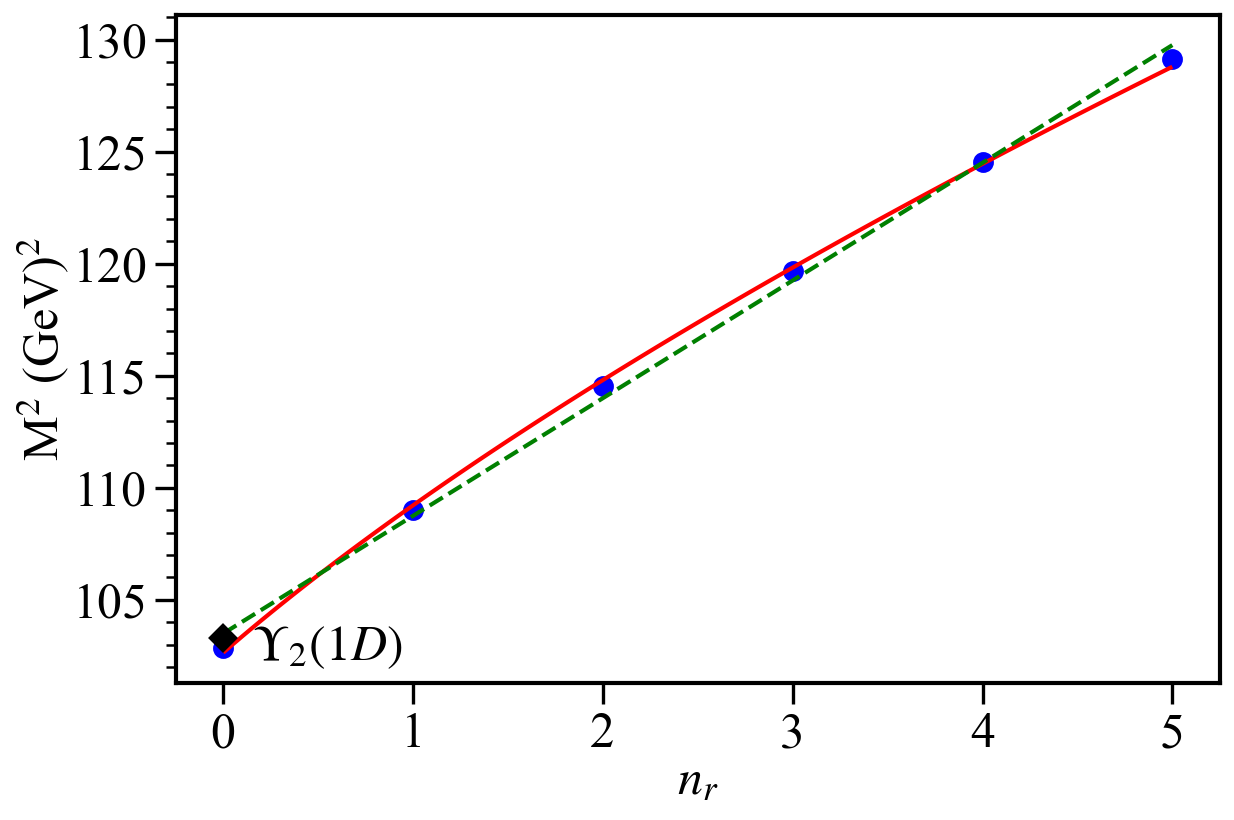}\label{fig:Bottom_nMsq_D_upsilon2}}
     \qquad
     \subfigure[$\Upsilon_{3}(n^3D_3)$ states.]{\includegraphics[width=.472\textwidth]{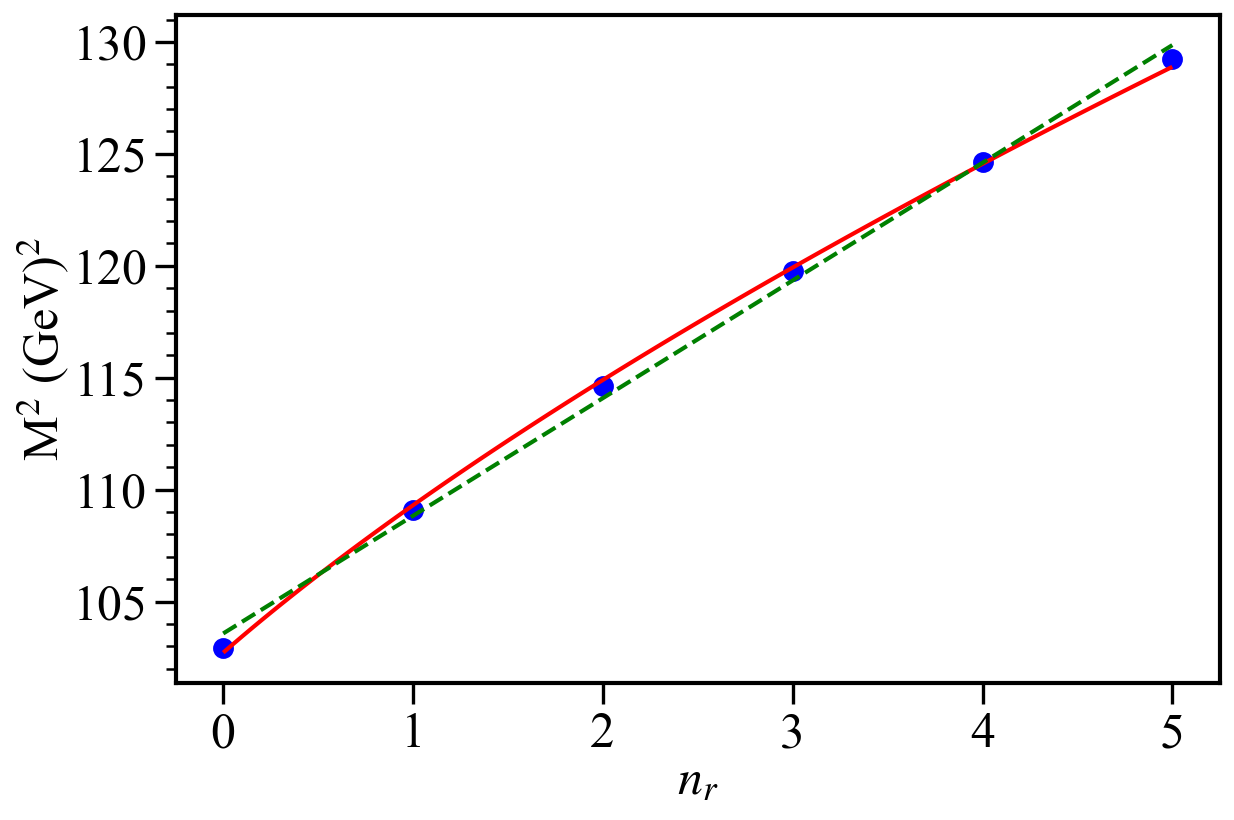}\label{fig:Bottom_nMsq_D_upsilon3}}
        \caption{Radial Regge trajectories of $D$-wave bottomonium. Legend same as in Fig.~\ref{fig:1_Charm_radial_regge_S}.}
        \label{fig:7_Bottom_radial_regge_D}
\end{figure}

\begin{figure}
     \centering
     \subfigure[States with $J^P = 0^-, 1^+$, and $2^-$.]{\includegraphics[width=.472\textwidth]{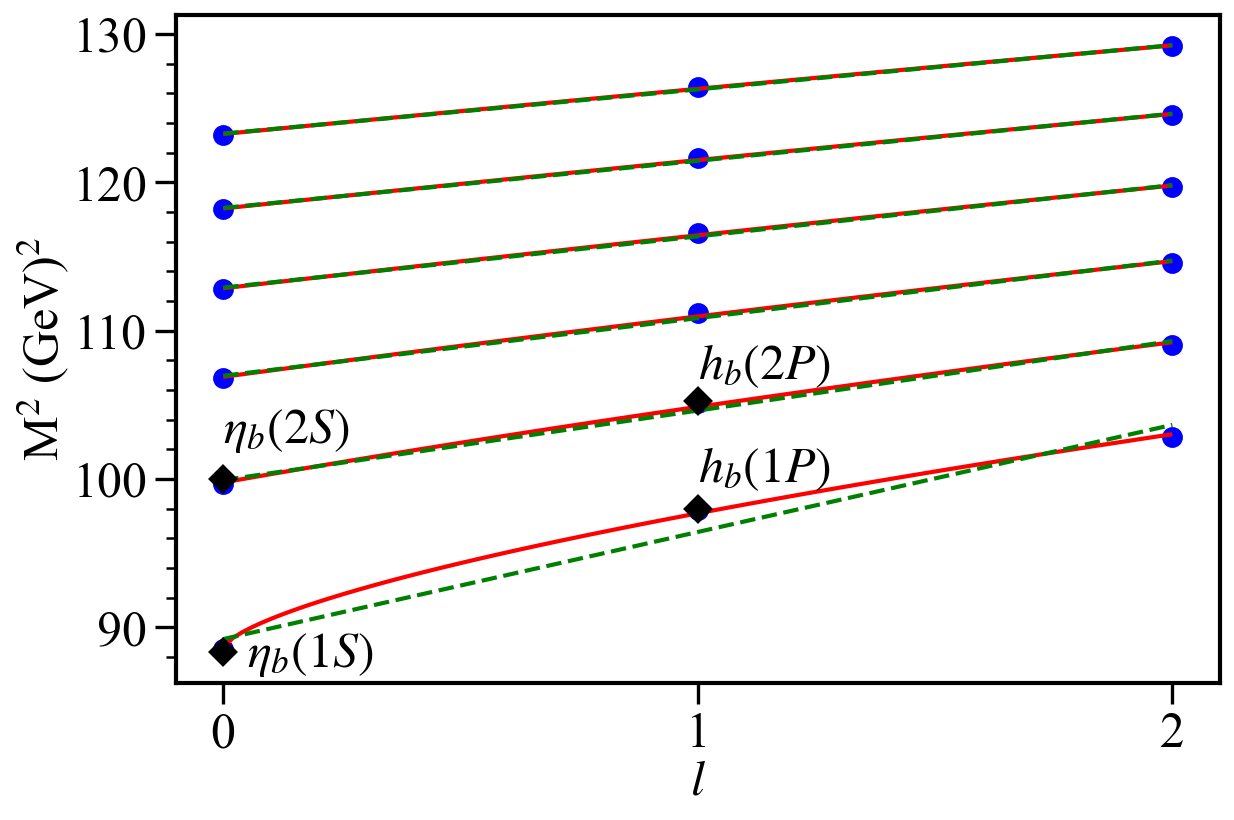}\label{fig:Bottom_lMsq_singlets}}
     \qquad
     \subfigure[States with $J^P = 1^-, 2^+$, and $3^-$.]{\includegraphics[width=.472\textwidth]{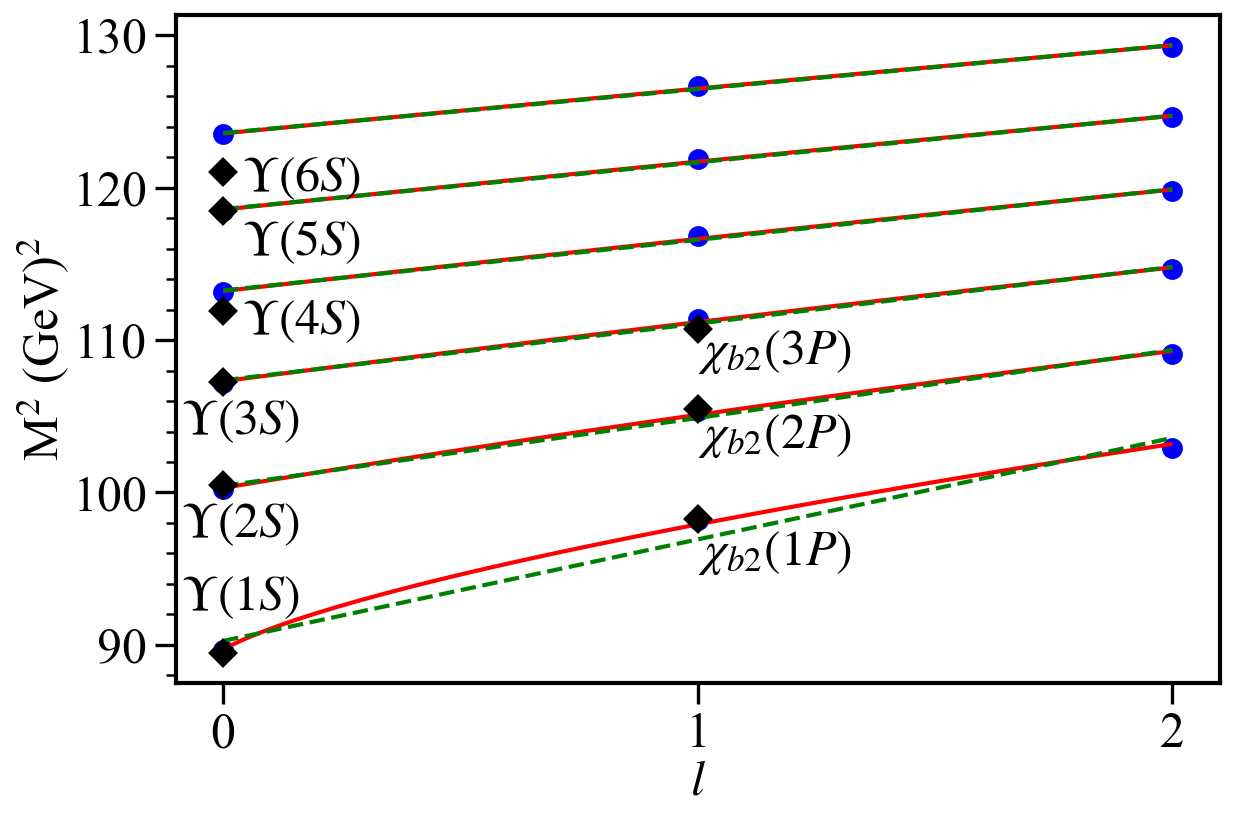}\label{fig:Bottom_lMsq_triplets}}
        \caption{Parent and daughter orbital Regge trajectories of bottomonium. Legend same as in Fig.~\ref{fig:1_Charm_radial_regge_S}.}
        \label{fig:8_Bottom_orbital_regge}
\end{figure}

We study the radial and orbital Regge trajectories of $c\bar{c}$- and $b\bar{b}$-systems by fitting our calculated mass values to Eqs.~\eqref{eqn:11_regge_linear} and \eqref{eqn:12_regge_chennonlinear} to obtain linear and non-linear trajectories, respectively. The fitting is performed through iminuit, which is a Jupyter-friendly Python interface for the Minuit2 C++ library \cite{iminuit, James:1975dr}. In the case of non-linear Regge trajectories (Eq.~\eqref{eqn:12_regge_chennonlinear}), the universal parameter $\beta$ for charmonium is calculated by fitting the radial Regge trajectory of $J/\psi$ and its excited states, and is found to be $2.3481$. Further, $c_l$ is obtained by fitting the orbital Regge trajectory containing the states $\eta_c(1S)$, $h_c(1P)$ and $\eta_{c2}(1D)$, which is equal to $1.8511$. Similarly for $b\bar{b}$-system, we obtain the parameter $\beta = 5.3961$ from the $\Upsilon$ and its excited states, whereas we calculate $c_l = 2.1815$ from $\eta_b(1S)$, $h_b(1P)$ and $\eta_{b2}(1D)$ states. However, $c_0$ and $c_1$ vary for individual trajectories in both systems. It may be noted that the parameters of linear Regge trajectories are determined by fitting the respective masses of states using Eq.~\eqref{eqn:11_regge_linear}. These parameters, \textit{i.e.}, $\alpha_{lin}$, $\beta_{lin}$, and $C_{lin}$ are determined independently for each trajectory. Finally, we plotted the non-linear and linear Regge trajectories in comparison to the available experimental values, as shown in Figs.~\ref{fig:1_Charm_radial_regge_S} $-$~\ref{fig:8_Bottom_orbital_regge}. We observe the following from the radial and orbital Regge trajectories of charmonium and bottomonium.

\begin{enumerate}
    \item In $S$-wave charmonium (Fig.~\ref{fig:1_Charm_radial_regge_S}), we notice that our calculated masses depart from linearity and favor non-linear behavior. However, it is notable that Eq.~\eqref{eqn:12_regge_chennonlinear} is unable to completely encapsulate the higher excited states.

    \item In the case of $P$-wave charmonium (Fig.~\ref{fig:2_Charm_radial_regge_P}), our computed values are in better agreement with the linear behavior of the Regge trajectories.

    \item Similarly, our $D$-wave charmonium (Fig.~\ref{fig:3_Charm_radial_regge_D}) Regge trajectories favor linear behavior. This shows that the linearity of radial Regge trajectories increases with the orbital quantum number.

    \item In the orbital Regge trajectories of charmonium (Fig.~\ref{fig:4_Charm_orbital_regge}), the parent trajectory tends to be non-linear, whereas the daughter trajectories are linear. Further, the orbital Regge trajectory governed by Eq.~\eqref{eqn:12_regge_chennonlinear} shows an increase in linearity with $n_r$.

    \item In the bottomonium system, the radial Regge trajectories of $S$-wave states (Fig.~\ref{fig:5_Bottom_radial_regge_S}) exhibit profound non-linearity in contrast to the charmonium system, the fit given by Eq.~\eqref{eqn:12_regge_chennonlinear} accommodates excited states as well. As observed in charmonia, the non-linearity in Regge trajectories tends to decrease with increasing mass, as shown by Fig.~\ref{fig:6_Bottom_radial_regge_P} for the bottomonium $P$-wave states.

    \item In $D$-wave bottomonium, as shown in Fig.~\ref{fig:7_Bottom_radial_regge_D}, due to large mass of the states the non-linearity of the fit Eq.~\eqref{eqn:12_regge_chennonlinear} lean towards a linear behavior; thus, both linear and non-linear Regge trajectories result in roughly similar fit to the calculated masses. Nevertheless, the non-linear trajectory accommodates the states better.

    \item As shown in Fig.~\ref{fig:8_Bottom_orbital_regge}, the orbital Regge trajectories of bottomonium behave in a manner similar to those of charmonium, \textit{i.e.}, the daughter trajectories are linear and parallel to each other, while the parent trajectory exhibits a distinct non-linear behavior.
\end{enumerate}

Thus, in our fits for charmonium and bottomonium, the non-linear nature of the Regge trajectories is observed when $n_r$ and $l$ are small, and the linear nature takes over for excited states. Notably, the non-linearity is more pronounced in bottomonium system. The behavior of our trajectory fits can be explained based on our chosen potential Eq.~\eqref{eqn:1_cornel_pot}. The linear component of the potential grows stronger with increase in distance. Thus, for $P$-wave and $D$-wave, the influence of this linear component is more profound and therefore, our calculated masses fit the linear Regge trajectory better. On the other hand, both the linear and Coulomb parts of the potential play equally important roles in the case of the ground and lowest excited states, which could be responsible for non-linearity. This reasoning also explains the stronger non-linearity seen in lower bottomonium states compared with lower charmonium states. It may be noted that our predictions for masses are in good agreement with the experimental values, and the obtained fits for the Regge trajectories further emphasize the reliability of our numerical values.

\subsection{Square of Radial Wave function at the Origin}
\label{subsec:B_ND_wfo}

\begin{figure}
     \centering
     \subfigure[$S$-wave $c\bar{c}$ states.]{\includegraphics[width=.478\textwidth]{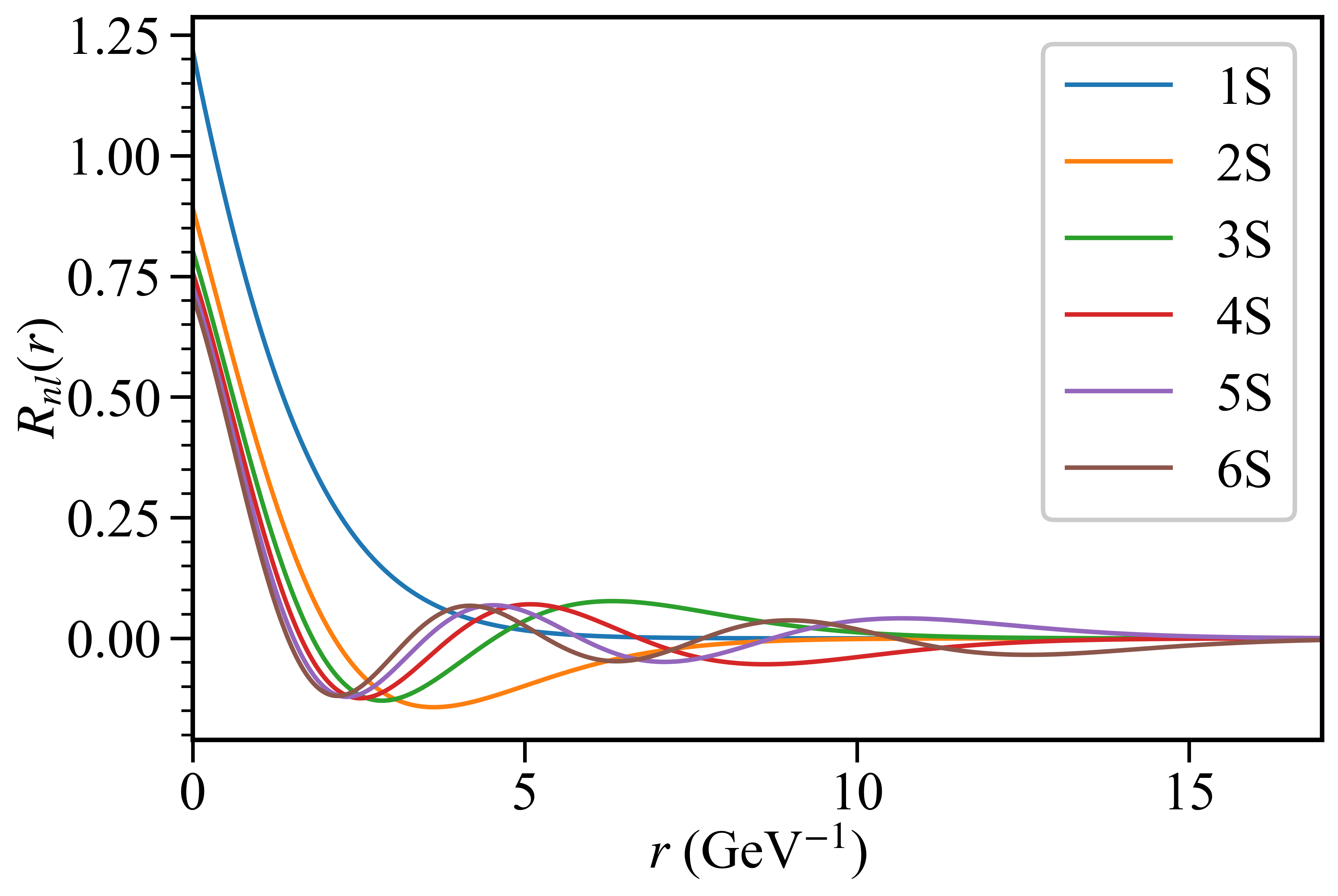}\label{fig:charm_s_wfn_Rr_y}}
     \qquad
     \subfigure[$S$-wave $b\bar{b}$ states.]{\includegraphics[width=.462\textwidth]{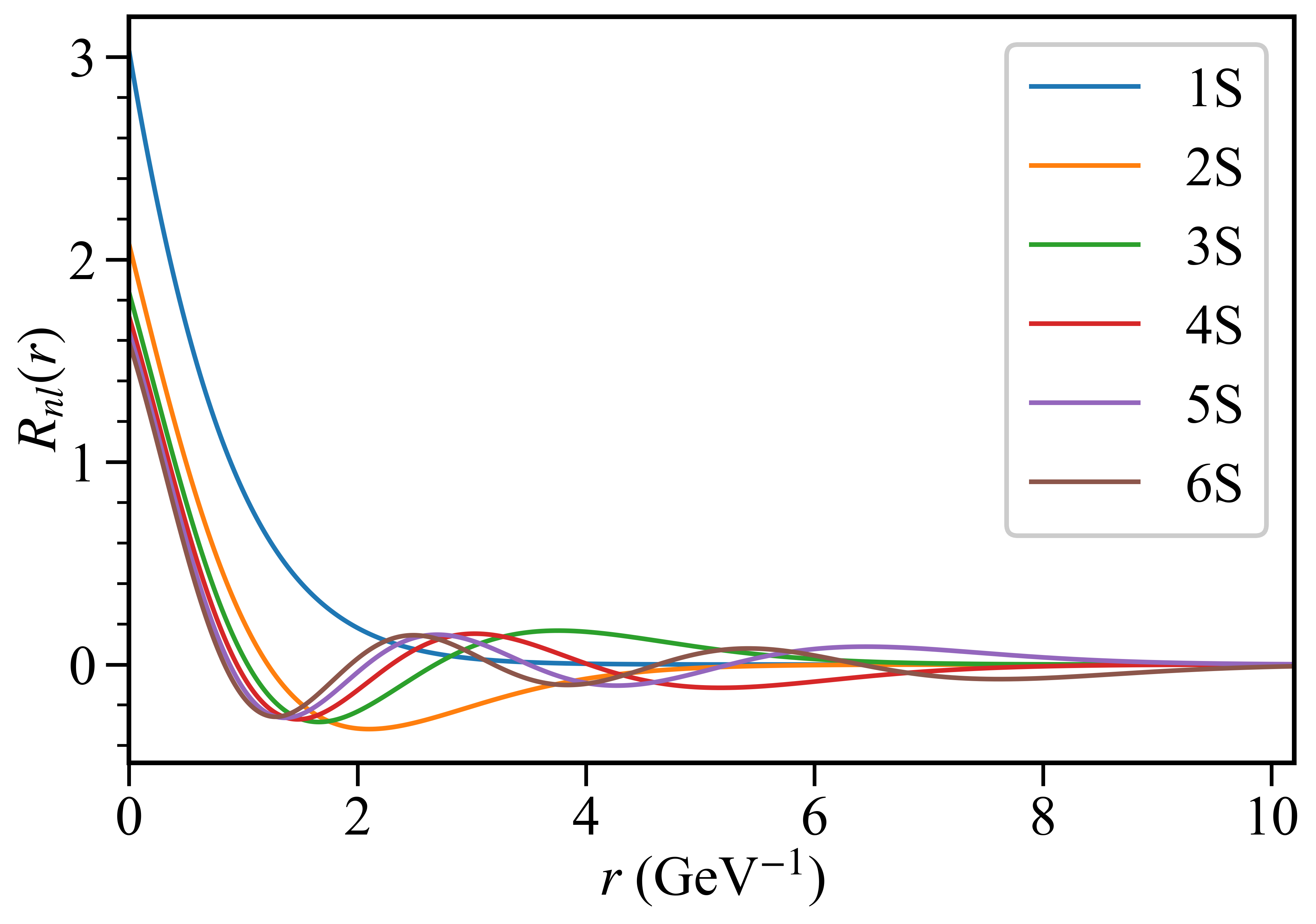}\label{fig:bottom_s_wfn_Rr_y}}
     \subfigure[$P$-wave $c\bar{c}$ states.]{\includegraphics[width=.471\textwidth]{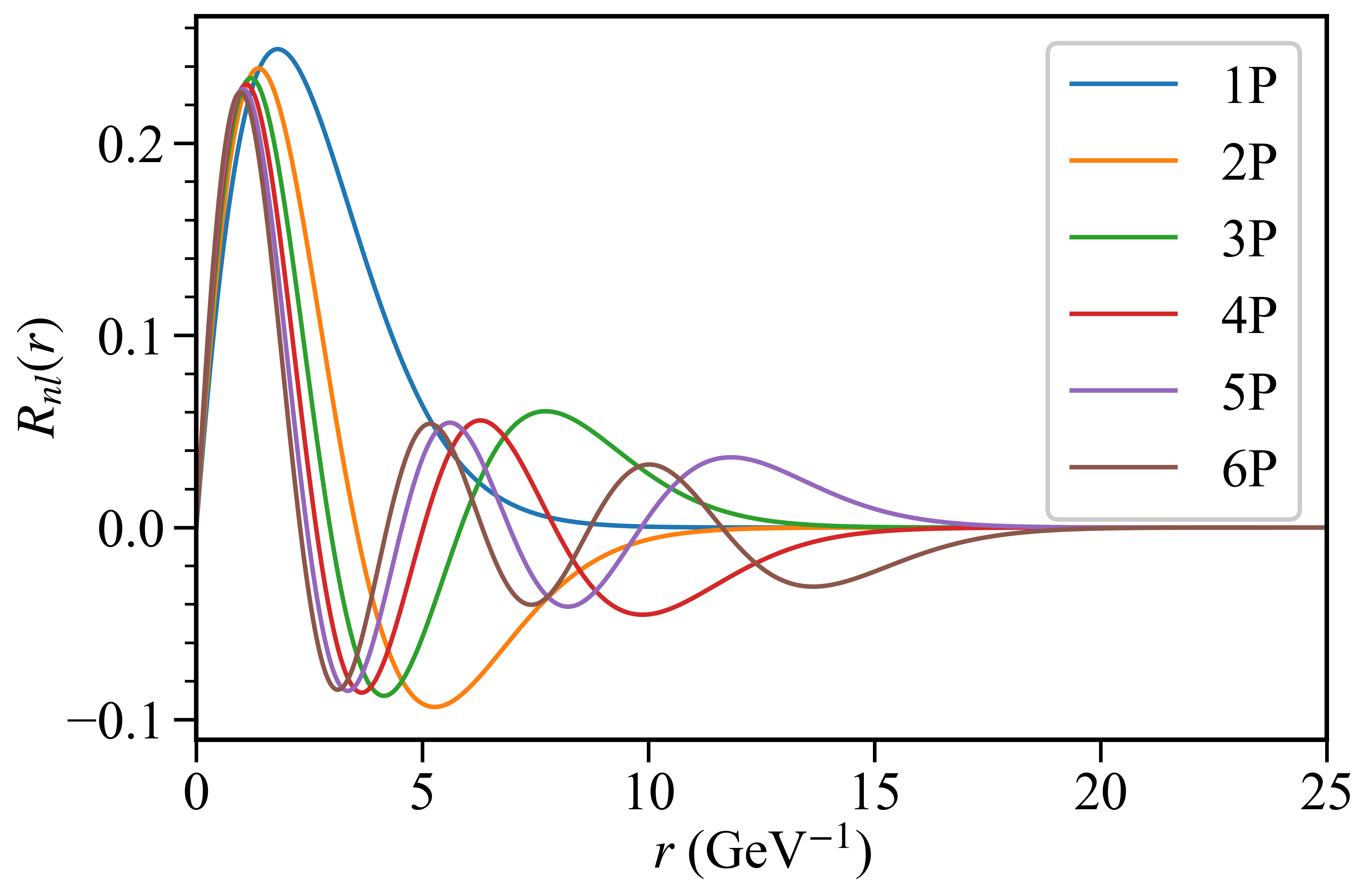}\label{fig:charm_p_wfn_Rr_y}}
     \qquad
     \subfigure[$P$-wave $b\bar{b}$ states.]{\includegraphics[width=.475\textwidth]{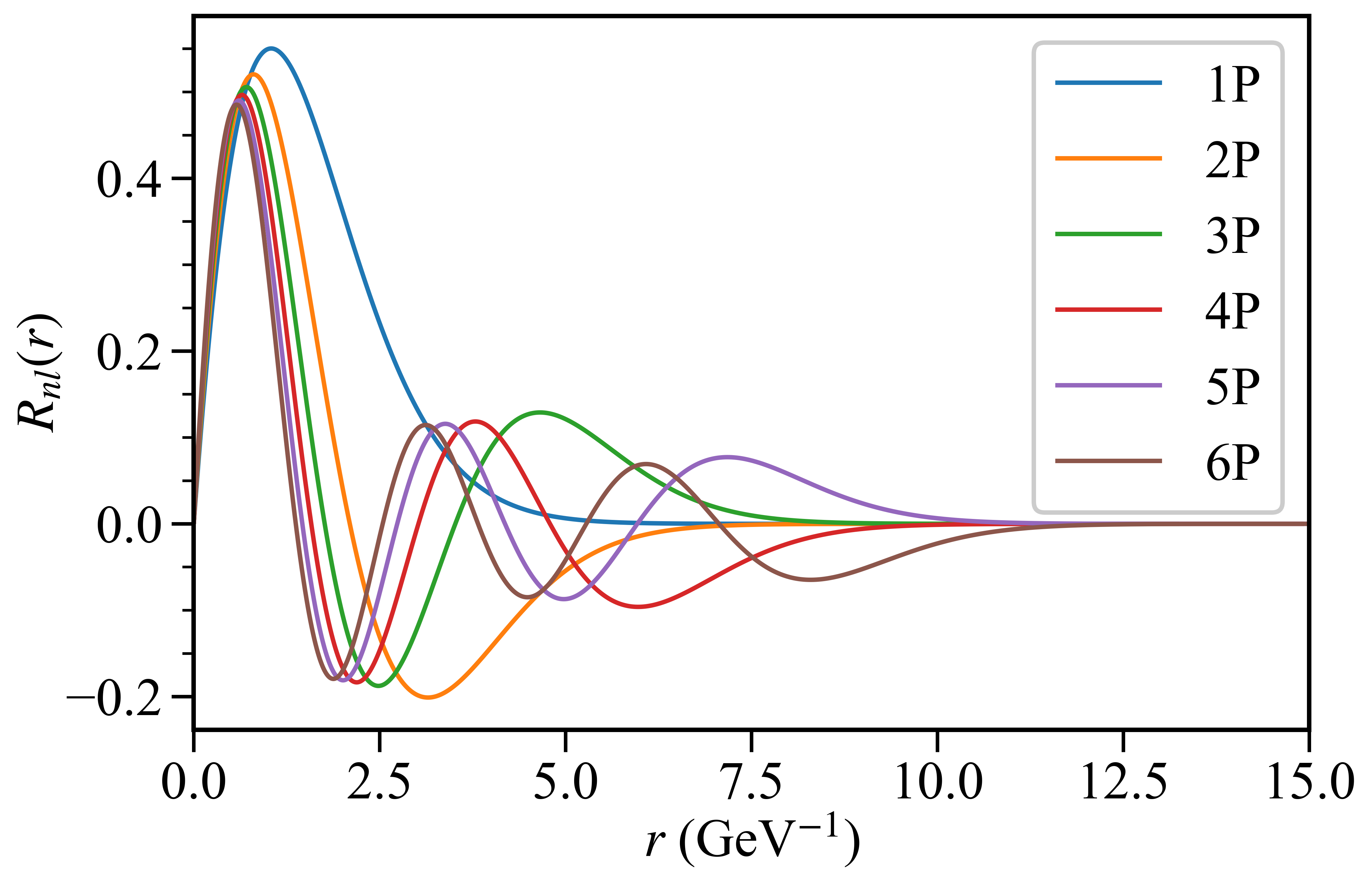}\label{fig:bottom_p_wfn_Rr_y}}
     \subfigure[$D$-wave $c\bar{c}$ states.]{\includegraphics[width=.472\textwidth]{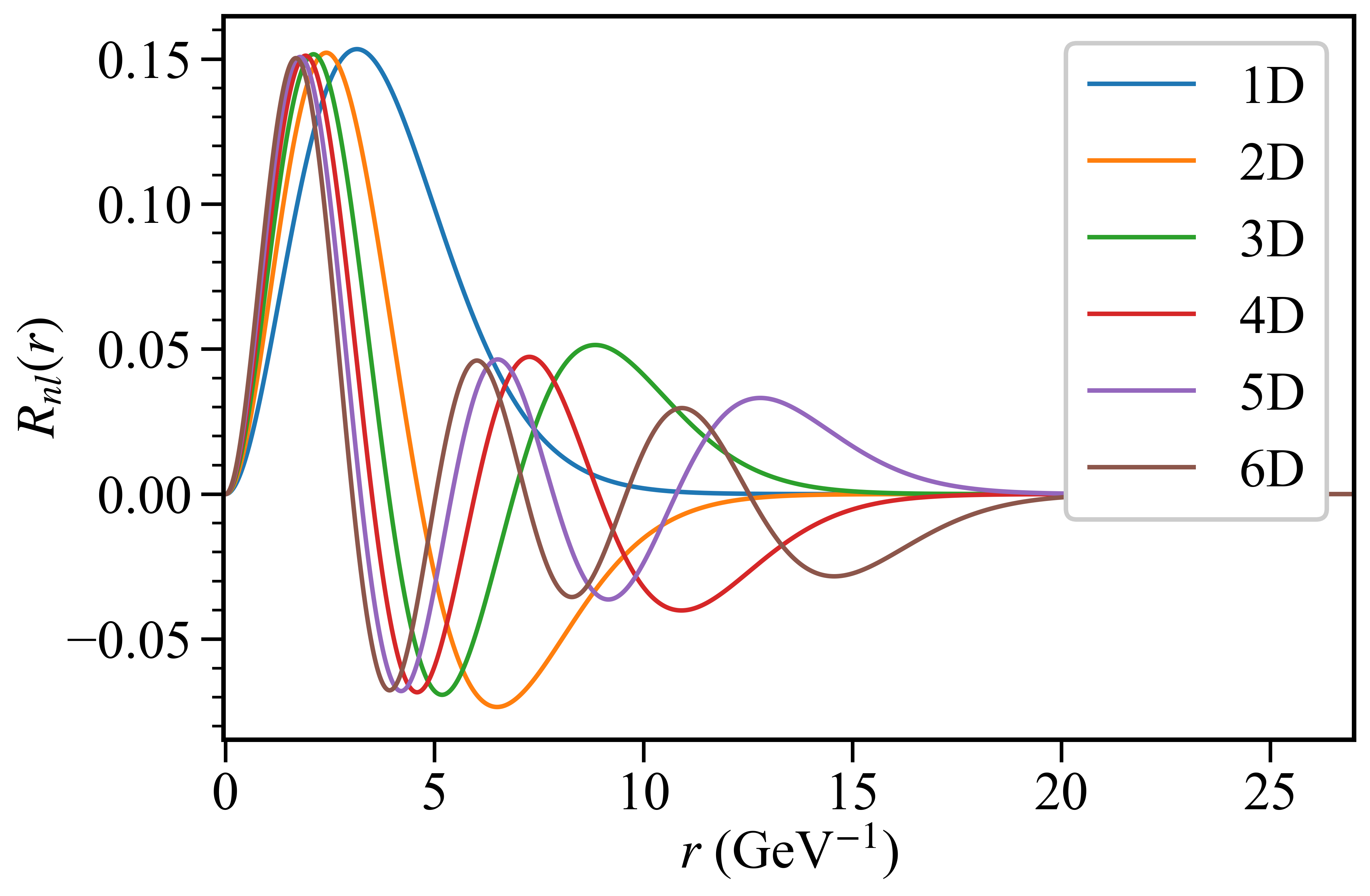}\label{fig:charm_d_wfn_Rr_y}}
     \qquad
     \subfigure[$D$-wave $b\bar{b}$ states.]{\includegraphics[width=.472\textwidth]{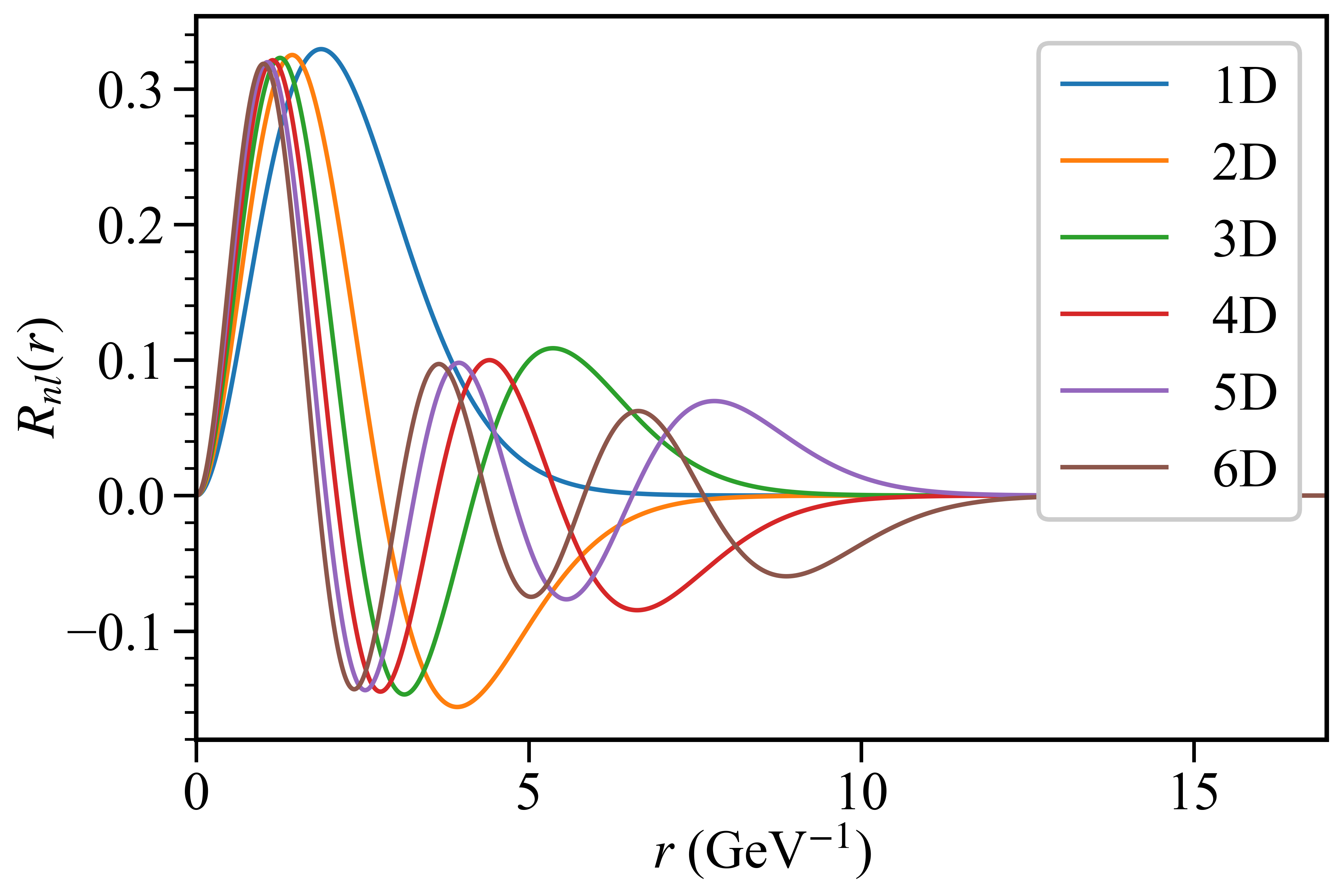}\label{fig:bottom_d_wfn_Rr_y}}
        \caption{Radial wave functions of charmonium (left panel) and bottomonium (right panel).}
        \label{fig:9_c_b_wfn_Rr_y}
\end{figure}

We estimate the radial wave function at the origin for $S$-wave states by extrapolating the corresponding numerical eigenfunction values at $r = 0$. For the orbitally excited $P$-wave states, we obtain the first derivative of their numerical eigenfunctions. We then extrapolate the derivative of the radial wave function, $R'_{nP}(r)$, to estimate its value at the origin ($r = 0$) for the corresponding $P$-wave states. Similarly, we find the second derivative of the numerical eigenfunctions of $D$-wave states and extrapolate them at $r = 0$. Subsequently, these estimates are squared to obtain the square of the radial wave function at the origin for $S$-wave states and its first non-vanishing derivative at the origin for $P$- and $D$-wave states, \textit{i.e.}, $|R^{(l)}_{nl}(0)|^2$. For $S$-wave quarkonia, we also establish the same using Eq.~\eqref{eqn:14_wfo_swave}. Our estimates for the charmonium and bottomonium states are listed in column $2$ of Tables~\ref{tab:4_c_states_wfo} and \ref{tab:5_b_states_wfo}, respectively. We also compare our numerical estimates with those of other theoretical approaches \cite{Liao:2014rca, Eichten:2019hbb} and give them in columns $3$ and $4$ of the above mentioned tables. Furthermore, we present the radial wave function plots of the charmonium and bottomonium states in Fig.~\ref{fig:9_c_b_wfn_Rr_y}. The following are our observations.

\begin{enumerate}
    \item We observe that the $|R^{(l)}_{nl}(0)|^2$ values decrease with an increase in the orbital quantum number ($l$). This suggests that the size of the bound state increases with $l$. We observe the same trend in our estimates for both charmonium and bottomonium states.

    \item For $S$-wave charmonium and bottomonium states, the $|R^{(l)}_{nl}(0)|^2$ estimates decrease with an increase in the principal quantum number ($n$). On the other hand, the $|R^{(l)}_{nl}(0)|^2$ values increase with $n$ for $P$- and $D$-wave states. This occurs due to the energy distribution of the states.

    \item We noticed that the $|R^{(l)}_{nl}(0)|^2$ depends on the phenomenological parameters ($\alpha_s$ and $\sigma$) of the potential, \textit{i.e.}, its numerical value increases with increasing $\alpha_s$ and $\sigma$. We wish to remark that $R_{nl}(r)$ is determined along with the masses of $c\bar{c}$ and $b\bar{b}$ states from the potential defined in Sec.~\ref{sec:2_methodology}. As shown in Sec.~\ref{subsec:A_ND_Mass_spectra}, we obtained reliable mass predictions which are in good agreement with the available experimental masses for both charmonium and bottomonium systems corresponding to the inputs defined in Table~\ref{tab:1_parameters}. Specifically, the parameter $\alpha_s$ plays a crucial role in the prediction of mass states, as it appears in all the interaction terms in the Hamiltonian. Consequently, $\alpha_s$ takes a distinct value and is fixed uniquely for determining the masses of charmonium and bottomonium states. Therefore, the estimated $|R^{(l)}_{nl}(0)|^2$ is essentially fixed corresponding to masses. However, $\alpha_s$ being scale dependent may lead to some complications in different processes. Therefore, the choice of $\alpha_s$ in different processes, for example, the masses and decay widths, requires different numerical values of $\alpha_s$ \cite{Kwong:1987ak}. This is evident when we move from the charmonium to bottomonium system.
    
    \item The radial wave function plots for bottomonium are narrow as compared to charmonium, although they show similar behavior corresponding to $r$, as shown in Fig.~\ref{fig:9_c_b_wfn_Rr_y}. Further, the radial wave function peaks for the $P$- and $D$-wave states are shifted from the origin in the positive $r$ direction as expected.
\end{enumerate}

We compare our $|R^{(l)}_{nl}(0)|^2$ estimates with those of Refs.~\cite{Liao:2014rca, Eichten:2019hbb} as shown in Tables~\ref{tab:4_c_states_wfo} and \ref{tab:5_b_states_wfo} for $c\bar{c}$- and $b\bar{b}$-systems, respectively. It is to be noted that $|R^{(l)}_{nl}(0)|^2$ is model dependent, corresponding to the choice of the potential. In addition, differences in the input parameters will lead to variations in the estimates of $|R^{(l)}_{nl}(0)|^2$ quoted by the various models. In the case of charmonium, our numerical results for $S$-wave estimates agree well with \cite{Eichten:2019hbb} for most states except for $S$-wave ground state, which is larger. However, we observe that our results are in general smaller than the estimates of \cite{Liao:2014rca}. The disagreement between the models is more pronounced for higher excited states. In the case of bottomonium, the disagreement between different models increases further due to substantial difference in the input parameters, as stated before. Thus, each model presents a different picture but at the same time also provides a reasonable range for $|R^{(l)}_{nl}(0)|^2$. Furthermore, despite the disagreement in the exact numerical estimates of $|R^{(l)}_{nl}(0)|^2$ for the heavy quarkonium states, the models show a similar behavior of $|R^{(l)}_{nl}(0)|^2$ with the variation of $n$ and $l$. This shows that the estimation of the square of the radial wave function at the origin is non-trivial. Additionally, the scale dependence of the wave function arising through $\alpha_s$ may also have different influence on different processes. Therefore, one shall expect variation of $|R^{(l)}_{nl}(0)|^2$ in different processes, for example, in annihilation decay widths, which we will consider in the next section.

\subsection{Annihilation Decay Widths}
\label{subsec:C_ND_Anni}
We calculated the decay widths of the leptonic and diphoton decays of various charmonium and bottomonium states by using our estimates of $|R^{(l)}_{nl}(0)|^2$ in the respective Eqs.~\eqref{eqn:16_leptonic_S}, \eqref{eqn:17_leptonic_D}, \eqref{eqn:19_diphoton_S} $-$ \eqref{eqn:21_diphoton_3P2}. Consequently, we list our predictions for the leptonic and diphoton decay widths in Tables~\ref{tab:6_charm_lep_dw},~\ref{tab:7_bottom_lep_dw},~\ref{tab:8_charm_2photon_dw}, and~\ref{tab:9_bottom_2photon_dw} for $c\bar{c}$- and $b\bar{b}$-systems, respectively.
\subsubsection*{Scale dependent correction factor f}
In Sec.~\ref{subsec:B_ND_wfo}, we remarked that the $|\Psi(0)|^2$ ($= \frac{|R_{nS}(0)|^2}{4\pi}$) is scale dependent and sensitive to variations in the phenomenological parameters, which makes it non-trivial to estimate. Further, as shown in Eq.~\eqref{eqn:16_leptonic_S}, $\alpha_{sc}$ and $|\Psi(0)|^2$ appear as a product and are usually treated as a constant in the decay width expression. However, in the calculation of decay widths, we fixed $\alpha_{sc}$ to a value different from that used for the determination of masses, which in turn yielded $|\Psi(0)|^2$. It may be noted that multiple studies have in the past exercised and argued in favor of different values of running strong coupling constant in masses and decay widths. On the other hand, $|\Psi(0)|^2$ determined from masses is kept constant for decay width calculations, where one shall expect a correction due to the variation of $|\Psi(0)|^2$ with respect to the scale corresponding to the strong coupling constant. At the same time, we should not ignore the fact that the determined $|\Psi(0)|^2$ provide reliable predictions for the masses when compared to the experiment, therefore, should be treated as the best estimate to begin with for the calculations of decay widths. In addition, the scale dependence of $\alpha_{sc}$ pose serious challenges to the estimation of its value at different mass scales. The decay width formula, Eq.~\eqref{eqn:16_leptonic_S}, is modified to include the scale dependence of the strong coupling constant and the correction factors are introduced as shown in \cite{Kwong:1987ak}. The $\alpha_{sc}$ is better understood at higher energy scales, \textit{i.e.}, $\sim\mathcal{O}(m_b^2)$, and its estimation is non-trivial at lower energy scale $\le 1$ GeV \cite{QuarkoniumWorkingGroup:2004kpm}. Therefore, a usual solution to explain the decay widths is by varying $\alpha_{sc}$ with fixed $|\Psi(0)|^2$ to compare with experimental numbers. However, the scale dependence of $|\Psi(0)|^2$ needs to be addressed. It faces a dual challenge owing to the fact that $|\Psi(0)|^2$ that explains the masses is different from the $|\Psi(0)|^2$ involved in the interaction processes and the scale dependence of $|\Psi(0)|^2$ appearing through the strong coupling constant in the short distance part of the potential. Thus, a phenomenological correction in $|\Psi(0)|^2$ can be used to understand the order of such scale dependent effects and to improve the decay width predictions. Therefore, we introduce the scale dependent correction factor $f$ to estimate effective $|\Psi(0)|^2$ involved in interaction processes. We determine $f$ independently for the charmonium and bottomonium systems by comparing the ratio of the experimental to the theoretical leptonic decay width of the lowest lying state. As shown in Eq.~\eqref{eqn:18_f_flav_fac}, all the factors in the ratio will cancel, except for $|\Psi(0)|^2$, resulting in a correction factor corresponding to the decay width. Similarly, for $S$-wave charmonium, we utilize Eq.~\eqref{eqn:19_diphoton_S} to compare the experimental and theoretical diphoton decay widths and estimate the corresponding scale dependent factor $f_\gamma$ from their ratio. The experimental leptonic decay widths of $J/\psi{(1 {}^3S_1)}$ and $\Upsilon{(1 {}^3S_1)}$ are $(5.53 \pm 0.10)$~keV and $(1.340 \pm 0.018)$~keV, respectively, which yield correction factors $f = 0.768$ for charmonium and $f = 0.836$ for bottomonium, corresponding to their theoretical values using Eq.~\eqref{eqn:18_f_flav_fac}, \textit{i.e.},
\[
    f = \frac{\Gamma_{expt}(J/\psi{(1 {}^3S_1)} \rightarrow e^+e^-)}{\Gamma_{theo}(J/\psi{(1 {}^3S_1)} \rightarrow e^+e^-)}  = 0.768 \text{ for $J/\psi$ states, and }
\]
\[
    f = \frac{\Gamma_{expt}(\Upsilon{(1 {}^3S_1)} \rightarrow e^+e^-)}{\Gamma_{theo}(\Upsilon{(1 {}^3S_1)} \rightarrow e^+e^-)}  = 0.836 \text{ for $\Upsilon$ states}.
\]
Additionally, we obtain $f = 8.078$ from the ratio of the experimental value $\Gamma(\psi_1{(1 {}^3D_1)} \rightarrow e^+e^-) = (0.256 \pm0.016)$ keV to the theoretical prediction for $D$-wave charmonium. Likewise, the experimental diphoton decay widths of low-lying states $\Gamma(\eta_c{(1 {}^1S_0)} \rightarrow \gamma \gamma) = (5.4 \pm0.4$)~keV, $\Gamma(\chi_{{c0}}{(1 {}^3P_0)} \rightarrow \gamma \gamma) = (2.2 \pm 0.2)$~keV, and $\Gamma(\chi_{{c2}}{(1 {}^3P_2)} \rightarrow \gamma \gamma) = (0.56 \pm 0.03)$~keV and their corresponding theoretical predictions are used to estimate the scale dependent correction factors $f_\gamma = 0.384$ for $\eta_c$, $f_\gamma = 0.382$ for $\chi_{{c0}}$ and $f_\gamma = 0.757$ for $\chi_{{c2}}$ states, respectively. Thus, the scale dependent factors, estimated from the low-lying states, will introduce a uniform correction to match the decay with of low-lying states with the experiment. Subsequently, we utilize the same scale dependent correction factors to predict the decay widths of higher excited charmonium and bottomonium states. 

\subsubsection{Leptonic Decay Widths}
In this subsection, we discuss the leptonic decay widths of the charmonium and bottomonium states. We list our predictions for the leptonic decay widths of charmonia and bottomonia, corresponding to Eqs.~\eqref{eqn:16_leptonic_S}$-$\eqref{eqn:17_leptonic_D} for the lowest order (without QCD corrections), with the QCD corrections, and with the scale dependent factor in columns $2-4$, as shown in Tables~\ref{tab:6_charm_lep_dw} and \ref{tab:7_bottom_lep_dw}. Furthermore, we compare our predictions of charmonium decay widths with the available experimental decay widths \cite{ParticleDataGroup:2022pth} and predictions of other theoretical works, \textit{i.e.}, SP \cite{Li:2009zu}, NR \cite{Soni:2017wvy}, NRC \cite{Kher:2018wtv}, and SR \cite{Radford:2007vd}, listed in columns $5-9$ of Table~\ref{tab:6_charm_lep_dw}. Note that the experimental state $\psi(4040)$ is taken as the theoretical $\psi(3^3S_1)$ state and its decay width is used in our comparison. We observe the following:

\begin{enumerate}
    \item Our $S$-wave decay widths without QCD correction are more than twice of their corresponding experimental values, whereas inclusion of the QCD correction term improves the consistency with respect to the experimental results. We note that the QCD correction term introduces a uniform reduction throughout the leptonic decay widths of charmonium roughly by $50\%$. This reduction in the decay widths corresponds to the QCD corrections obtained at $\alpha_{sc} = 0.3$. However, a discrepancy between the experimental and theoretical numbers still exists. It maybe noted that a larger numerical value of $\alpha_{sc} \sim (0.37 - 0.44)$, can lead to a better agreement with the experimental numbers for the observed processes. However, the past studies have suggested that $\alpha_{sc}$ ranges from  $0.2 - 0.3$ to explain the leptonic decay widths in the charmonium system~\cite{Lansberg:2009xh}. It may further be noted that to explain the masses, $\alpha_{s}$ is taken to be very large as compared to the decay widths, which affects the determination of $|\Psi(0)|^2$. A similar observation can be made based on the lattice calculations. Therefore, the variation of $\alpha_s$ itself is not sufficient to explain the experimental measurements.

    \item It can be pointed out that the aforementioned product $\alpha_{sc}|\Psi(0)|^2$ (which appears when including QCD the correction term) is crucial to resolve the disagreement between theory and experiment. Both being scale dependent shall be treated independently. Therefore, we phenomenologically express the scale dependent correction factor $f$ to address the variation in $|\Psi(0)|^2$ corresponding to scale while keeping $\alpha_{sc}$ fixed with respect to the available information in the literature. As mentioned before, the other analysis of the charmonium decay widths in the literature favors a smaller value of $\alpha_{sc}$ as used in the present calculation, therefore, a decrease in $\alpha_{sc}$ will need a larger correction factor with respect to $|\Psi(0)|^2$ to explain the experimental numbers.

    \item The scale dependent factor $f$, which is calculated from the lowest lying charmonium state, introduces a correction of $\sim 25\%$ to match with experiment. We introduce the same correction factor to predict the decay widths of the excited states of charmonium which agree well with available experimental results, except for the decay of $\psi(3^3S_1)$ state as shown in column $4$ of Table \ref{tab:6_charm_lep_dw}. The discrepancy generally exists for $\Gamma(\psi(3^3S_1) \rightarrow e^+e^-)$ corresponding to all the other theoretical models which predict larger decay widths for the $\psi(3^3S_1)$ decay. Further, it may be noted that most recently Mo \textit{et al.} \cite{Mo:2010bw} has calculated the range of this decay from $(0.6 - 1.4)$ keV corresponding to four-fold ambiguity due to relative phases, which agrees well with our predictions.
    
    \item Similarly, in $D$-wave decay of $\psi(1^3D_1) \rightarrow e^+e^-$, we notice that none of the predictions including the lowest order and first-order QCD correction could match the experimental observation. The comparison with the experiment yields a large correction of $\sim 8.078$, using which we predict the decay widths of the remaining $D$-wave excited states. It is interesting to note that the leptonic decay widths of $D$-wave states increase with $n$, whereas they decrease for $S$-wave states with an increase in $n$.
\end{enumerate}

We compare our predictions for $S$-wave charmonium with those of other theoretical works and notice considerable variations among the models. Our values are in agreement with the SP \cite{Li:2009zu} approach, where their values for the decay widths without and with correction are roughly similar to our predictions, particularly for the latter. However, their predictions for higher excited states, beyond $\psi(3^3S_1)$, are smaller than ours. The same is observed when we consider the predictions with correction factor. Throughout the results, we observe that the predictions from the NR \cite{Soni:2017wvy} approach are smaller than ours and the experimental results. Similarly, NRC \cite{Kher:2018wtv} quotes smaller values than ours both with and without the QCD correction. Finally, the SR \cite{Radford:2007vd} approach predictions are smaller than ours, except in the case of $\psi(4^3S_1)$. In the case of $D$-wave charmonia, a discrepancy in the order of the decay width values is noted in all other theoretical approaches except for NRC \cite{Kher:2018wtv}. Our results with the correction factor agree well with those of NRC \cite{Kher:2018wtv} predictions without QCD corrections. Their predictions are smaller than ours with correction factor $f$. However, our results are in general larger than the predictions of SP \cite{Li:2009zu} for the cases without correction factor $f$.

We observe that our predictions for the $S$-wave charmonium improve drastically with the inclusion of the scale factor $f$; however, these results may suffer uncertainties corresponding to variation in the numerical value of $\alpha_{sc}$. Moreover, the corresponding uncertainties will also affect the numerical estimates of wave function at the origin in different models. Additionally, we notice that the $D$-wave $|R^{(l)}_{nl}(0)|^2$ estimates fall well short of experimental expectations to have larger ambiguities than the $S$-wave states. It may be emphasised that the wave function of quarkonia being non-zero at long distance may experience non-perturbative effects. In addition, there may be other QCD corrections in higher powers of $\alpha_{sc}$ to the decay width expressions, but these corrections are expected to be smaller in magnitude. Therefore, owing to such corrections and numerical uncertainties arising from the model dependence of wave function and strong coupling constant, the decay width predictions shall be treated as estimates \cite{Godfrey:1985xj, Godfrey:2015dia}. Nonetheless, improvements of numerical decay width estimates can not be ignored with inclusion of phenomenological scale dependent factors.

Now, we shift our focus to our predictions for the leptonic decay widths of bottomonium states listed in columns $2-4$ of Table~\ref{tab:7_bottom_lep_dw}. For the sake of comparison, we have given the available experimental values \cite{ParticleDataGroup:2022pth} and predictions from other theoretical models, \textit{i.e.}, GI \cite{Godfrey:2015dia}, MGI \cite{Wang:2018rjg}, CQM \cite{Segovia:2016xqb}, SP \cite{Li:2009nr}, NR \cite{Soni:2017wvy}, IMWC \cite{Pandya:2021vby}, and NRC \cite{Kher:2022gbz}, given in columns $5-12$ of Table~\ref{tab:7_bottom_lep_dw}. Our observations are as follows.

\begin{enumerate}
    \item Comparing our $S$-wave decay widths with the experimental values, we observe that the decay widths for the lowest order (without QCD correction) are overestimated; however, the inclusion of the QCD correction term reduces the numerical values roughly by $35\%$. As observed in the case of charmonium, the choice of $\alpha_{sc} = 0.2$ for the bottomonium case considerably improves our decay width predictions. It may be noted that the margin of variation in the numerical value of $\alpha_{sc}$ for the bottomonium sector is very small, \textit{i.e.}, literature supports a robust value of around $\alpha_{sc} \sim 0.2$ \cite{Kwong:1987ak, McNeile:2010ji, Chakraborty:2014aca}. However, similar to the charmonium decay widths, the experimental observations can be explained by and large with $\alpha_{sc}$ ranging from $0.27 - 0.33$ for most of the observed decay widths. Here again, the $\alpha_{s}$ used for masses is considerably different from its value for decay widths, which should affect the scale dependence of $|\Psi(0)|^2$. Therefore, similar to charmonium, we consider the incorporation of the scale dependent factor $f$ in the bottomonium case.

    \item We observe that our predictions for bottomonium leptonic decay widths with the correction factor $f=0.836$ are in good agreement with the experimental values for all $S$-wave states, except for $\Upsilon{(6 {}^3S_1)}$ which is overestimated. We wish to point out that the $\Upsilon{(6 {}^3S_1)}$ decay width requires an extremely large value of $\alpha_{sc} \sim 0.43$ without correction factor to match the experimental observation. Compared to the results with the correction factor in charmonium $S$-wave states, we observe that the bottomonium $S$-wave predictions with the correction factor are more consistent with the experiment. This suggests that the phenomenological estimation of the scale dependent factor $f$ with respect to $|\Psi(0)|^2$ improves our results in comparison to the experiment for both charmonium and bottomonium.

    \item In the absence of experimental observations, we restrict our predictions to the lowest order and first-order QCD corrections for the leptonic decay widths of $D$-wave states. However, we observe that the leptonic decay widths for the $D$-wave states show an increasing trend with respect to the increase in radial excitations, in contrast to those of $S$-wave states.
\end{enumerate}

Comparing our $S$-wave leptonic decay width predictions inclusive of QCD correction term with those of other theoretical models \cite{Godfrey:2015dia, Wang:2018rjg, Segovia:2016xqb, Li:2009nr, Soni:2017wvy, Pandya:2021vby, Kher:2022gbz}, we observe that all the model predictions are of the same order as compared to the experiment \cite{ParticleDataGroup:2022pth}, however, they differ in magnitude. On the other hand, for $D$-wave our results for the low-lying states are smaller as compared to all the other models. In particularly, the results of IMWC \cite{Pandya:2021vby} are larger in magnitude than those of the other models, even with the inclusion of QCD corrections. Thus, it may be pointed out that the scale dependent factor $f$ for the charmonium and bottomonium states improves the agreement between theory and experiment for a particular numerical value of the strong coupling constant favored by the literature. As pointed out before, the decay width estimates may have additional higher order QCD corrections, including non-perturbative effects; however, the magnitude of these corrections would be smaller for the bottomonium system.

\subsubsection{Diphoton Decay Widths}

Our predictions for diphoton decay widths are listed in Tables~\ref{tab:8_charm_2photon_dw} and \ref{tab:9_bottom_2photon_dw} for the pseudoscalar (${}^1S_0$), scalar (${}^3P_0$), and tensor (${}^3P_2$) states of charmonium and bottomonium, respectively. We compare our predictions with available experimental values and other theoretical results. In Table~\ref{tab:8_charm_2photon_dw}, we list our results for charmonium diphoton decay widths for the lowest order and with QCD correction in columns $2$ and $3$, respectively. In column $4$, we provide our predictions for the charmonium diphoton widths corresponding to the correction factor $f_\gamma$. As done for leptonic decay widths, $f_\gamma$ for $S$-wave is obtained from the ratio of the lowest lying experimental decay width of $\eta_c({}^1S_0)$ to its theoretical value with QCD correction. We compare our predictions with the experimental results \cite{ParticleDataGroup:2022pth} and the theoretical predictions of SP \cite{Li:2009zu}, NR \cite{Soni:2017wvy}, and NRC \cite{Kher:2018wtv} given in columns $5-8$ of Table~\ref{tab:8_charm_2photon_dw}. Our observations are as follows.

\begin{enumerate}
    \item In the case of pseudoscalar states decaying into photons, we observe that our prediction for the $\eta_{{c}}{(1 {}^1S_0)}$ state is nearly four-fold the experimental value at the lowest order. The inclusion of the QCD correction, for $\alpha_{sc}$ fixed at $0.3$, yields numerical values that are smaller by $\sim 30\%$ than those for the lowest order. However, the numerical result of $\Gamma(\eta_c{(1 {}^1S_0)} \rightarrow \gamma \gamma)$ is yet larger as compared to experimental observation roughly by a factor of $2$. It may be noted that diverse values of $\alpha_{sc}$ are used in the literature to explain this result, despite of which the theoretical predictions are larger than the experimental result. In our case, \textit{e.g.}, if we use $\alpha_{sc}$ as large as $0.56$ in the QCD correction term, our numerical value becomes smaller in magnitude but fails to match the experiment. Further, $\alpha_{sc}=0.56$ introduces a uniform reduction in the decay width predictions $\sim \mathcal{O}(42\%)$ as compared to results at $\alpha_{sc} = 0.3$. It is important to note that even a larger correction corresponding to higher values of $\alpha_{sc}$ is insufficient to explain the experimental results. Therefore, the introduction of a scale dependent correction factor corresponding to $|\Psi(0)|^2$ can be justified. Furthermore, it can be argued that the scale dependence of $\alpha_{sc}$ corresponding to different processes cannot be ignored; however, at the same time, the scale dependence of $|\Psi(0)|^2$ should also have a considerable effect \cite{Kwong:1987ak}. Therefore, we include the scale dependent phenomenological correction factor, $f_\gamma = 0.384$, corresponding to the experiment. Utilizing this correction factor, we predict the diphoton decay widths for higher excitations of $S$-wave charmonium, as shown in column $4$ of Table~\ref{tab:8_charm_2photon_dw}.
    
    \item Our decay width prediction for $P$-wave scalar state $\chi_{{c0}}{(1 {}^3P_0)}$ is larger by a factor of more than $2$ as compared to the experimental observation. It is worth pointing out that the QCD correction term in the diphoton decay width expression of $P$-wave scalar states is positive; however, it is marginally affected by the scale dependent variation of $\alpha_{sc}$.  Therefore, analogous to the diphoton decays of pseudoscalar states, a scale dependent correction is required to obtain the experimental decay width. Thus, we determine the scale correction factor $f_\gamma = 0.382$ from the ratio of the experimental decay width $\Gamma(\chi_{{c0}}{(1 {}^3P_0)} \rightarrow \gamma\gamma)$ and the theoretical prediction (with QCD correction), which is used to predict the decay widths of higher excited $P$-wave scalar states.

    \item In the case of the tensor state, $\chi_{{c2}}{(1 {}^3P_2)}$, we observe that our prediction with the QCD correction term is marginally, \textit{i.e.}, $\sim 25\%$, larger than the experimental value. Therefore, we estimate the scale correction factor $f_\gamma$ as $0.757$ for the tensor states. Note that our estimates for the wave function at the origin (as shown in calculation for determination of masses) do not discriminate between the $P$-wave states. Therefore, the observation of such decay widths can shed more light on the effects of scale dependence which influence the decay dynamics in such processes.

    \item We observe that the diphoton decay widths of the scalar and tensor states increase in magnitude as we move from ground to higher excited states, whereas the decay widths of the pseudoscalar states exhibit the opposite trend.
\end{enumerate}

In regards to other theoretical approaches, we notice that our predictions for the pseudoscalar states decaying into photons are larger than those of all other theoretical works, namely, SP \cite{Li:2009zu}, NR \cite{Soni:2017wvy}, and NRC \cite{Kher:2018wtv}. It is interesting to note that all the theoretical estimates are larger than the experimental value, which in turn highlights the differences among the various models corresponding to multiple parameters. In addition, our results show a uniform increase or decrease in the decay widths corresponding to decay rate formulae~\eqref{eqn:19_diphoton_S}$-$\eqref{eqn:21_diphoton_3P2}; however, a few of the models, \textit{i.e.}, SP \cite{Li:2009zu} and NR \cite{Soni:2017wvy}, deviate from this trend, as shown in Table~\ref{tab:8_charm_2photon_dw}. Furthermore, the aforementioned larger value of $\alpha_{sc}$, \textit{e.g.}, $0.6$, although shows a relatively better agreement with the experiment for the lowest lying states, fails to explain the decay widths of excited states by a huge margin.

In Table~\ref{tab:9_bottom_2photon_dw}, we present our predictions for the diphoton decay widths of bottomonium, as shown in columns $2$ and $3$. It is worth mentioning that experimental results for diphoton decay widths are unavailable for the bottomonium system; therefore, we compare our predictions with those of other theoretical models, such as GI \cite{Godfrey:2015dia}, MGI \cite{Wang:2018rjg}, CQM \cite{Segovia:2016xqb}, SP \cite{Li:2009nr}, NR \cite{Soni:2017wvy}, IMWC \cite{Pandya:2021vby}, and NRC \cite{Kher:2022gbz}, given in columns $4-10$. The following are our observations.

\begin{enumerate}
    \item In the case of $S$-wave states, we observe that the QCD correction term with $\alpha_{sc} = 0.2$ leads to $\sim 20\%$ reduction in the decay width predictions as compared to the lowest order results. We wish to remark that the results for the leptonic decay widths of bottomonium show lesser discrepancies with respect to the experiment than charmonium. Therefore, we expect that the numerical results for the diphoton decay widths should be reliable with QCD corrections. In addition, in the absence of experimental results, the scale dependent correction factor $f_\gamma$ for the bottomonium states is difficult to estimate. However, the comparison between the charmonium and bottomonium sectors may provide a reasonable explanation of the range for the scale correction factor. We wish to remark that the ratio between the leptonic (Eq.~\eqref{eqn:16_leptonic_S}) and diphoton (Eq.~\eqref{eqn:19_diphoton_S}) decay width expressions for the $S$-wave ground states yields a dynamical relation based on proportionality, \textit{i.e.}, 
    \[\frac{\Gamma({1}^3S_1 \rightarrow e^+e^-)}{\Gamma({1}^1S_0 \rightarrow \gamma\gamma)} \propto \frac{4}{9}\frac{m_Q^2}{M_{1S}^2 e^2_Q}.\] 
    The square of the quark charge in this relation leads to a notable change in the dynamical factor, which means that bottomonium diphoton decay widths are expected to be substantially enhanced as compared to that of charmonium. 
    In fact, we observe that the ratio 
    \[\frac{\Gamma({1}^3S_1 \rightarrow e^+e^-)}{\Gamma({1}^1S_0 \rightarrow \gamma\gamma)} = 0.511 ~\text{for}~ \alpha_{sc} = 0.3\] 
    in the case of charmonium decays, while we get,  
    \[\frac{\Gamma({1}^3S_1 \rightarrow e^+e^-)}{\Gamma({1}^1S_0 \rightarrow \gamma\gamma)} = 2.599 ~\text{with}~ \alpha_{sc} = 0.2\] 
    for bottomonium. 
    The ratio of the experimental to theoretical decay widths can be related to scale dependent factors as 
    \[\frac{\Gamma({1}^3S_1 \rightarrow e^+e^-)}{\Gamma({1}^1S_0 \rightarrow \gamma\gamma)} \times \frac{\Gamma_{expt}({1}^1S_0 \rightarrow \gamma\gamma)}{\Gamma_{expt}({1}^3S_1 \rightarrow e^+e^-)} = \frac{f_\gamma}{f},\] 
    which gives 
    $\frac{f_\gamma}{f} \sim 0.5$ for charmonium decays where the ratio 
    \[\frac{\Gamma_{expt}({1}^1S_0 \rightarrow \gamma\gamma)}{\Gamma_{expt}({1}^3S_1 \rightarrow e^+e^-)} \sim 1.\] 
    This explains the reduction in the diphoton decay width predictions for charmonium, inclusive of QCD correction, corresponding to the scale correction factor $f_\gamma$. Similarly, for bottomonium decays, we can obtain a reasonable range of diphoton decay widths in the absence of experimental results. Based on the trend of various theoretical models, one may expect the diphoton decay width to be smaller than the leptonic decay width of the bottomonium vector ground state; accordingly, we assume 
    \[\frac{\Gamma_{expt}({1}^1S_0 \rightarrow \gamma\gamma)}{\Gamma_{expt}({1}^3S_1 \rightarrow e^+e^-)} \lesssim 1.\] 
    Therefore, we believe that $\frac{f_\gamma}{f} \lesssim 2.6$ may be considered as a phenomenological upper bound for the diphoton decay of pseudoscalar bottomonium in our results, for a given value of $f$. Thus, assuming that the diphoton decay width of the pseudoscalar state is equal to the leptonic decay width of the lowest lying vector state with $f= 0.836$, we get $f_{\gamma} = 2.17$, leading to $\Gamma(\Upsilon({1}^1S_0) \rightarrow \gamma\gamma )= 1.34$ keV. On the other hand, for the leptonic decay width to be twice as large as the diphoton decay width, we get $f_{\gamma}=1.09$ yielding $\Gamma(\Upsilon({1}^1S_0) \rightarrow \gamma\gamma)= 0.67$ keV. Nevertheless, both of these assumptions lead to an estimate of the diphoton decay width comparable to the predictions of other theoretical models. Consequently, unlike charmonium, the diphoton decay width predictions of bottomonium pseudoscalar states are expected to be enhanced corresponding to the scale dependent correction factor. Therefore, we believe that if in case a phenomenological correction factor is required from prospective experimental results, the correction factor will act as an enhancement rather than a reduction as noted in charmonium.

    \item It may be noted that we observe a similar decay width pattern in bottomonium diphoton decays as observed for charmonium. As expected, the value of decay widths decreases for the $S$-wave states and increases for the $P$-wave states as we move from ground to higher excited states. Aforementioned, this behavior can be attributed to the corresponding change in wave function at the origin. The QCD correction terms, likewise, decrease the decay width values for the pseudoscalar and tensor states while increasing the decay widths of the scalar states. We also observe that the diphoton decay predictions of bottomonium states are smaller than those of their charmonium counterparts.
\end{enumerate}

As observed in the case of charmonium, here again, the model dependence and various parameters lead to a large spectrum of decay width predictions among other theoretical works. Nevertheless, most of the predictions are of the same order with a few exceptions. For the pseudoscalar states, we observe that the predictions of the relativistic models of GI \cite{Godfrey:2015dia} and MGI \cite{Wang:2018rjg} are larger than those of all the other models, including ours, for the lower states, while being mostly comparable for the higher excitations. The approaches of SP \cite{Li:2009nr}, CQM \cite{Segovia:2016xqb}, and NR \cite{Soni:2017wvy} compare well with our predictions, with some exceptions. IMWC \cite{Pandya:2021vby} and NRC \cite{Kher:2022gbz} give predictions smaller than ours for all pseudoscalar states. In the case of the diphoton decay widths of scalar states, we observe that the GI \cite{Godfrey:2015dia} and MGI \cite{Wang:2018rjg} models predict much larger decay widths than ours. Interestingly, in contrast to our decay widths, which show a gradual increase, the GI \cite{Godfrey:2015dia} and MGI \cite{Wang:2018rjg} predictions show an opposite trend. These opposite trends among various models are mainly due to the varying numerical values of the wave function at the origin. For the tensor states decaying into photons, we observe that our predictions are in agreement with the relativistic models of GI \cite{Godfrey:2015dia} and MGI \cite{Wang:2018rjg}. Along with our model, both relativistic models show an increasing trend in the decay width values. It may be noted that CQM \cite{Segovia:2016xqb}, SP \cite{Li:2009nr}, and NRC \cite{Kher:2022gbz} predictions are exceedingly small as compared to the rest of the models.

Thus, for all states, we observe that our predictions fall within the extremes of other theoretical approaches. A consensus between the various model predictions for the decay widths can be attained with the availability of experimental data. However, precise estimation of the scale dependence will remain a challenge. We expect that our results will shed light on the discrepancy between theory and experiment. Given the ambiguities in the theoretical methodologies as well as experimental observations, the study of decay widths requires huge efforts from both theory and experiment to understand the underlying physics.

\section{Summary and Conclusions}
\label{sec:4_sum_conclu}

In the present work, we studied charmonium and bottomonium spectroscopy using Cornell potential modified by an additional constant potential term in a non-relativistic framework. We predicted masses and estimated the square of the radial wave function at the origin by solving the Schrodinger wave equation. First, we obtained the degenerate masses of the heavy quarkonium ground and excited states. Later, we calculated the hyperfine splitting between the respective $S$-, $P$-, and $D$-wave states through the perturbative inclusion of spin-dependent correction terms, which resulted in the non-degenerate masses of corresponding $n^{2S+1}L_{J}$ states of charmonium and bottomonium (up to $n=6$ and $L=2$). We optimized our potential parameters to provide reliable predictions of the mass spectrum and obtained the corresponding estimates of the wave function at the origin. In addition, we predicted the spin splitting among various states and compared our results with experiment. We then studied the Regge trajectories of charmonium and bottomonium systems and discussed the non-linear nature of the trajectories observed in low-lying states. Furthermore, we investigated the annihilation decay widths of charmonium and bottomonium, although we restricted our analysis to leptonic and diphoton decay widths. We predicted the decay widths of heavy quarkonium decays involving leptonic and diphoton decay processes to both lowest order and with QCD correction terms involving strong coupling constant $\alpha_s$. Additionally, we observed the scale dependence of $\alpha_s$ on these processes by using a different numerical value and extended our phenomenological analysis to study the scale dependence of the wave function at the origin for the same. Following which, we predicted the leptonic and diphoton annihilation decay widths for both the heavy quarkonium systems inclusive of phenomenological scale dependent factor. We list our major findings and conclusions as follows:
\begin{itemize}
    \item We observed that our masses are in very good agreement with experimental results, including predictions of the excited states of charmonium and bottomonium. We note that the margin of error as compared to experimental results is lesser in bottomonium than charmonium.
    \item We found that hyperfine mass splitting between the lowest lying pseudoscalar and vector states, \textit{i.e.}, $M_{1^3S_1} - M_{1^1S_0}$ and other higher order mass splittings, namely, $M_{\psi(2S)} - M_{J / \psi(1S)}$, $M_{\Upsilon(2S)} - M_{ \eta_{b}(2S)}$, ${M}_{\Upsilon(3S)} - {M}_{\Upsilon(2S)}$ are in excellent agreement with experimental results. 
    \item We analyzed the Regge trajectories of quarkonia that follow non-linear behavior for low-lying states. The non-linearity is more pronounced in the bottomonium system than in charmonium. The non-linearity in the trajectories indicates that both the linear and Coulomb parts of the potential play equally important roles in the case of the ground and lowest excited states. 
    \item In order to study annihilation decay widths, we obtained the wave function at the origin, $|R^{(l)}_{nl}(0)|^2$, for $c\bar{c}$- and $b\bar{b}$-systems. We observed that the $|R^{(l)}_{nl}(0)|^2$ depends on the phenomenological parameters ($\alpha_s$ and $\sigma$) of the potential. We found that $|R^{(l)}_{nl}(0)|^2$ is model dependent corresponding to the choice of the potential, and $\alpha_s$, particularly, plays a significant role in mass predictions. However, the $|R^{(l)}_{nl}(0)|^2$ and $\alpha_s$ both appear in the decay widths as a product; therefore, the scale dependence of $\alpha_s$ may lead to a scale dependent correction in the wave function at origin (through $\alpha_s$).
    \item We proposed a phenomenological correction factor to account for the scale dependence of leptonic and diphoton widths and predicted the decay widths of heavy quarkonia. We notice that the scale dependence of $\alpha_{s}$ pose serious challenges to these processes as the determination of the suitable numerical value of $\alpha_{s}$ at different mass scales is non-trivial. We also observe that the variation of $\alpha_s$ itself is not sufficient to explain the experimental measurements, given the fact that $\alpha_{s}$ takes larger values to explain masses as compared to the decay widths. Thus, we introduce the scale dependent correction factor $f$ obtained from the experimental decay width of the lowest lying state decay process to estimate effective $|\Psi(0)|^2$ involved for a fixed value of $\alpha_{s}$ in concerned interaction process. We predicted the decay widths of higher excitations of charmonium and bottomonium through the incorporation of the scale dependent factor, which match well with available experimental results. Therefore, we treated  $\alpha_{s}$ and $|\Psi(0)|^2$ independently to estimate the scale dependent effects, knowledge of which could be crucial to bridge the gap between theory and experiment.
\end{itemize}
It may be noted that the decay width predictions for leptonic and diphoton decays are usually treated as estimates, owing to the model dependent uncertainties and the QCD corrections \cite{Godfrey:1985xj, Godfrey:2015dia}. However, we emphasize that any additional corrections involving higher powers of the strong coupling constant would be smaller in magnitude for heavy quarkonium systems. Besides this, the scale dependent effects are more prominent in the charmonium system than bottomonium. The ambiguities in the charmonium system could be influenced by its energy region, which may have dominant non-perturbative effects that are reflected in the wave function estimates. Regardless, our phenomenological scale correction factors considerably offset these ambiguities. Thus, we believe that our predictions of the properties of charmonium and bottomonium states are reliable and can prove useful in experimental searches.

\section*{Acknowledgments}
The authors gratefully acknowledge the financial support by the Department of Science and Technology (SERB:CRG/2018/002796), New Delhi, India.

\newpage

\begin{table}
    \centering
    \caption{Potential parameters in charmonium and bottomonium systems.}
    \label{tab:1_parameters}
\begin{tabular}{ccc}
        \toprule
        Parameters              & Charmonium    & Bottomonium \\
        \midrule
        $m_Q$ (in GeV)          & 1.5         & 4.8 \\
        $\alpha_s$              & 0.56         & 0.345 \\
        $\sigma$ (in GeV${}^2$) & 0.14         & 0.19 \\
        $\rho$ (in GeV)         & 1.19         & 3.54 \\
        $V_c$ (in GeV)          &     $-0.014$        & $-0.107$ \\
        \bottomrule
    \end{tabular}
\end{table}

\begin{longtable}{@{}cccccccccccc@{}}
\caption{Mass spectra of charmonium in MeV.\label{tab:2_c_mass_long}}\\
\toprule
State                      & Ours    & PDG\cite{ParticleDataGroup:2022pth} & NR\cite{Barnes:2005pb} & GI\cite{Barnes:2005pb} & NR\cite{Deng:2016stx} & SP\cite{Deng:2016stx} & SP\cite{Li:2009zu} & NR\cite{Soni:2017wvy} & NRC\cite{Kher:2018wtv} & SR\cite{Radford:2007vd} & LQCD\cite{Kalinowski:2015bwa} \\* \midrule
\endhead
\bottomrule
\endfoot
\endlastfoot
$\eta_{{c}}{(1 {}^1S_0)}$  & 2983.74 & $2983.9 \pm0.4$                     & 2982                   & 2975                   & 2983                  & 2984                  & 2979               & 2989                  & 2995                  & 2980.3  & $2884 \pm 50$                \\
$J/\psi{(1 {}^3S_1)}$      & 3096.68 & $3096.900 \pm0.006$                 & 3090                   & 3098                   & 3097                  & 3097                  & 3097               & 3094                  & 3094                  & 3097.36 & $3056 \pm 50$                \\
$\eta_{{c}}{(2 {}^1S_0)}$  & 3629.76 & $3637.7 \pm1.1$                     & 3630                   & 3623                   & 3635                  & 3637                  & 3623               & 3602                  & 3606                  & 3597.1  & $3535 \pm 55$                \\
$\psi{(2 {}^3S_1)}$        & 3676.89 & $3686.10 \pm0.06$                   & 3672                   & 3676                   & 3679                  & 3679                  & 3673               & 3681                  & 3649                  & 3685.5  & $3662 \pm 56$                \\
$\eta_{{c}}{(3 {}^1S_0)}$  & 4039.69 & -                                   & 4043                   & 4064                   & 4048                  & 4004                  & 3991               & 4058                  & 4000                  & 4014.0    & -                \\
$\psi{(3 {}^3S_1)}$        & 4072.73 & $4039 \pm1$                         & 4072                   & 4100                   & 4078                  & 4030                  & 4022               & 4129                  & 4036                  & 4094.9   & -               \\
$\eta_{{c}}{(4 {}^1S_0)}$  & 4376.70  & -                                   & 4384                   & 4425                   & 4388                  & 4264                  & 4250               & 4448                  & 4328                  & -        & -               \\
\bottomrule
\end{longtable}

\begin{longtable}{@{}ccccccccccc@{}}
\caption{Mass spectra of bottomonium in MeV.\label{tab:3_b_mass_long}}\\
\toprule
State                      & Ours     & PDG\cite{ParticleDataGroup:2022pth} & GI\cite{Godfrey:2015dia} & MGI\cite{Wang:2018rjg} & CQM\cite{Segovia:2016xqb} & SP\cite{Deng:2016ktl} & SP\cite{Li:2009nr} & NR\cite{Soni:2017wvy} & IMWC\cite{Pandya:2021vby} & NRC\cite{Kher:2022gbz} \\* \midrule
\endhead
\bottomrule
\endfoot
\endlastfoot
$\eta_{{b}}{(1 {}^1S_0)}$  & 9408.90  & $9398.7 \pm 2$                      & 9402                     & 9398                   & 9455                      & 9390                  & 9389               & 9428                  & 9412.22                   & 9423                   \\
$\Upsilon{(1 {}^3S_1)}$    & 9470.06  & $9460.4 \pm 0.1$                   & 9465                     & 9463                   & 9502                      & 9460                  & 9460               & 9463                  & 9460.75                   & 9463                   \\
$\eta_{{b}}{(2 {}^1S_0)}$  & 9984.03  & $9999 \pm 4 $                       & 9976                     & 9989                   & 9990                      & 9990                  & 9987               & 9955                  & 9995.48                   & 9983                   \\
$\Upsilon{(2 {}^3S_1)}$    & 10010.15 & $10023.4 \pm 0.5  $                & 10003                    & 10017                  & 10015                     & 10015                 & 10016              & 9979                  & 10026.22                  & 10001                  \\
$\eta_{{b}}{(3 {}^1S_0)}$  & 10334.96 & -                                   & 10336                    & 10336                  & 10330                     & 10326                 & 10330              & 10338                 & 10339.00                     & 10342                  \\
$\Upsilon{(3 {}^3S_1)}$    & 10354.28 & $10355.1 \pm 0.5$                   & 10354                    & 10356                  & 10349                     & 10343                 & 10351              & 10359                 & 10364.65                  & 10354                  \\
$\eta_{{b}}{(4 {}^1S_0)}$  & 10620.33 & -                                   & 10623                    & 10597                  & -                         & 10584                 & 10595              & 10663                 & 10572.49                  & 10638                  \\
\bottomrule
\end{longtable}

\begin{table}
\setlength{\tabcolsep}{10pt}
\centering
\caption{{Square of the radial wave function at the origin $|R^{(l)}_{nl}(0)|^2$ for charmonium states in GeV${}^{3+2l}$.}}
\label{tab:4_c_states_wfo}
\begin{tabular}{cccc}
\toprule
State &  Ours & \cite{Liao:2014rca} & \cite{Eichten:2019hbb} \\
\midrule
$1S$ &      1.4856 &    0.974 & 1.0952 \\
$2S$ &      0.7956 &    0.889 & 0.6966 \\
$3S$ &      0.6455 &    0.821 & 0.5951 \\
$4S$ &      0.5752 &    0.807 & 0.5461 \\
$5S$ &      0.5326 &    0.812 & 0.5160 \\
$6S$ &      0.5034 &    0.842 &   -    \\
\midrule
$1P$ &      0.1007 &    0.091 & 0.1296 \\
$2P$ &      0.1338 &    0.169 & 0.1767 \\
$3P$ &      0.1572 &    0.244 & 0.2106 \\
$4P$ &      0.1764 &    0.320 & 0.2389 \\
$5P$ &      0.1933 &    0.376 &   -    \\
$6P$ &      0.2085 &     -    &   -    \\
\midrule
$1D$ &      0.0159 &    0.036 & 0.0329 \\
$2D$ &      0.0331 &    0.104 & 0.0692\\
$3D$ &      0.0510 &    0.202 & 0.1074 \\
$4D$ &      0.0696 &    0.328 &   -    \\
$5D$ &      0.0886 &     -    &   -    \\
$6D$ &      0.1080 &     -    &   -    \\
\bottomrule
\end{tabular}
\end{table}

\begin{table}
\centering
\setlength{\tabcolsep}{10pt}
\caption{{Square of the radial wave function at the origin $|R^{(l)}_{nl}(0)|^2$ for bottomonium states in GeV${}^{3+2l}$.}}
\label{tab:5_b_states_wfo}
\begin{tabular}{cccc}
\toprule
State &  Ours   & \cite{Liao:2014rca} & \cite{Eichten:2019hbb} \\
\midrule
 $1S$ &      9.1909 &   9.140 & 5.8588 \\
 $2S$ &      4.3130 &   4.771 & 2.8974 \\
 $3S$ &      3.3768 &   3.901 & 2.2496 \\
 $4S$ &      2.9557 &   3.499 & 1.9645 \\
 $5S$ &      2.7059 &   3.324 & 1.7990 \\
 $6S$ &      2.5363 &   3.183 & 1.6885 \\
 \midrule
 $1P$ &      1.5663 &   1.218 & 1.6057 \\
 $2P$ &      1.9741 &   1.667 & 1.8240 \\
 $3P$ &      2.2554 &   1.961 & 1.9804 \\
 $4P$ &      2.4870 &   2.325 & 2.1175 \\
 $5P$ &      2.6904 &   2.613 & 2.2430 \\
 $6P$ &      2.8748 &   2.886 & 2.3600 \\
 \midrule
 $1D$ &      0.6325 &   0.732 & 0.8394 \\
 $2D$ &      1.2798 &   1.632 & 1.5572 \\
 $3D$ &      1.9357 &   2.643 & 2.2324 \\
 $4D$ &      2.6014 &   3.753 & 2.8903 \\
 $5D$ &      3.2772 &   5.164 & 3.5411 \\
 $6D$ &      3.9632 &    -    &  -     \\
\bottomrule
\end{tabular}
\end{table}

\begin{table}
    \centering
    \caption{Leptonic decay widths (in units of keV) of charmonium states at lowest order, with QCD correction, and with scale dependent factor. For other theoretical models, the values inclusive of QCD corrections are given in parentheses where available.}
    \label{tab:6_charm_lep_dw}
    \begin{tabular}{@{}c|ccc|ccccc@{}}
\toprule
State                 & $\Gamma_{e^+e^-}$ & $\Gamma_{e^+e^-}^{corr}$ & $f\Gamma_{e^+e^-}^{corr}$ & PDG\cite{ParticleDataGroup:2022pth} & SP\cite{Li:2009zu} & NR\cite{Soni:2017wvy} & NRC\cite{Kher:2018wtv} & SR\cite{Radford:2007vd} \\ \midrule
$J/\psi{(1 {}^3S_1)}$ & 14.6742           & 7.2007                   & 5.5301                    & $5.53 \pm 0.10$\footnote{Used as input to calculate $f$ for $S$-wave states.}                     & 11.8(6.60)          & 2.925                 & 8.335(3.623)           & 4.28                    \\
$\psi{(2 {}^3S_1)}$   & 5.5742            & 2.7353                   & 2.1007                    & $2.29 \pm 0.06$                     & 4.29(2.40)          & 1.533                 & 2.496(1.085)           & 2.25                    \\
$\psi{(3 {}^3S_1)}$   & 3.6862            & 1.8088                   & 1.3892                    & $0.86 \pm 0.07$\footnote{$\psi(3^3S_1) \equiv \psi(4040)$}                     & 2.53(1.42)         & 1.091                 & 1.722(0.748)           & 1.66                    \\
$\psi{(4 {}^3S_1)}$   & 2.8101            & 1.3789                   & 1.0590                     & -                                   & 1.73(0.97)         & 0.856                 & 1.378(0.599)           & 1.33                    \\
$\psi{(5 {}^3S_1)}$   & 2.2869            & 1.1222                   & 0.8618                    & -                                   & 1.25(0.70)          & 0.707                 & 1.168(0.508)           & -                       \\
$\psi{(6 {}^3S_1)}$   & 1.9333            & 0.9487                   & 0.7286                    & -                                   & 0.88(0.49)         & 0.602                 & 1.017(0.442)           & -                       \\
\bottomrule
\end{tabular}
\end{table}

\begin{table}
    \centering
    \caption{Leptonic decay widths of bottomonium at lowest order, with QCD correction, and with scale dependent factor ($\Gamma({n}^3S_1 \rightarrow e^+e^-)$ in keV and $\Gamma({n}^3D_1 \rightarrow e^+e^-)$ in eV). For other theoretical models, the values inclusive of QCD corrections are given in parentheses where available.}
    \label{tab:7_bottom_lep_dw}
\scalebox{0.9}{
\begin{tabular}{@{}c|ccc|cccccccc@{}}
\toprule
State                     & $\Gamma_{e^+e^-}$ & $\Gamma_{e^+e^-}^{corr}$ & $f\Gamma_{e^+e^-}^{corr}$ & PDG \cite{ParticleDataGroup:2022pth} & GI\cite{Godfrey:2015dia} & MGI\cite{Wang:2018rjg} & CQM\cite{Segovia:2016xqb} & SP\cite{Li:2009nr} & NR\cite{Soni:2017wvy} & IMWC\cite{Pandya:2021vby} & NRC\cite{Kher:2022gbz} \\ \midrule
$\Upsilon{(1 {}^3S_1)}$   & 2.4268            & 1.6028                   & 1.3399                    & $1.340 \pm0.018$\footnote{Used as input to calculate $f$ for $S$-wave states.}                     & 1.44                     & 1.65                   & 0.71                      & 2.31(1.60)          & 1.098                 & 1.1191(0.7700)              & 1.074(0.582)           \\
$\Upsilon{(2 {}^3S_1)}$   & 1.0192            & 0.6732                   & 0.5628                    & $0.612 \pm0.011$                     & 0.73                     & 0.821                  & 0.37                      & 0.92(0.64)         & 0.67                  & 0.7859(0.5442)            & 0.363(0.197)           \\
$\Upsilon{(3 {}^3S_1)}$   & 0.7458            & 0.4926                   & 0.4118                    & $0.443 \pm0.008$                     & 0.53                     & 0.569                  & 0.27                      & 0.64(0.44)         & 0.541                 & 0.6171(0.4288)            & 0.274(0.149)           \\
$\Upsilon{(4 {}^3S_1)}$   & 0.6186            & 0.4086                   & 0.3416                    & $0.272 \pm0.029$                     & 0.39                     & 0.431                  & 0.21                      & 0.51(0.35)         & 0.47                  & 0.5096(0.3549)            & 0.238(0.129)           \\
\bottomrule
\end{tabular}
}
\end{table}

\begin{table}
    \caption{Diphoton decay widths (in units of keV) of pseudoscalar (${}^1S_0$), scalar (${}^3P_0$), and tensor (${}^3P_2$) charmonium states at lowest order, with QCD correction, and with scale dependent factor. For other theoretical models, the values inclusive of QCD corrections are given in parentheses where available.}
    \label{tab:8_charm_2photon_dw}
    \centering
\begin{tabular}{@{}c|ccc|cccc@{}}
\toprule
State                      & $\Gamma_{\gamma\gamma}$ & $\Gamma_{\gamma\gamma}^{corr}$ & $f_{\gamma}\Gamma_{\gamma\gamma}^{corr}$ & PDG\cite{ParticleDataGroup:2022pth} & SP\cite{Li:2009zu} & NR\cite{Soni:2017wvy} & NRC\cite{Kher:2018wtv} \\ \midrule
$\eta_{{c}}{(1 {}^1S_0)}$  & 20.8470                  & 14.0785                        & 5.4061                         & $5.4 \pm0.4$\footnote{Used as input to calculate $f_{\gamma}$ for $\eta_c$ states.}                      & 8.50                & 7.231                 & 10.351(6.621)          \\
$\eta_{{c}}{(2 {}^1S_0)}$  & 11.1646                 & 7.5397                         & 2.8952                          & -                                   & 2.40                & 5.507                 & 4.501(2.879)           \\
$\eta_{{c}}{(3 {}^1S_0)}$  & 9.0582                  & 6.1172                         & 2.3490                          & -                                   & 0.88               & 4.971                 & 3.821(2.444)           \\
$\eta_{{c}}{(4 {}^1S_0)}$  & 8.0710                   & 5.4506                         & 2.0930                         & -                                   & -                  & 4.688                 & 3.582(2.291)           \\
$\eta_{{c}}{(5 {}^1S_0)}$  & 7.4738                  & 5.0473                         & 1.9382                          & -                                   & -                  & 4.507                 & 3.460(2.213)            \\
$\eta_{{c}}{(6 {}^1S_0)}$  & 7.0633                  & 4.7700                         & 1.8317                          & -                                   & -                  & 4.377                 & 3.378(2.161)           \\
\bottomrule
\end{tabular}
\end{table}

\begin{table}
    \caption{Diphoton decay widths (in units of keV) of pseudoscalar (${}^1S_0$), scalar (${}^3P_0$), and tensor (${}^3P_2$) bottomonium states at lowest order and with QCD correction. For other theoretical models, the values inclusive of QCD corrections are given in parentheses where available.}
    \label{tab:9_bottom_2photon_dw}
    \centering
\begin{tabular}{@{}c|cc|ccccccc@{}}
\toprule
State                      & $\Gamma_{\gamma\gamma}$ & $\Gamma_{\gamma\gamma}^{corr}$ & GI\cite{Godfrey:2015dia} & MGI\cite{Wang:2018rjg} & CQM\cite{Segovia:2016xqb} & SP\cite{Li:2009nr} & NR\cite{Soni:2017wvy} & IMWC\cite{Pandya:2021vby} & NRC\cite{Kher:2022gbz} \\ \midrule
$\eta_{{b}}{(1 {}^1S_0)}$  & 0.7872                  & 0.6168                                                   & 0.94                     & 1.05                   & 0.69                      & 0.527              & 0.387                 & 0.3782(0.3035)            & 0.3364(0.2361)         \\
$\eta_{{b}}{(2 {}^1S_0)}$  & 0.3694                  & 0.2894                                                   & 0.41                     & 0.489                  & 0.36                      & 0.263              & 0.263                 & 0.2636(0.2122)            & 0.1270(0.0896)         \\
$\eta_{{b}}{(3 {}^1S_0)}$  & 0.2892                  & 0.2266                                                   & 0.29                     & 0.323                  & 0.27                      & 0.172              & 0.229                 & 0.2067(0.1668)            & 0.1029(0.0726)         \\
$\eta_{{b}}{(4 {}^1S_0)}$  & 0.2531                  & 0.1984                                                   & 0.20                      & 0.237                  & -                         & 0.105              & 0.212                 & 0.1705(0.1378)            & 0.0942(0.0666)         \\
$\eta_{{b}}{(5 {}^1S_0)}$  & 0.2318                  & 0.1816                                                   & 0.17                     & 0.192                  & -                         & -                  & 0.201                 & 0.1455(0.1176)            & 0.0900(0.0636)         \\
\bottomrule
\end{tabular}
\end{table}

\FloatBarrier

\end{document}